\documentclass[12pt]{article}
\usepackage{amsfonts}
\usepackage{amsmath}
\usepackage{amssymb}
\usepackage{graphicx}
\usepackage{rotating}
\usepackage[round]{natbib}
\usepackage{setspace}
\usepackage{multirow}
\usepackage{array}
\usepackage{geometry}
\usepackage{bezier}
\usepackage{booktabs}
\usepackage{tabularx}
\usepackage{subcaption}
\usepackage[usenames,dvipsnames]{color}
\usepackage[table]{xcolor}

\usepackage[pdftex, bookmarksopen=true, bookmarksnumbered=true,
pdfstartview=FitH, breaklinks=true, urlbordercolor={0 1 0}, citebordercolor={0 0 1}]{hyperref}
  
\newcommand{\ind}{\mathbb{I}}

\newcommand{\by}{\mathbf{y}}
\newcommand{\bY}{\mathbf{Y}}
\newcommand{\btheta}{\boldsymbol{\theta}}
\newcommand{\blambda}{\boldsymbol{\lambda}}
\newcommand{\bM}{\mathbf{M}}
\newcommand{\bA}{\mathbf{A}}
\newcommand{\bb}{\mathbf{b}}
\newcommand{\bmu}{\boldsymbol{\mu}}
\newcommand{\bL}{\mathbf{L}}
\newcommand{\bU}{\mathbf{U}}

\title{Measurement That Matches Theory: Theory-Driven Identification in IRT Models}

\author{Marco Morucci\thanks{New York University, marco.morucci@nyu.edu}\and Margaret J. Foster \thanks{Duke University, margaret.foster@duke.edu}\and Kaitlyn Webster \thanks{Independent scholar, kmwebster819@gmail.com}\and So Jin Lee \thanks{Harvard University, sojinlee@hks.harvard.edu}\and David A. Siegel \thanks{Duke University, david.siegel@duke.edu}
\thanks{The authors thank Miranda Ding, Cat Jeon, Yangfan Ren, and Priya Subramanian for excellent research assistance. This material is based upon work supported by the National Science Foundation under grant no. SES-1727249.}}

\begin{document}
\maketitle
\begin{abstract} 
Measurement is the weak link between theory and empirical test. Complex concepts such as ideology, identity, and legitimacy are difficult to measure; yet, without measurement that matches theoretical constructs, careful empirical studies may not be testing that which they had intended. Item Response Theory (IRT) models offer promise by producing transparent and improvable measures of latent factors thought to underlie behavior. Unfortunately, those factors have no intrinsic substantive interpretations. Prior solutions to the substantive interpretation problem require exogenous information about the units, such as legislators or survey respondents, which make up the data; limit analysis to one latent factor; and/or are difficult to generalize. We propose and validate a solution, IRT-M, that produces multiple, potentially correlated, generalizable, latent dimensions, each with substantive meaning that the analyst specifies before analysis to match theoretical concepts. We offer an R package and step-by-step instructions in its use, via an application to survey data.
\end{abstract}
\doublespacing
\section*{Introduction}

Modern social science is theoretically rich and methodologically sophisticated. Measurement, the bridge between theory and empirics, lags behind. That is understandable given the complexity of social-scientific theoretical concepts. Democracy, ideology, legitimacy, power, identity: none of those are easy to pin down with simple proxies. Yet, without measurement that matches theory, careful empirical studies may not truly be testing what they were intended to test (e.g., \citealt{adcock2001measurement, barber2021comparing, pietryka2021anes, reuning2019exploring}). Moreover, without measures that maintain their meanings across time and place, scholars working with the same theoretical concepts but in different systems cannot build on each others' work. We aim to improve the connection between theory and measurement by proposing and validating a solution, IRT-M, that derives multiple, potentially correlated, generalizable, latent dimensions from data. Each dimension IRT-M produces possesses a substantive meaning that the analyst specifies before analysis to match theoretical concepts, ensuring the link between theory and measurement.

Ameliorating the measurement problem requires transparently constructed and improvable measures of theoretical concepts. Dimensional-reduction techniques, such as Item Response Theory (IRT) models, assign to each unit in the data a position along each of one or more latent dimensions. Units can range from survey respondents, to legislators, to nations. Positions in latent space are chosen by the model to best predict each unit's responses to a series of items. Thus, positions along latent dimensions predict individuals' behavior. IRT models are transparently constructed as long as their inputs are public. They are improvable as new data become available, via Bayesian estimation. Perhaps for those reasons, IRT models are increasingly employed in the social sciences as measurement tools. While they most frequently have been applied to legislative or judicial ideal point estimation \citep{poole1985, martin2002dynamic, bailey2018two}, they have also been used to derive a variety of measures relating to regime traits \citep{marquardt2019makes} or state capacity \citep{hanson2013leviathan}, qualities of human rights \citep{schnakenberg2014dynamic, hill2016avoiding} or wartime sexual violence \citep{kruger2020latent}, interstate hostility \citep{terechshenko2020hot}, states’ preferences over investor protection \citep{montal2020states}, peace frameworks \citep{williams2019latent}, leaders' willingness to use force \citep{carter2020framework}, state trade legislation \citep{lee2019exports}, international norms \citep{girard2021reconciling}, media freedom \citep{solis2020measuring}, and women's inclusion, rights, and security \citep{karim2018gender}.

Unfortunately, dimensional-reduction techniques do not fully solve the measurement problem because they don't intrinsically capture theoretical concepts: the latent dimensions that best predict the data need not match the theoretical concepts of interest, or even have any clear substantive meaning at all. One way the latter might occur would be if the latent dimensions found by an IRT model were complex combinations of clear theoretical concepts, so that each dimension would be difficult to interpret on its own. That is particularly likely in low-dimensional latent spaces that are intended to explain complicated social behavior, a point raised in \cite{aldrich2014} and to which we return in our fourth section.

The problem of substantive interpretation is exacerbated by the lack of modeled correlations between dimensions in typical IRT models, since theoretical concepts are often correlated. For example, we may believe theoretically that voting is driven by ideological positions along economic and social dimensions. If so, in order to test the theory we would need a two-dimensional IRT model to return values along those two ideological dimensions. However, if partisanship strongly predicts voting behavior along both economic and social ideological dimensions, the IRT model might instead return one latent dimension that is a combination of economic and social ideological positions, and a second latent dimension, uncorrelated with the first, that either has no clear theoretical meaning or that carries a meaning unrelated to the theory being tested. We discuss that point further in the analysis of roll call data that forms part of our model validation.

The problem of mismatched theory and measure is particularly evident when traditional IRT methods are used to analyze responses to surveys designed with the specific intent of measuring theoretical concepts. In such cases, the survey design includes information about the manner in which survey questions tie to concepts of interest. Yet, that information is not used when computing latent dimensions from survey responses in a traditional IRT model. In fact, traditional IRT methods can estimate latent dimensions that capture none of the carefully established links between theoretical concepts and survey responses, simply because there exists a model that fits the data better than one that takes the question design into account.

Prior solutions to the problem of substantive interpretation---that is, solutions intended to avoid incorrectly characterizing the substantive meaning of latent dimensions derived from IRT and other dimensional-reduction techniques---have generally taken three forms. The first involves the use of additional information about units in the data, information exogenous to the data source from which the latent dimensions will be derived. One common example of that occurs in the placement of legislators in a two-dimensional space. Legislative voting patterns are only one of many data sources that can speak to legislators' ideological positions; others include speeches and donor behavior. Analysts can use such exogenous information to fix the positions of a small subset of well-known legislators; voting behavior is then sufficient to discern the others' positions. The second form of solution uses complementary methods to extract more information from the same data source from which the latent dimensions will be derived. One example of that makes use of LDA topic models to associate bill, judicial decision, or metadata text with issues, and those issues with latent dimensions revealed by voting patterns \citep{gerrish2011predicting, gerrish2012they, lauderdale2014scaling}.

Both forms of solution represent advances over the standard IRT approach in that they enable substantive interpretation of the discovered latent dimensions. However, both are limited in their range of applicability, for related reasons.

First, the data requirements for the application of either solution are not always, or even often, met. The first solution requires exogenous information on units in the data, and as such is often not available. The second requires that the conditions necessary for application of the complementary method be met. In the case of LDA topic modeling, that means either long texts or a limited issue space. The typical anonymous survey---a common source of individual-level data---usually fails to satisfy either requirement.

Second, neither solution guarantees accurate characterization of the substantive meaning of the discovered latent dimensions. In the case of exogenous information, one might use public speeches, for example, to specify that two legislators are positioned at opposite extremes in each dimension of a two-dimensional latent space capturing economic and social ideology. However, both the IRT's output and the public speeches would also be consistent with a two-dimensional latent space in which one dimension corresponded to partisanship and the other to any other topic over which the two legislators had opposite positions. In that case, \textit{neither} dimension might capture social or economic ideological positions. In the case of complementary methods, one is limited by the inherent limitations of the complementary method. For example, nearly all text models suffer from the difficulty of identifying stance and tone \citep{bestvater2023sentiment}.

Third, the meanings of the dimensions suggested by both exogenous information and complementary methods can change over time or differ by place or data source. For example, in the context of exogenous information, two latent dimensions of ideology at some time and place might capture economic and civil rights concerns \citep{poole1985}, security and religion \citep{schofield2005multiparty}, or agreement with the Western liberal order and North-South conflict \citep{bailey2018two}. But those can change over time, as did the link between the second latent dimension and civil rights in Congress or the North-South dimension in U.N. voting \citep{bailey2018two}. Should there be common units across time, place, or data source, then one can standardize measures via those units, but that requires the additional assumption that those units are not changing latent positions over time, an assumption not available in many contexts. In the context of complementary methods, many such methods lack a way to standardize across contexts. For instance, topic models inductively label scales based on what the topic model finds in the data. The degree to which the models shape the uncovered topics varies based on the degree of supervision as well as other preprocessing decisions \citep{denny2018text}, and the topics that best predict texts might vary across time, space, and data source, leading to different impressions of ground truth \citep{foster2023change}. 

The third form of solution to the problem of substantive interpretation avoids many of those concerns by carefully selecting and employing response data that is thought to be related only to a single latent dimension. That approach has been used productively, such as in measuring leaders' willingness to use force \citep{carter2020framework}, the strength of norms \citep{girard2021reconciling}, respect for human rights \citep{schnakenberg2014dynamic}, and peace agreements \citep{williams2019latent}. Unfortunately, no matter how careful the selection, it comes with one significant limitation: analysts cannot allow their response data to be generated by more than one latent dimension. There are many cases in which one dimension might be sufficient, particularly as a first approximation. For example, if partisanship theoretically were to drive most voting behavior, a one-dimensional measure of partisan attachment might be sufficient to predict most voting behavior. In that case, a one-dimensional IRT model run on voting data might produce a good measure of partisan attachment. \cite{fariss2014respect} makes such an argument with respect to physical integrity rights.

However, very often the theories we want to test contain multiple theoretical concepts that are imperfectly correlated. One could attempt to measure multiple latent dimensions by estimating one dimension at a time, with each measurement using a different subset of the data. The subsets would have to be non-overlapping in that case: otherwise the assumption that one latent dimension is sufficient to predict the data would be violated for the overlapping indicators. We view that level of precision across subsets as rare. Further, even should there be such non-overlapping subsets, that estimation procedure does not model correlation between latent dimensions, rendering it poorly able to capture imperfectly correlated theoretical concepts. As imperfectly correlated theoretical concepts are common in the social sciences---e.g., legitimacy and democracy---we view the range of applicability of the one-dimensional solution as somewhat limited.

We offer a novel solution to the measurement problem via a model that enables transparent substantive interpretation across multiple, possibly correlated, latent dimensions. It does so without needing to leverage either exogenous information about individual units in the data or complementary methods. Our solution, which we call IRT-M, is a semi-supervised approach based on Bayesian Item Response Theory. In the spirit of the one-dimensional solution, it employs careful selection of response data. In IRT-M, the analyst identifies before the analysis how each latent dimension is supposed to affect each response to each item, and the model uses that information to output a set of theoretically-defined latent dimensions that may also be imperfectly correlated. Thus, the range of applicability of IRT-M is considerably broader than that of existing solutions to the problem of substantive interpretation.

The IRT-M model takes as input data in which each of a set of units (e.g., survey respondents; legislators; peace treaties) respond to an array of items (e.g., answers to survey questions; votes on bills; elements of treaties). The present iteration of IRT-M requires dichotomous items, with any non-dichotomous item needing to be reformatted into a series of dichotomous ones, but extensions to more general data, as we discuss further, are straightforward. In the pre-analysis step, the analyst hand-codes each item according to its connection to each latent dimension, recalling that latent dimensions capture specific theoretical concepts from one's theory (e.g., perception of threat from immigration; ideological position; degree to which a peace treaty captures minority rights and security). In that coding, each item-latent dimension pair is assigned a value of $1,-1,0$ or \texttt{NA}. A $1$ ($-1$) indicates that a unit with greater (lesser) values along that dimension is more likely to respond positively to that item. A $0$ indicates that a unit's value along that dimension does not predict that unit's response to that item. An \texttt{NA} indicates that the analyst has no prior belief about whether that dimension influences one's response to that item. The M in IRT-M represents the constraint matrices that capture those coding rules; the rules capture the theory, ensuring that the measurement matches the theory.

Thus, one's theory plays a key role in the IRT-M approach. If the theory captures some aspect of the process that generated the data used to derive the latent dimensions, and if the coding is applied consistently, the model will produce measures of relevant theoretical concepts that are constant in meaning across disparate data sources and across time and place. For example, a second ideological dimension coded to capture civil rights will maintain that substantive meaning across time, even if it becomes less predictive of variation in the data. In contrast, if the theory does not capture the data generating process well, then the coding should reveal that by assigning values of $0$ or \texttt{NA} to many item-latent dimension pairs. The model therefore ensures that measurement matches theory both positively, by constraining the IRT model to connect items to theoretically-determined latent dimensions, and negatively, by not using items in analysis that are unrelated to the theory at hand.

IRT-M can be used in any setting in which one would use an unsupervised IRT model, as coding all item-dimension pairs as \texttt{NA} when there is no theory turns IRT-M into an unsupervised model. However, it is particularly well-suited when there are clear underlying theoretical mechanisms driving the responses of units to items. We use a motivating example drawn from \cite{kentmen2017anti} both to make that point and to provide an intuitive introduction and step-by-step guide to IRT-M. \cite{kentmen2017anti} describe how in surveys of Europeans, attitudes toward immigration relate to perceptions of different threats that might be induced by immigration, including economic, cultural, and religious threats. We coded the February-March 2021 wave of Eurobarometer for questions related to those three threats, an additional health threat, since the survey took place during the Covid-19 pandemic, and the two outcome attitudes of support for immigration and support for the European Union. We then employed IRT-M to derive posterior distributions of survey respondents' positions along each of the six latent dimensions. We show that, given sufficient related questions, our measures of latent dimensions possess construct validity and our latent threat dimensions are correlated with each other. We also find some correlation of threats to attitudes, particularly support for the EU.

The section following our motivating example offers a more technical presentation of IRT-M in brief and suggests how it may be easily extended to non-dichotomous items. Our fourth section presents two forms of validation of IRT-M. The first employs simulated data to show that IRT-M yields a reduction in error as compared to established approaches to dimensional reduction while still possessing similar convergence properties. Further, that error remains comparable or lower to established approaches even under substantial misspecification of the model. In other words, one need not perfectly code one's theoretical concepts into a constraint matrix in order to accurately measure them. 

The second form of validation applies IRT-M to roll call votes in both the $85^{th}$ and $109^{th}$ House and Senate. We hand-code all votes in order to employ IRT-M, rather than use exogenous information on a subset of legislators, as does DW-NOMINATE. In doing so, we recover legislators' placements on two theoretically meaningful latent dimensions that are broadly consistent with positions derived from DW-NOMINATE, particularly along the first, theoretically clearer, latent dimension corresponding roughly to economic ideology. At the same time, we also produce a theoretically clear second latent dimension that is usually strongly correlated with the first. That correlation suggests a role for partisan voting behavior. We also, by varying our coding rules, illustrate the manner in which the theory drives the measurement. Together, our two validation exercises indicate that although one need not be perfect in coding the constraint matrix to make use of IRT-M, straying too far from the latent concepts in question risks measuring different concepts entirely.

Though using IRT-M entails an up-front cost in coding a constraint matrix based on one's theory, it produces measurements of positions on conceptually-meaningful latent dimensions while eliminating the need for exogenous information about units in the data to identify the model. That opens up numerous applications, which we briefly discuss in the conclusion. We also provide an R package that will allow analysts to employ the IRT-M framework and model in their own analyses.\footnote{Package to be made available subsequent to this manuscript's acceptance for publication.}

\section*{Applying the IRT-M Model}

The IRT-M model produces a set of latent dimensions whose substantive meaning derives from the theoretical concepts underlying pre-analysis coding rules. Thus, applying the IRT-M model begins with identifying a theory that possesses a set of theoretical concepts that one desires to measure. IRT-M will translate each concept to a corresponding latent dimension and compute a latent position for each data unit on each dimension. To illustrate application of IRT-M, we use a motivating example drawn from \cite{kentmen2017anti}, which offers an overview of anti-immigration attitudes in Europe. \cite{kentmen2017anti} also calls explicitly, in the abstract, for scholars to ``employ methodological techniques that capture the underlying constructs associated with attitude and public opinion,'' a task for which IRT-M is ideally suited.

\cite{kentmen2017anti} elaborate on the manner by which perceptions of threat partially determine one's attitudes toward both immigration and the European Union. We draw three distinct dimensions of threat from their discussion of the literature: a sense of economic threat, a sense of cultural threat, and a sense of religious threat. Together with the two attitudes those threat perceptions are theorized to influence, we thus have five theoretical concepts: economic threat, cultural threat, religious threat, support of immigration, and support of the EU. Each of those five concepts corresponds to a substantively meaningful latent dimension, and the theory indicates that people's values on the first three latent dimensions influence their values on the last two.

The next step in applying the IRT-M model is to identify a data source comprising items that one believes would be informative about individual data units' positions along the latent dimensions identified by the theory. In the context of our example, we require data that are informative about each of the three threats and the two attitudes. We identified the Eurobarometer survey for that purpose, and specifically chose the February-March 2021 (94.3) wave due to availability \citep{EB9432021}. In reading through the survey codebook, it became clear that there was a fourth sense of threat, specific to the Covid-19 pandemic, for which the survey was informative and which might have also influenced attitudes toward immigration and the EU along the same lines as those discussed in \cite{kentmen2017anti}: health threat. Thus, we added that latent dimension to our list, giving us six in all.

After identifying the latent theoretical dimensions of interest and a candidate data source, the next step is to produce the constraint matrix. We accomplish that here by coding the items in the data according to their expected relationships to the underlying theoretical dimensions. Each item in the data is an opportunity for the unit to express its positions along the latent dimensions specified in the theory. In the context of the Eurobarometer survey, an item is a possible answer to a survey question, and a positive response to that item is choosing that answer to the question. Each answer to a question is an item as all items must be dichotomous. Converting all non-dichotomous options to a set of dichotomous items is straightforward. For instance, a five-level feeling thermometer survey question would be broken down into five separate dichotomous items. That is known in the machine-learning literature as One Hot encoding. Similarly, a continuous variable would be made dichotomous through the use of thresholds. In the next section we discuss how to generalize our model for more general items. In other contexts, responses to items may be votes on bills or resolutions, or the presence of particular elements of treaties, constitutions, or other documents.

After creating an array of dichotomous items, one assigns to each item-latent dimension pair one of four possible values: $1, -1$, $0$, or \texttt{NA}. One assigns a $1$ if a positive response to that item would be predicted by a greater value along that latent dimension. Conversely, one assigns a $-1$ if a positive response to that item would be predicted by a lesser value along that latent dimension. If one believes that values on that latent dimension do not influence the likelihood of a positive response to that item, one assigns a $0$. Finally, if one truly has no prior belief about whether and how values on that latent dimension influence the likelihood of a positive response to that item, IRT-M allows the user to assign \texttt{NA} to the item-latent dimension pair.

Returning to our motivating example, consider the latent dimension of economic threat. We would code a survey question that asks about one's personal job situation as relevant to that underlying dimension. If the survey question had four possible responses, ranging from ``very good'' to ``very bad,'' then there would be four item-latent dimension pairs to code. We would code the two ``bad'' responses as $1$ and the two ``good'' responses as $-1$, since they would be predicted by more and less perception of threat, respectively. In contrast, a survey question asking for one's opinion about a common European Asylum system would be coded as $0$ along the economic threat dimension, since we do not view responses to that question as influenced appreciably by one's sense of economic threat.

In coding the Eurobarometer survey wave, we are implicitly assuming that the theoretical concepts we are trying to measure are constant across both space and time, at least during the time the survey was fielded. We are also assuming that the connection between those concepts and the survey questions is constant across both space and time. We view those assumptions as fair for our motivating example, but what would happen if either were violated?

The assumption of constant theoretical concepts is a necessary one for the application of IRT-M. Should concepts not apply equally across space or should they change over time, then the application of IRT-M should be restricted to those regions or times in which the concepts are uniform. Otherwise, it is not clear what concept one would be measuring. In contrast, it is not a problem for IRT-M if the tie between concepts and items varies across space or time, as long as the coding rules capture that variation. In some cases, there may be appreciable variation in coding rules across time and space \citep{fariss2014respect, fariss2019yes}. For example, a survey question regarding discrimination might change in meaning across time due to shifts in a country’s demographics, or might mean different things in different countries. In such cases, the coding rules should assign a value to each item-latent dimension pair appropriate to each country, at each time. In other cases, particularly when items are presented to samples of the same population in the same context over time, as they often are for repeated surveys, coding rules may be constant across time. In other words, coding rules should be time and space dependent when necessary.

While we expect exact coding rules to vary between coders---connecting items to theoretical concepts allows for subjective assessment---our simulations, described subsequently, illustrate that IRT-M's performance is not significantly reduced even when the constraints are only $50-75\%$ correct. Thus, based on our simulation, the model can still return reasonable estimates even with partially mis-specified constraints.

\begin{figure}
     \centering
     \begin{subfigure}[b]{0.48\textwidth}
         \centering
         \includegraphics[width=\textwidth]{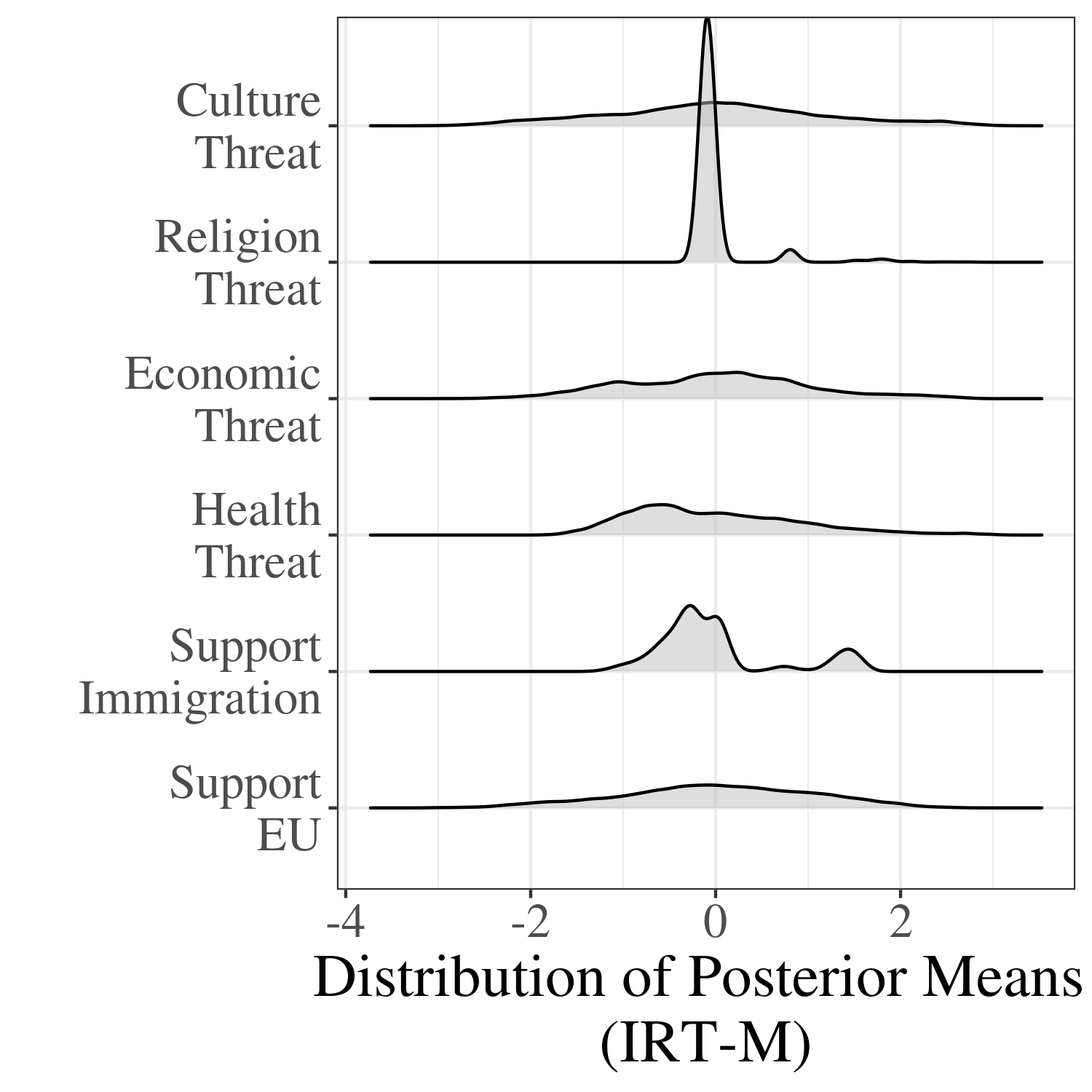}
     \end{subfigure}%
     \begin{subfigure}[b]{0.48\textwidth}
         \centering
         \includegraphics[width=\textwidth]{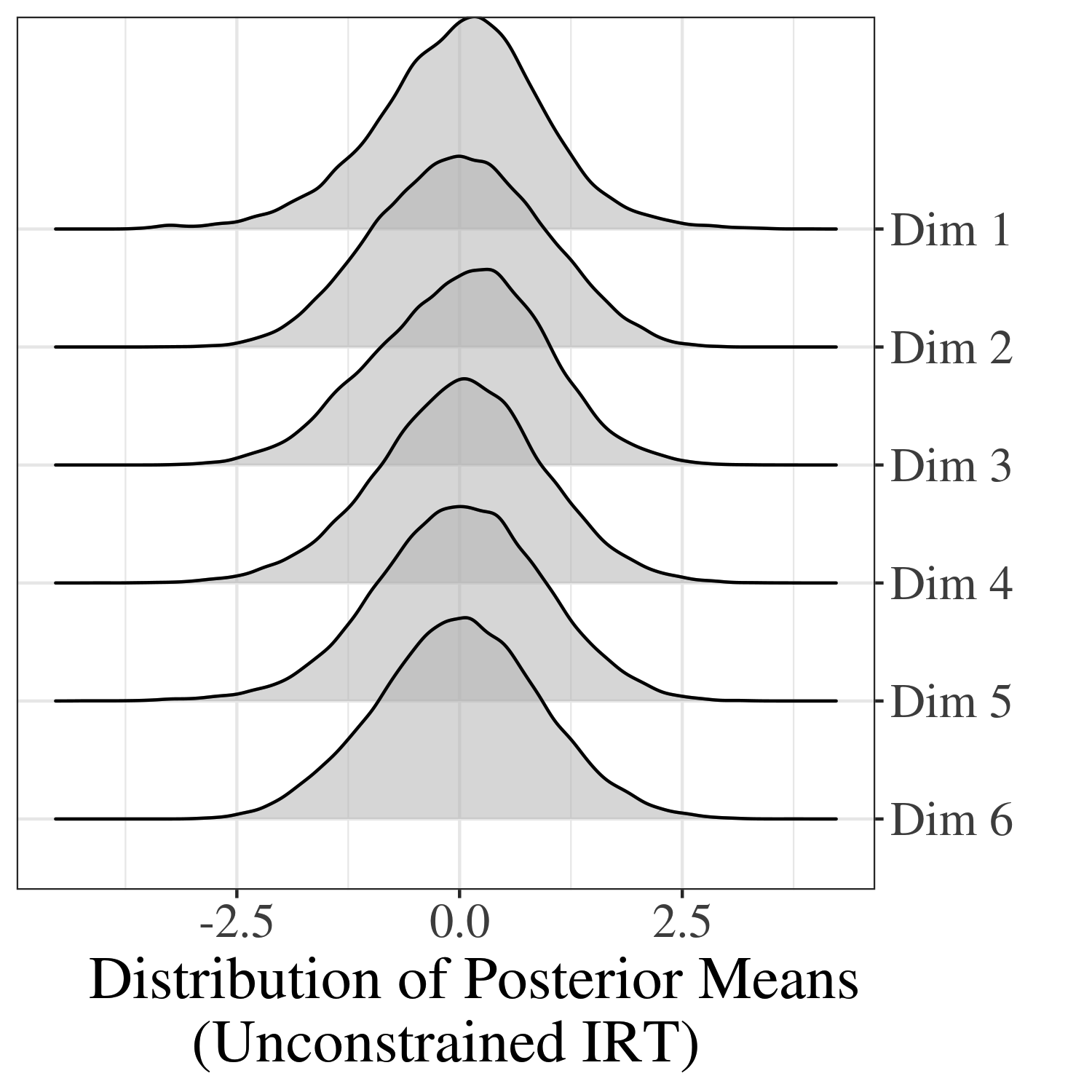}
     \end{subfigure}
        \caption{Comparison of Distributions of Latent Dimension Posterior Means. \\Left: IRT-M output, with named dimensions. Right: unconstrained IRT output.}
        \label{fig:comp1}
        \footnotesize\textbf{Note:} These figures are obtained by applying the traditional kernel density estimator with automatic bandwidth selection to the posterior means obtained by each methods for each the $N$ respondents.
\end{figure}

\begin{figure}
     \centering
     \begin{subfigure}[b]{\textwidth}
         \centering
         \includegraphics[width=\textwidth]{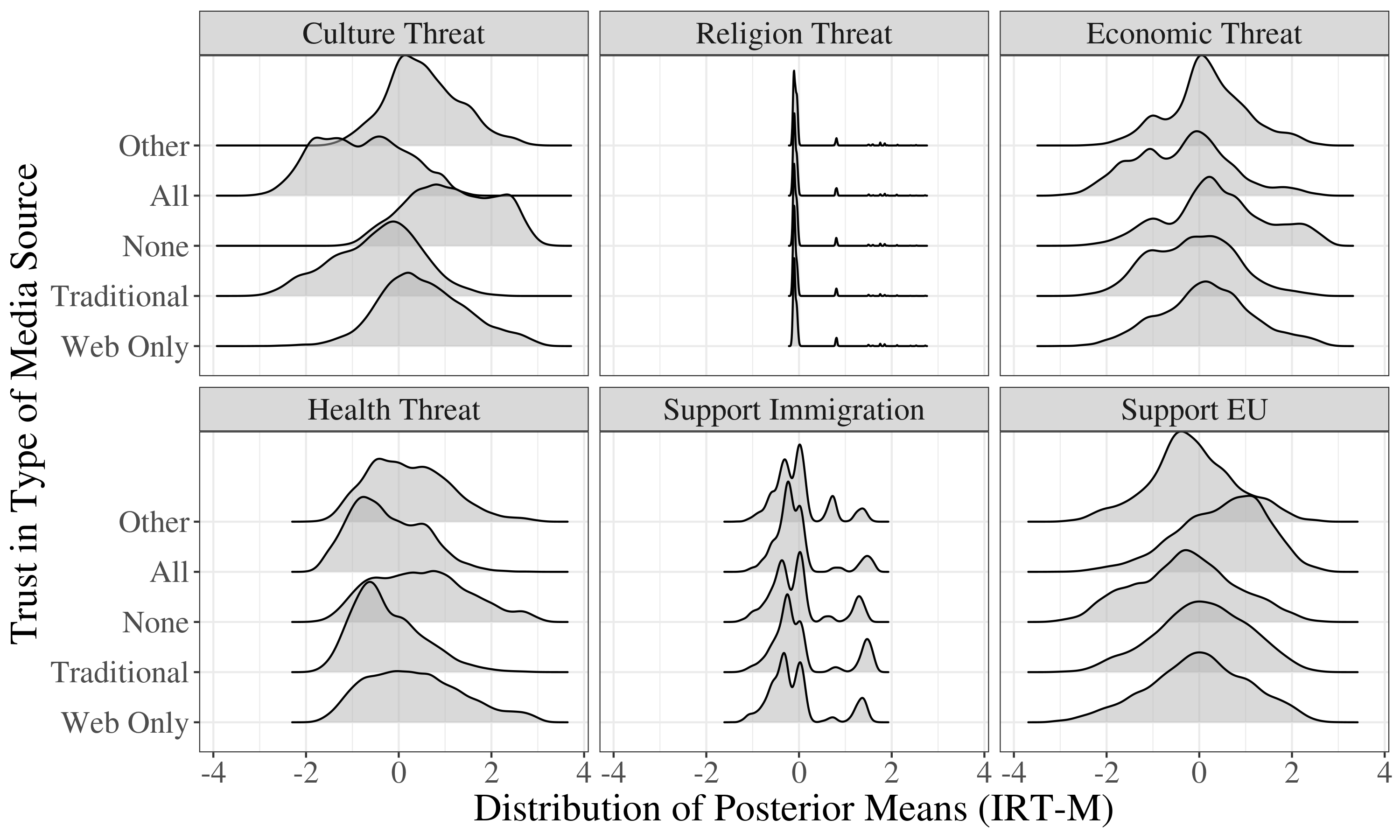}
     \end{subfigure}
     \hfill
     \begin{subfigure}[b]{\textwidth}
         \centering
         \includegraphics[width=\textwidth]{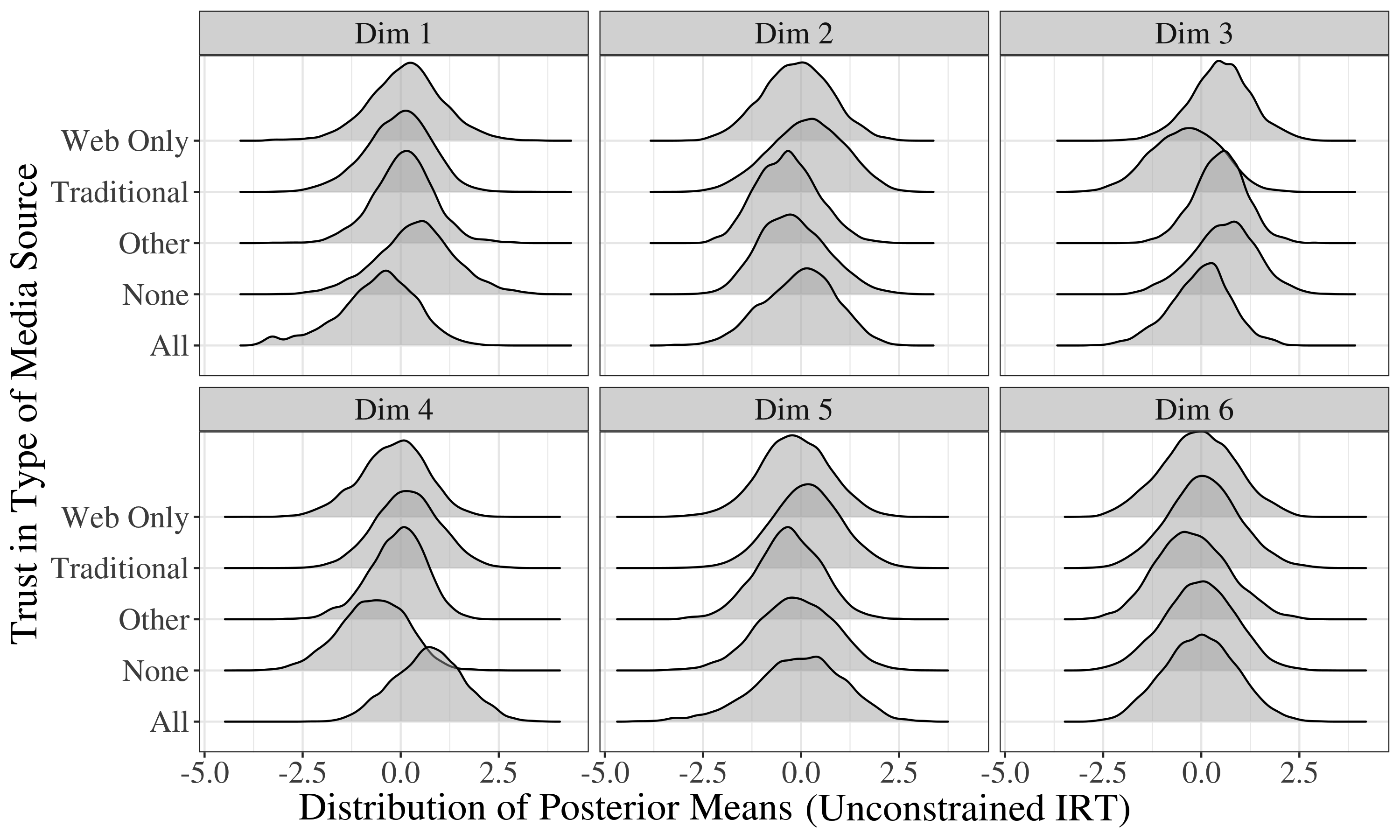}
     \end{subfigure}
        \caption{Comparison of Distributions of Posterior Means of Latent Dimensions by Trust in Media. \\Top: IRT-M output, with named dimensions. Bottom: unconstrained IRT output.}
        \label{fig:comp2}
                \footnotesize\textbf{Note:} These figures are obtained by applying the traditional kernel density estimator with automatic bandwidth selection to the posterior means obtained by each methods for each the $N$ respondents.
\end{figure}

Our set of constraints captures coding for all item-latent dimension pairs in the model, and its use distinguishes IRT-M from unsupervised IRT models. Once coding is complete, one inputs the constraints (in the form of a matrix) and the data into the IRT-M package \texttt{IRTM}. The output of IRT-M is a user-definable number of draws from the posterior distribution of each unit's position along each latent dimension. In other words, IRT-M maps out the posterior probability distribution capturing each unit's position in the latent space defined by the theoretical concepts encoded in the constraints. One can average those draws---taking the expected value of the posterior distribution---along each latent dimension for each unit. Doing so produces a distribution of the units' expected positions in latent space, as in Figures \ref{fig:comp1} and \ref{fig:comp2}. That captures the distribution of the theoretical concepts in the sample. One can then compare the distributions of different concepts within a sample, or of the same concepts across samples. Individual units' positions in latent space can also be used in subsequent analyses as individual-level direct measures of theoretical concepts.

We can use our motivating example to illustrate IRT-M's output. While there is a great deal one could do with our six latent dimensions---and we provide coding rules, data, and complete model output for those who might want to explore further---we focus on three figures that showcase both what IRT-M can do and some differences between IRT-M and an unconstrained IRT model.

Figure \ref{fig:comp1} compares the six posterior distributions of units' positions along the latent dimensions arising from IRT-M (on the top) to those arising from an unconstrained six-dimensional IRT model (on the bottom). It is clear that the distributions of latent positions in the sample seem to vary significantly across theoretical concepts in IRT-M, but do not appreciably vary in the unconstrained IRT model. We'll return to that point shortly, but first we highlight the distribution arising from IRT-M most different from the rest, that corresponding to religious threat. Unlike the other five dimensions, the distribution of religious threat is sharply peaked near zero with only a small rise away from that peak. Is it the case that, unlike each of the other perceptions of threat, almost no one in the sample perceives religious threat? To answer that, we can turn to the constraints and the coding rules from which they were derived. What we find is that the February-March 2021 wave of Eurobarometer is a poor data source for measuring religious threat. Only three questions, all regarding whether or not terrorism is the most important issue, are reasonably related to religious threat. Further, during a pandemic, very few people felt that terrorism was the most important issue, leading to little variation in response to those questions. Consequently, we should have little confidence in our measure of the religious threat concept. Instead that latent dimension is capturing a more narrow concept of issue importance. That highlights a point made in the previous section: IRT-M does not find latent dimensions that best predict the data. Rather, it computes positions along a latent dimension using only those items tied closely to the theoretical concept connected to that latent dimension. When there are few or no such items, IRT-M will not produce a good measure of that concept. That is a feature, not a bug, of the approach, and we left that dimension in the analysis to reiterate that important point.

What about the other measures of threat? One way to assess the level of confidence we should have in them is by exploring their construct validity. One way to do that is to disaggregate each distribution so as to enable comparisons to behavioral expectations. Figure \ref{fig:comp2} displays results for one possible disaggregation, in which the population is split according to trust in different media sources. We would expect, all else equal, that those who don't trust the media would be more likely to perceive threat, particularly cultural threat, whereas those who trust the traditional media would be less likely to perceive threat. That is what we see in the distributions derived from IRT-M, particularly for cultural and health threat, less strongly for economic threat. Further, those who trust all media sources generally perceive threat more similarly to those who trust traditional sources than to those who trust no sources. In contrast, only one of the dimensions derived from the unconstrained IRT model, that for Theta3, approaches that behavior. One can repeat that exercise for different subgroups. For example, we find (not shown) that upper class respondents generally perceive less threat than lower class respondents across the three meaningfully measured threat dimensions. That pattern is not observed in the unconstrained model application. Figure \ref{fig:comp2} thus reiterates another key point: where there are sufficient informative items in the data, IRT-M produces latent dimensions that possess substantive validity.

\begin{figure}
     \centering
     \includegraphics[width=0.6\textwidth]{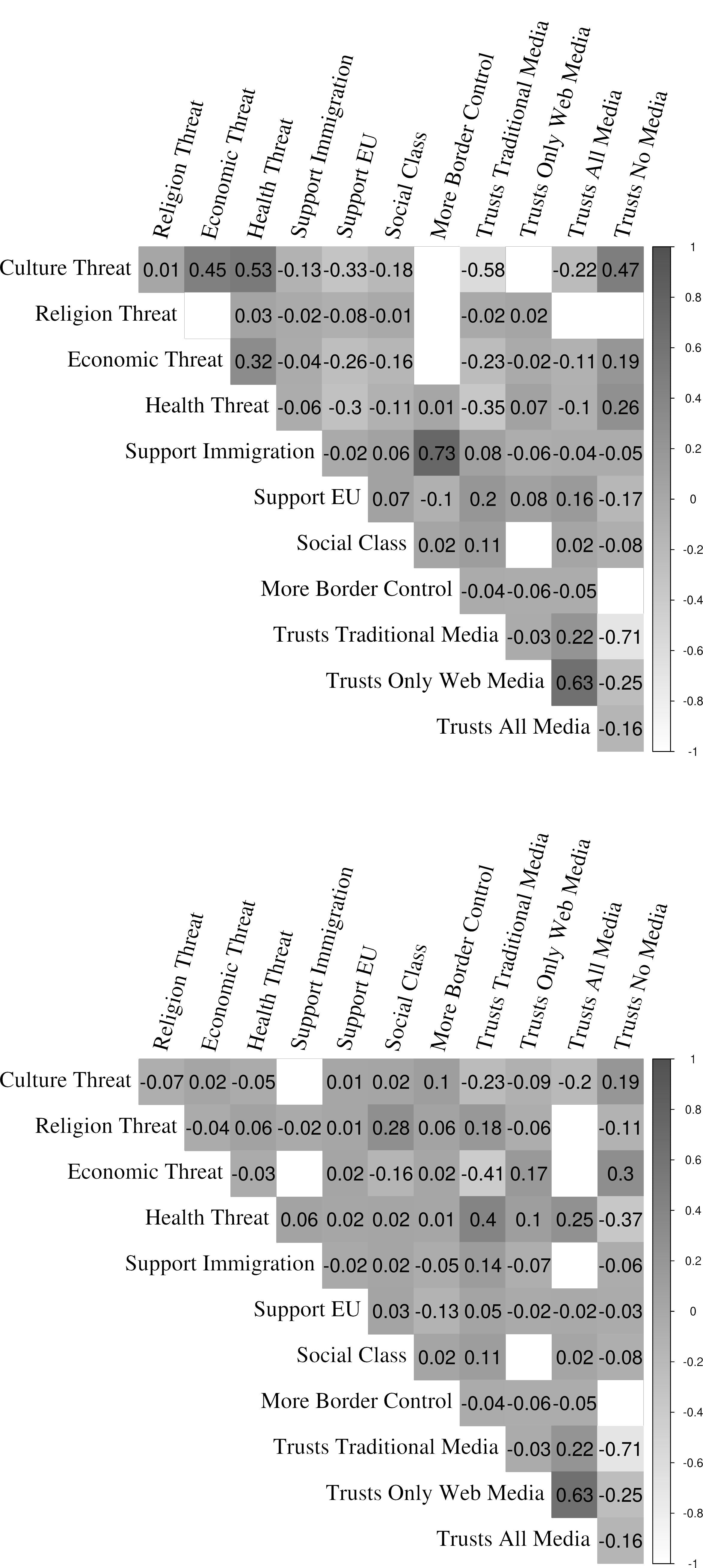}
        \caption{Correlations: Latent Dimensions, Media Trust, and Support for Border Controls. \\Top: IRT-M output, with named dimensions. Bottom: unconstrained IRT output.}
        \label{fig:compcorrs}
\end{figure}

Figure \ref{fig:compcorrs} extends the point about validity to the attitude measures. For both IRT-M and an unconstrained IRT, it displays a correlation matrix linking the six latent dimensions, as well as four values of trust in the media, a social class variable, and one question directly asking about more border controls. The border control question is coded so that higher values imply less desire for more border controls. We would expect it to be strongly correlated with the Support for Immigration latent variable, as it is both substantively related and one of about ten survey questions that are informative about that latent dimension and coded as such in the constraints. The figure bears out that expectation strongly, while the same is not true for any of the latent dimensions arising from the unconstrained IRT model. Thus, whatever is being captured by those unconstrained dimensions, it is not directly interpretable as support for immigration, one outcome variable of interest.

Figure \ref{fig:compcorrs} also speaks to another characteristic of IRT-M: it models correlated latent dimensions, in order to capture correlated theoretical concepts. IRT models typically assume independent priors over the latent dimensions; yet, there are often theoretical reasons to believe that the underlying latent dimensions of interest are correlated. Explicitly modeling the correlation between dimensions, as does IRT-M, avoids a fundamental disconnect between theory, data, and model. That comes through in the figure: the three meaningful threats are all positively correlated with each other, in addition to a lack of trust in the media, while being negatively correlated with trust in traditional media. Those correlations are substantively important, and a model without them would not be able to capture different aspects of threat. In contrast, in the unconstrained IRT the latent dimensions are not substantially correlated with each other, which forces the concepts they represent to be largely unrelated.

Finally, Figure \ref{fig:compcorrs} speaks to the idea in \cite{kentmen2017anti} that perceived threats are linked to attitudes about immigration and the EU. We find that the three meaningful threats are all negatively correlated with support for both immigration and the EU, though the correlations with the former are weak. 

We present the technical specification of IRT-M in the next section, but before doing so it is worth highlighting one more difference between IRT-M and unconstrained IRT models that is more technical in nature. In an unconstrained IRT model, identification and scaling of the latent dimensions occurs via the use of exogenous information relating to units in the data. For example, ideological ideal points might be taken as given for a small number of legislators. In IRT-M, the constraints both directly allow for model identification and indirectly scale the latent dimensions. The latter occurs automatically in the course of running the model via the creation of anchor points. Anchor points are artificial units that possess extreme positions along the latent dimensions. The constraints characterize behavior consistent with extreme values: if someone has a positive response to every item coded as $1$ and a negative response to every item coded as $-1$ for some latent dimension, then it is impossible for a person to show up as more extreme along that latent dimension in the data at hand. Accordingly, the extreme anchor points we construct serve as reference points for the bounds of the space spanned by the latent dimensions. Other, real, units are measured relative to those anchor points, aiding in interpretation of their values along each latent dimension. No exogenous information relating to the units is needed: all required information arises from the pre-analysis step of coding the constraints, and so relates only to the items used by IRT-M to compute latent positions and is theoretically rather than empirically driven. Anchor points are automatically removed prior to output.

\section*{Methodology}
We now describe our approach more formally. There are several key points of divergence from existing IRT-based models such as \cite{pscl}. The first is our pre-analysis coding of the constraints, described earlier. That coding entails conversion of the data to a series of binary (yes/no) items, followed by assignment of a $1, -1$, $0$, or \texttt{NA} for each item-latent dimension pair. Each assignment is based on whether or not the theoretically defined latent dimension positively, negatively, or doesn't predict at all the value of that item. The second point of divergence is that identification in our model is accomplished via our constraints, as we will describe. In contrast, typically multi-dimensional ideal-point-estimation models require for identification purposes some exogenous information relating to the units in the data, usually the ideal points of several units. The third point of divergence is that we explicitly model the covariance of the latent dimensions in our model. IRT models typically assume independent priors for the the latent dimensions and do not model their covariance; yet, researchers often have theoretical reasons to believe that the underlying latent dimensions of interest are correlated. By modeling covariance, we allow for closer connections between theory, data, and model.

Formally, begin by considering a set of $i = 1, \dots, N$ units each responding to $k=1, \dots, K$ binary (e.g., yes/no) items. Let $y_{ik}\in\{0,1\}$ denote unit $i$'s response to item $k$, and $\bY \in \{0, 1\}^{N\times K}$ be a binary matrix in which row $i$ is unit $i$'s set of responses to all items, and column $k$ is the vector of all units' responses to item $k$, so that each entry corresponds to $y_{ik}$ as just defined. We would like to model units' responses to items as a function of values along $j=1,\dots,d$ latent dimensions, denoted for unit $i$ by the vector $\btheta_i$. We will also refer to each $\theta_{ji}$ as a factor $j$. We choose a 2-parameter model \citep{rasch1960} for this task, as it is commonly used in applied political science \citep[e.g.,][]{clinton2004, tahk2018}:
\begin{align}
    \Pr(y_{ik}=1) &= g(\blambda^T_k\btheta_i - b_k),\label{Eq:ModelNoM}
\end{align}
where $\blambda$ is a $d$-length vector of loadings, $b_k$ is a negative intercept, and $g$ is a link function, mapping from $\mathbb{R}$ to $[0, 1]$. The intercept is commonly understood as a difficulty parameter in the psychometric testing literature, and can be seen here as an average rate of ``yes'' responses to item $k$. Each entry in the vector $\blambda_k$ is a real value, denoted by $\lambda_{kj}$. The sign of $\lambda_{kj}$ represents whether a larger factor $j$ (i.e., a larger $\theta_{ji}$) will increase or reduce the likelihood of a ``yes'' response to item $k$, while the magnitude of $\lambda_{kj}$ represents the overall influence of factor $j$ on the likelihood of responding ``yes'' to item $k$. For example, a large positive $\lambda_{kj}$ tells us that units with a large factor $j$ are much more likely to respond ``yes'' to item $k$.

The main objective of this paper, phrased formally, is to introduce a strategy to encode theoretical information linking latent dimensions and items in the model above, while at the same time resolving its identification issues. We do so by introducing a set of $K$ matrices: $\bM_1, \dots, \bM_K$. There is one matrix for each item, and each matrix is $d\times d$ and diagonal. Entries in the diagonal of each matrix are denoted by $m_{kjj}$ and are allowed to either take values in $\mathbb{R}$ or to be missing, where missingness is denoted by $m_{kjj} = \texttt{NA}$. We then retain the same model in Eq. \eqref{Eq:ModelNoM}, but constrain the loadings $\blambda$ as indicated in Table \ref{tab:mconstraints}.
\begin{table}[!htbp]
    \centering
    \begin{tabular}{c|c|l}
        Entry in diagonal $j$ of matrix $k$ & Constraint on $\lambda_{kj}$ & Prior belief  \\
        \hline\hline
        $m_{kjj} = 0$ & $\lambda_{kjj} = 0$ & Factor $j$ does not affect answers\\
        & &  to item $k$\\
        \hline
        $m_{kjj} > 0$ & $\lambda_{kjj} \in (0, \infty)$ & Factor $j$ \textit{positively} affects likelihood of \\
        & & answering yes to item $k$. \\
        \hline
        $m_{kjj} < 0$ & $\lambda_{kjj} \in (-\infty, 0) $ & Factor $j$ \textit{negatively} affects likelihood of \\
        & & answering yes to item $k$. \\
        \hline
        $m_{kjj} = \texttt{NA}$ & $\lambda_{kjj}  \in (-\infty, \infty)$ & No prior on whether and how factor $j$ affects\\
        & & answers to item $k$. 
    \end{tabular}
    \caption{Constraints on Loadings $\blambda$ from Theory}
    \label{tab:mconstraints}
\end{table}
As Table \ref{tab:mconstraints} shows, we constrain our model by (potentially) pre-specifying whether the relationship between the response to a item and each latent dimension is positive, negative, zero, or not defined. That accomplishes two fundamental goals. First, it lets one introduce known theoretical connections between items and latent dimensions into the modeling framework. Second, it allows for identification of the model. Before expanding on how our framework accomplishes both of those goals, we give a full description of the model we employ. We choose to adopt a Bayesian specification for our model in Eq. \eqref{Eq:ModelNoM}: we place priors on all of its parameters and express the constraints imposed on $\blambda$ by $\bM$ as prior items on the former.

The full hierarchical model is as follows:
\begin{align}
    Y_{ik} &= \ind_{[\mu_{ik} > 0]} \label{Eq:Ymodel}\\
    \mu_{ik} &\sim \mathcal{N}(\blambda_k^T\btheta_i - b_k, 1) \label{Eq:Mumodel}\\
    \lambda_{kj} &\sim 
    \begin{cases}
    \mathcal{N}_{[0, \infty]}(0, m_{kjj}^2) &\mbox{ if } m_{kjj} > 0\\
    \mathcal{N}_{[-\infty, 0]}(0, m_{kjj}^2) &\mbox{ if } m_{kjj} < 0\\
    \delta_{[\lambda_{kj} = 0]} &\mbox{ if } m_{kjj} = 0\\
    \mathcal{N}(0, 5) &\mbox{ if } m_{kjj} = \texttt{NA}\\
    \end{cases}&& k=1, \dots, K\label{Eq:lambdaPrior}\\
    b_k &\sim \mathcal{N}(0, 1)\label{Eq:bPrior}\\
    \btheta_i &\sim \mathcal{N}_d(\boldsymbol{0}, \boldsymbol{\Sigma})&& i=1, \dots, N\label{Eq:thetaPrior}\\
    \Sigma &\sim \mathcal{IW}_d(\nu_0, \mathbf{S}_0),&&\label{Eq:SigmaPrior}
\end{align}
where $\ind$ is the indicator function, $\mathcal{N}_P(\boldsymbol{\mu}, \boldsymbol{\Sigma})$ denotes the multivariate normal distribution of size $d$, with mean vector $\boldsymbol{\mu}$ and covariance matrix $\boldsymbol{\Sigma}$, $\mathcal{N}_{[a, b]}(\mu, \sigma^2)$ denotes the univariate normal density truncated between real valued constants $a$ and $b$, and $\mathcal{IW}_d$ denotes the Inverse-Wishart distribution with $\nu_0$ degrees of freedom and positive, semi-definite matrix $\mathbf{S}_0$. Finally, the vector of length $d$ that is 0 at all positions is denoted by $\boldsymbol{0}$. 

We choose a probit specification for our link function $g$ as this is standard practice in Bayesian IRT modeling \citep[e.g., ][]{pscl}; however, the model above can also be extended to logistic IRT by changing the distribution of $\mu_{ij}$. We express the constraints imposed on the loading vector, $\blambda$, by the M-matrix by varying which prior is chosen for this vector depending on $\bM$ (Eq. \eqref{Eq:lambdaPrior}).  Both $\nu_0$ and $\mathbf{S}_0$ are hyperparameters fixed at standard weakly-informative \citep[see, e.g., ][]{gelman2008} values in our model. That specification is motivated partly by our application, and partly by the fact that it allows us to solve the identification problems associated with the model in \eqref{Eq:ModelNoM} in conjunction with the $\bM$ matrices. Additional discussion of our model specification is available in Appendix A. 

A major contribution of our model specification is that it models the covariance between latent factors as a free parameter and the covariance between loadings as the identity matrix. 
Model identification while allowing correlation between factors requires shifting the assumption of independence to the item loadings in the $M$ matrices. This is, of course, a consequential modeling assumption. However, in the context of many applications of interest to social scientists, underlying factors---such as dimensions of ideology---are often interlinked. In contrast, items---such as survey question answers or elements of texts---are often functionally independent, frequently by design. By modeling independence at the item loading level, we leverage the data generating process's tendency to produce nearly-independent loadings. Our resulting IRT model better fits with many applications of interest. We will expand more on the effect of covariance parameters on model identification and estimation shortly. First, we elaborate on the role of our constraint matrix, $M$.

\vspace{-.1in}
\subsection*{Using $\mathbf{M}$ to Link Model and Theoretical Knowledge}

As indicated in Table \ref{tab:mconstraints}, in our approach $M$-matrices are used to model the extent to which we expect a given item $k$ to load on a factor $j$. Those loadings are encoded by changing the value of $m_{kjj}$. A value of $0$ indicates that factor $j$ does not influence the likelihood of a ``yes'' response to item $k$, and therefore item $k$ should not load on that factor. An absolute value of $|m_{kjj}|=1$ indicates that we believe that factor $j$ affects item $k$, but have no strong belief about the magnitude of its effect. A larger absolute value of $|m_{kjj}| > 1$ indicates that factor $j$ should have an effect stronger than the average factor on the likelihood of response to item $k$. Values of $m_{kjj}$ that are greater (less) than $0$ indicate that we believe increasing factor $j$ leads to a greater (lesser) likelihood of a ``yes'' response to item $k$. All those items are visibly encoded in our model by the prior items in Eq. \eqref{Eq:lambdaPrior}. Those priors are important for both model identification and substantive interpretation of results. On the latter point, they are crucial in assigning a clearly defined meaning to each dimension: dimensions are theoretically prior in our approach, and they may be more or less relevant for understanding a unit's responses to particular items. On the former point, as we will discuss further, constraints embedded in the matrices allow for model identification in the absence of exogenous information about the units.

That does not mean, however, that a user needs to specify all $m_{kjj}$ in our model. Only a small number ($d(1-d)$, as we will see) of $m_{kjj}$ must be set to zero for identification. Further, even that requirement could be weakened if one desired to use exogenous information on units. Whenever a loading is not specified, $m_{kjj}$ is set to missing, and the relevant parameter, $\lambda_{kj}$, is drawn from a standard normal distribution, as is normally done in IRT modeling. The most likely scenario in practice is one in which analysts have good knowledge of the loadings of only \textit{some} items, and this model specification allows them to encode their knowledge without having to put strong priors on items about which they do not have strong knowledge.

\vspace{-.1in}
\subsection*{Using $\mathbf{M}$ for Model Identification}
Identification issues arise when dealing with models such as that in \eqref{Eq:ModelNoM}. Here, we show how diagonal matrices such as our constraint matrices can be used to solve those issues, without the need for additional prior distributions or more complex model specifications. 

\vspace{-.3in}\paragraph{Location and Scale Invariance} Models such as that in \eqref{Eq:ModelNoM} are not identified because, for any value of the parameters $\mu_{ik} = \blambda_k^T\btheta_i$ and $b_k$, a different set of parameter values that gives rise to the same likelihood could be constructed with $\mu'_{ik} = \blambda_k^T\btheta_i + a$, and $b_k' = b_k - a$, for some real value $a$. Similarly, for any value of $\blambda^T_k\btheta_i$, we could construct a different set of parameters that gives rise to the same likelihood with $\blambda_k' = \blambda_k\cdot a$ and $\btheta_i' = \btheta_i\cdot\frac{1}{a}$. Those issues are known as location and scale invariance, respectively, and can be prevented by fixing the mean of $\btheta_i$ and $b_k$, and the variance of $\blambda_k$. We implement this with our model prior choices in \eqref{Eq:lambdaPrior} -- \eqref{Eq:SigmaPrior}.

\vspace{-.3in}\paragraph{Rotation Invariance} One other identification issue that arises is that, for any $\blambda_{k}$ and $\btheta_{i}$, the same exact value of the product $\blambda_{k}^T\btheta_{i}$ can be obtained with the product of different parameter values $\blambda_k'=\blambda_{k}\bA$ and $\btheta_i'=\bA\btheta_{i}$, where $\mathbf{A}$ is an orthogonal rotation matrix. This implies that $\blambda_{k}^T\btheta_{i} = \blambda_{k}^{'T}\btheta_{i}'$. There are several possible ways to prevent this issue, commonly known as rotation invariance. In general, $d(d-1)$ constraints must be imposed on the model to prevent this issue \citep{howe1955}. A researcher may want to impose more than $d(d-1)$ constraints if perceived theoretical clarity allows for this; we discuss this setting in Appendix \ref{app:model}.  Applications of unconstrained IRT generally achieve identification by specifying, before analysis, values along the latent dimensions for a set of units in the data\citep[see e.g., ][]{pscl}. Though that approach does encode links between latent dimensions and theoretical concepts, it also requires that the analyst know the location of some units in the latent space ahead of time, which is not always possible.

Instead, IRT-M imposes the necessary $d(d-1)$ constraints to prevent rotation invariance via the $\bM$ matrices. There are two ways to do so. First, one can set at least $d(d-1)$ zeros in the diagonals of all the M-matrices, cumulatively. That is the method we describe above. Formally, this requirement can be stated as
    \begin{align*}
        \sum_{k=1}^K\sum_{j=1}^d \ind_{[m_{kjj}=0]} \geq d(1-d).
    \end{align*}
Setting $m_{kjj} = 0$ is equivalent to assuming that item $k$ does not load on factor $j$, or that factor $j$ does not influence the likelihood of responding ``yes'' to item $k$. Substantively, that means that one must have a certain amount of prior knowledge regarding the connection between items in the data and the latent dimensions. While one does not need any of the items to load on only one of the latent dimensions, a certain number of items must not load on all the latent dimensions.

Second, one can fix at least $d(1-d)$ of the $\btheta_i$. Rather than make assumptions about the locations of units in latent space as is commonly done, however, IRT-M instead creates anchor points from the $M$ matrices. Those anchor points rely on assumptions about links between items and latent dimensions, rather than about units, and so their use allows analysts to solve rotation invariance issues without having to make strong assumptions about the units themselves. Anchor points also set the scale for the latent space, as described previously. Relying solely on anchor points for identification requires the $\bM$ matrices to generate two anchors per dimension, according to the procedure described below. However, any combination of $\bM$ matrix constraints and anchor points allows for identification, as long as the total number of constraints is at least $d(d-1)$.

The intuition behind anchor points is that a hypothetical data unit that responds ``yes'' to all items that are positively influenced by one factor $j$, and ``no'' to all the items that are not will have a large value of factor $j$. The formal procedure to create a positive anchor point for one factor $j$, starting from a set of $M$-matrices, is as follows:
\begin{enumerate}
    \item Create a new data point, $\by^{new}$.
    \item For all items $\ell \in \{m_{kjj} > 0: k =1, \dots, K\}$, i.e., all items such that the direction of their loading on factor $j$ is known and positive, set $y^{new}_{\ell} = 1$.
    \item For all items $\ell \in \{m_{kjj} < 0: k =1, \dots, K\}$, i.e., all items such that the direction of their loading on factor $j$ is known and negative, set $y^{new}_{\ell} = 0$.
    \item For all the remaining items, $r$,  set $y^{new}_{r}$ to missing.
    \item Set $\theta^{new}_{j} = D$, where $D$ is some positive constant that is in an extreme of the latent space.
\end{enumerate}
To create a negative anchor for the same factor $j$, the procedure can be followed exactly, except by setting $y^{new}_{\ell} = 0$ at step 2, $y^{new}_{\ell} = 1$ at step 3, and by making $D$ a negative constant at step 5. In the rare occurrence that an anchor point has exactly the same answers as an actual respondent, then the $\btheta$ for that respondent should also be set to that of the anchor point just created. 

\vspace{-.1in}
\subsection*{Model Extensions}
In order to simplify exposition, we have presented our model in the context of a probit-based IRT model; however, our proposed approach can be easily extended to logistic IRT and multinomial IRT and, while we will not discuss it in detail, to standard factor analysis for a real-valued outcome as well. The hierarchical formulation of our model also allows for simple handling of missing outcome data. Finally, our model can be modified so that the co-variance matrix of the loadings, rather than the factors, is learned from the data. We discuss that last modification in Appendix A. All of those extended models allow Gibbs sampling formulations for their MCMC posterior sampling procedures, greatly improving convergence speed. Importantly, none of those extensions affect how analysts specify and use $M$-matrices in our approach, further supporting the flexibility and wide applicability of our proposed method.

\vspace{-.3in}\paragraph{Logistic and Multinomial IRT} Our model can be extended to logistic IRT and multinomial IRT by adopting the Polya-Gamma formulation of logistic regression of \citet{polson2013bayesian}. In a binary logistic IRT, we model our responses as $\Pr(Y_{ik}=1) = \frac{\exp(\psi_{ik})}{1 + \exp(\psi_{ik})}$, where $\psi_{ik} = \blambda_k^T\btheta_i - b_k$. \citet{polson2013bayesian} show that the likelihood for one item in this model is proportional to 
$$L_{ik}(\psi_{ik}) \propto \exp(\gamma_{ik}\psi_{ik})\int_{0}^{\infty}\exp(-\omega_{ik}\psi_{ik}^2/2)p(\omega_{ik})d\omega_{ik}, $$
where $\gamma_{ik} = Y_{ik} - 1/2$ and $\omega_{ik}$ follows a Polya-Gamma distribution with parameters $(1,0)$. If all the priors on the IRT parameters $\btheta_i, \blambda_k, b_k$ are left as they are in equations \eqref{Eq:lambdaPrior} -- \eqref{Eq:thetaPrior}, then that formulation for the likelihood leads to simple conditionally conjugate updates for all three parameters depending on $\omega_{ik}$, which also has a conditionally conjugate update. That allows for simple extension of the Gibbs sampler for our existing model to the logistic setting without loss of performance in terms of computation time. 

Additionally, that model is easily extended to multinomial IRT settings in which one item can have more options than a yes/no response. Suppose that $Y_{ik}$ can be one of $\ell = 1,\dots, L$ options. In that case, the outcome model becomes: $\Pr(Y_{ik}=\ell) = \frac{\exp(\psi_{ik\ell})}{\sum_{r=1}^L\exp(\psi_{ikr})}, $ where: $\psi_{ik\ell} = \blambda_{k\ell}^T\btheta_i - b_{k\ell}$. As \citet{polson2013bayesian} note, that model can also be expressed as a special case of the binary logistic model just discussed, leading to similar conditional updates for all the IRT parameters. Notably, the number of item loadings is now different for each item, as one loading vector $\blambda_{k\ell}$ has to be estimated for each possible response. That implies that it is possible to further constrain such response-level loadings with $M$-matrices: in that case, the analyst may specify a different $M_{k\ell}$ for each possible answer to a multiple-choice question. The prior on $\blambda_{k\ell}$ is then determined based on such a matrix in the same way that it is in the current model, for binary responses. 

\vspace{-.3in}\paragraph{Missing responses} Finally, our current formulation allows for simple inclusion of missing outcomes (responses) by modifying the Gibbs sampling update for $\mu_{ik}$: if a response is missing, then $\mu_{ik}$ can be drawn from an un-truncated normal distribution centered at the conditionally updated mean and variance. That is a common tool in Bayesian inference for probit models and it is explained in detail in, e.g.,  \citet{gelman1995bayesian}. That approach has the clear advantage of allowing analysts to directly deal with missing data without the need to specify other parameters; however, we note that there is evidence that values imputed this way are biased towards the conditional mean of their imputation distribution \citep{michel2021proper}. As we already have incorporated this model extension, we caution analysts who believe that imputation-induced bias might affect their estimates too strongly to make use of other imputation methods before employing IRT-M. 

\section*{Model Validation}
In this section, we present an in-depth empirical evaluation of our proposed methodology. We first study the performance of IRT-M on a set of simulated data, where ground truth is known and so can be coded directly into the $\bM$ matrices. Then, we apply our model to roll call data in the U.S. Congress in order to compare our results to those arising from a common IRT application, in a setting in which information on the positions of some legislators in latent ideological space is thought to be known.

\vspace{-.1in}
\subsection*{Simulated Data}
We generate 100 artificial data sets from the model specified in \eqref{Eq:Ymodel}--\eqref{Eq:SigmaPrior}. 
Parameter values for the main simulation parameters are held constant for each simulated dataset at the following values: $N=500$, $K=50$, $\mathbf{S}_0=\mathbf{I}_{3\times 3}$, $\nu_0 = 3$. We consider four different numbers of latent dimensions---$d=2, 3, 5,$ and $8$---in order to cover a range of substantively useful values. 

\paragraph{Comparison with other methods} We compare performance of our model (IRT-M) and a correctly specified set of $M$-matrices to two other methodologies: Principal Component Analysis (PCA) where the principal component associated with the largest eigenvalue is taken as the latent dimension of interest, and a model similar to the one in Model \eqref{Eq:ModelNoM}, but without any $M$-matrices or any constraints on the respondents, as well as uncorrelated factors and loadings. That second comparison model (IRT) is an implementation often used in applications in political science. We compare IRT-M both with and without correlated factors to those two models. For the IRT-M model with uncorrelated factors, we remove the prior in Eq. \eqref{Eq:SigmaPrior} from the model, and fix $\Sigma = \mathbf{I}_{d\times d}$. 

We first compare the mean squared error (MSE) of estimates obtained across the three methodologies. The MSE captures the average distance between the location of the true value of the theta (position along each latent dimension) for each observation in the data (each data unit) and the model's estimate of that observation's theta. As in other contexts, an MSE close to zero indicates that the model tends to estimate thetas close to their true values. The results, averaged across dimensions, are shown in Table \ref{tab:thetamse}. We note that assessing performance of latent factor models is a complex problem, and MSE can have problems as a metric \citep{michel2021proper}. Due to that, we randomly selected several points from each simulated dataset and inspected them to make sure that average MSE did indeed correspond to estimates closer to their true value: we found that to be the case. 

\begin{table}
\caption{RMSE for $\btheta$ -- averaged over dimensions, respondents, and simulations.}
\label{tab:thetamse}
\centering
\resizebox{0.98\textwidth}{!}{%
\begin{tabular}{rr|llll|llll|llll|llll}
\toprule
N & K &  \multicolumn{4}{c|}{d=2} & \multicolumn{4}{c|}{d=3} & 
\multicolumn{4}{c|}{d=5} & \multicolumn{4}{c}{d=8} \\
\midrule
&  & PCA & IRT & IRT-M & IRT-M & PCA & IRT & IRT-M  & IRT-M & PCA & IRT & IRT-M  & IRT-M  & PCA & IRT  & IRT-M  & IRT-M \\
\multicolumn{2}{c|}{Correlated $\btheta$?} & No & No & No & Yes & No &  No & No & Yes & No & No & No & Yes & No & No & No & Yes \\
\midrule
\cellcolor{gray!6}{10} & \cellcolor{gray!6}{10} & \cellcolor{gray!6}{1.953} & \cellcolor{gray!6}{4.145} & \cellcolor{gray!6}{0.024} & \cellcolor{gray!6}{\textbf{0.023}} & \cellcolor{gray!6}{2.431} & \cellcolor{gray!6}{3.849} & \cellcolor{gray!6}{0.028} & \cellcolor{gray!6}{\textbf{0.027}} & \cellcolor{gray!6}{2.576} & \cellcolor{gray!6}{2.625} & \cellcolor{gray!6}{0.031} & \cellcolor{gray!6}{\textbf{0.03}} & \cellcolor{gray!6}{--} & \cellcolor{gray!6}{--} & \cellcolor{gray!6}{--} & \cellcolor{gray!6}{--}\\
10 & 50 & 1.781 & 4.699 & \textbf{0.011} & 0.011 & 2.059 & 3.565 & \textbf{0.013} & 0.013 & 2.372 & 3.509 & 0.019 & \textbf{0.017} & -- & -- & -- & --\\
\cellcolor{gray!6}{10} & \cellcolor{gray!6}{100} & \cellcolor{gray!6}{2.284} & \cellcolor{gray!6}{10.131} & \cellcolor{gray!6}{\textbf{0.008}} & \cellcolor{gray!6}{0.008} & \cellcolor{gray!6}{2.092} & \cellcolor{gray!6}{5.828} & \cellcolor{gray!6}{\textbf{0.009}} & \cellcolor{gray!6}{0.009} & \cellcolor{gray!6}{2.309} & \cellcolor{gray!6}{4.298} & \cellcolor{gray!6}{0.013} & \cellcolor{gray!6}{\textbf{0.012}} & \cellcolor{gray!6}{--} & \cellcolor{gray!6}{--} & \cellcolor{gray!6}{--} & \cellcolor{gray!6}{--}\\
10 & 250 & 2.327 & 13.560 & 0.006 & \textbf{0.005} & 2.099 & 8.927 & \textbf{0.006} & 0.006 & 2.357 & 4.808 & 0.009 & \textbf{0.008} & -- & -- & -- & --\\
\cellcolor{gray!6}{10} & \cellcolor{gray!6}{500} & \cellcolor{gray!6}{2.082} & \cellcolor{gray!6}{58.273} & \cellcolor{gray!6}{\textbf{0.004}} & \cellcolor{gray!6}{0.004} & \cellcolor{gray!6}{2.141} & \cellcolor{gray!6}{17.066} & \cellcolor{gray!6}{\textbf{0.004}} & \cellcolor{gray!6}{0.004} & \cellcolor{gray!6}{2.183} & \cellcolor{gray!6}{4.144} & \cellcolor{gray!6}{0.007} & \cellcolor{gray!6}{\textbf{0.006}} & \cellcolor{gray!6}{--} & \cellcolor{gray!6}{--} & \cellcolor{gray!6}{--} & \cellcolor{gray!6}{--}\\
\addlinespace
50 & 10 & 2.034 & 3.167 & 0.1 & \textbf{0.099} & 2.103 & 3.105 & 0.106 & \textbf{0.104} & 2.332 & 2.630 & 0.128 & \textbf{0.126} & 2.521 & 2.370 & 0.148 & \textbf{0.146}\\
\cellcolor{gray!6}{50} & \cellcolor{gray!6}{50} & \cellcolor{gray!6}{2.256} & \cellcolor{gray!6}{3.348} & \cellcolor{gray!6}{0.028} & \cellcolor{gray!6}{\textbf{0.027}} & \cellcolor{gray!6}{2.254} & \cellcolor{gray!6}{3.250} & \cellcolor{gray!6}{0.034} & \cellcolor{gray!6}{\textbf{0.033}} & \cellcolor{gray!6}{2.486} & \cellcolor{gray!6}{2.625} & \cellcolor{gray!6}{0.044} & \cellcolor{gray!6}{\textbf{0.042}} & \cellcolor{gray!6}{2.553} & \cellcolor{gray!6}{2.489} & \cellcolor{gray!6}{0.060} & \cellcolor{gray!6}{\textbf{0.058}}\\
50 & 100 & 2.979 & 15.258 & 0.019 & \textbf{0.018} & 2.331 & 6.120 & \textbf{0.021} & 0.021 & 2.572 & 3.435 & 0.028 & \textbf{0.027} & 2.652 & 2.752 & 0.042 & \textbf{0.04}\\
\cellcolor{gray!6}{50} & \cellcolor{gray!6}{250} & \cellcolor{gray!6}{2.561} & \cellcolor{gray!6}{34.504} & \cellcolor{gray!6}{\textbf{0.011}} & \cellcolor{gray!6}{0.011} & \cellcolor{gray!6}{2.721} & \cellcolor{gray!6}{18.009} & \cellcolor{gray!6}{\textbf{0.011}} & \cellcolor{gray!6}{0.011} & \cellcolor{gray!6}{2.841} & \cellcolor{gray!6}{9.653} & \cellcolor{gray!6}{0.016} & \cellcolor{gray!6}{\textbf{0.015}} & \cellcolor{gray!6}{2.615} & \cellcolor{gray!6}{4.457} & \cellcolor{gray!6}{0.029} & \cellcolor{gray!6}{\textbf{0.028}}\\
50 & 500 & 2.529 & 69.705 & \textbf{0.007} & 0.007 & 2.769 & 80.276 & \textbf{0.007} & 0.007 & 2.753 & 16.239 & \textbf{0.01} & 0.01 & 2.897 & 7.970 & 0.023 & \textbf{0.022}\\
\addlinespace
\cellcolor{gray!6}{100} & \cellcolor{gray!6}{10} & \cellcolor{gray!6}{2.222} & \cellcolor{gray!6}{3.213} & \cellcolor{gray!6}{0.153} & \cellcolor{gray!6}{\textbf{0.15}} & \cellcolor{gray!6}{2.199} & \cellcolor{gray!6}{3.196} & \cellcolor{gray!6}{0.173} & \cellcolor{gray!6}{\textbf{0.167}} & \cellcolor{gray!6}{2.204} & \cellcolor{gray!6}{2.643} & \cellcolor{gray!6}{0.213} & \cellcolor{gray!6}{\textbf{0.208}} & \cellcolor{gray!6}{2.421} & \cellcolor{gray!6}{2.555} & \cellcolor{gray!6}{0.246} & \cellcolor{gray!6}{\textbf{0.243}}\\
100 & 50 & 2.640 & 3.296 & 0.045 & \textbf{0.043} & 2.667 & 2.643 & 0.052 & \textbf{0.048} & 2.682 & 2.635 & 0.065 & \textbf{0.059} & 2.621 & 2.378 & 0.087 & \textbf{0.08}\\
\cellcolor{gray!6}{100} & \cellcolor{gray!6}{100} & \cellcolor{gray!6}{2.461} & \cellcolor{gray!6}{8.781} & \cellcolor{gray!6}{0.025} & \cellcolor{gray!6}{\textbf{0.024}} & \cellcolor{gray!6}{2.710} & \cellcolor{gray!6}{3.376} & \cellcolor{gray!6}{0.031} & \cellcolor{gray!6}{\textbf{0.029}} & \cellcolor{gray!6}{2.948} & \cellcolor{gray!6}{2.860} & \cellcolor{gray!6}{0.038} & \cellcolor{gray!6}{\textbf{0.035}} & \cellcolor{gray!6}{2.812} & \cellcolor{gray!6}{2.470} & \cellcolor{gray!6}{0.050} & \cellcolor{gray!6}{\textbf{0.046}}\\
100 & 250 & 2.589 & 13.037 & 0.013 & \textbf{0.012} & 2.666 & 10.391 & 0.016 & \textbf{0.015} & 3.201 & 6.670 & 0.019 & \textbf{0.018} & 3.001 & 4.523 & 0.030 & \textbf{0.026}\\
\cellcolor{gray!6}{100} & \cellcolor{gray!6}{500} & \cellcolor{gray!6}{2.395} & \cellcolor{gray!6}{59.279} & \cellcolor{gray!6}{0.009} & \cellcolor{gray!6}{\textbf{0.008}} & \cellcolor{gray!6}{2.717} & \cellcolor{gray!6}{28.505} & \cellcolor{gray!6}{0.01} & \cellcolor{gray!6}{\textbf{0.009}} & \cellcolor{gray!6}{3.011} & \cellcolor{gray!6}{17.176} & \cellcolor{gray!6}{0.012} & \cellcolor{gray!6}{\textbf{0.011}} & \cellcolor{gray!6}{3.236} & \cellcolor{gray!6}{7.282} & \cellcolor{gray!6}{0.020} & \cellcolor{gray!6}{\textbf{0.018}}\\
\addlinespace
250 & 10 & 2.076 & 3.761 & 0.267 & \textbf{0.253} & 2.298 & 2.730 & 0.305 & \textbf{0.283} & 2.278 & 2.671 & 0.362 & \textbf{0.348} & 2.440 & 2.587 & 0.451 & \textbf{0.437}\\
\cellcolor{gray!6}{250} & \cellcolor{gray!6}{50} & \cellcolor{gray!6}{2.760} & \cellcolor{gray!6}{3.951} & \cellcolor{gray!6}{0.07} & \cellcolor{gray!6}{\textbf{0.064}} & \cellcolor{gray!6}{2.728} & \cellcolor{gray!6}{2.560} & \cellcolor{gray!6}{0.086} & \cellcolor{gray!6}{\textbf{0.075}} & \cellcolor{gray!6}{2.672} & \cellcolor{gray!6}{2.537} & \cellcolor{gray!6}{0.107} & \cellcolor{gray!6}{\textbf{0.09}} & \cellcolor{gray!6}{2.832} & \cellcolor{gray!6}{2.336} & \cellcolor{gray!6}{0.138} & \cellcolor{gray!6}{\textbf{0.115}}\\
250 & 100 & 2.099 & 3.995 & 0.038 & \textbf{0.036} & 2.896 & 3.425 & 0.045 & \textbf{0.042} & 3.291 & 2.758 & 0.059 & \textbf{0.05} & 2.971 & 2.167 & 0.072 & \textbf{0.06}\\
\cellcolor{gray!6}{250} & \cellcolor{gray!6}{250} & \cellcolor{gray!6}{2.418} & \cellcolor{gray!6}{5.269} & \cellcolor{gray!6}{0.017} & \cellcolor{gray!6}{\textbf{0.016}} & \cellcolor{gray!6}{3.117} & \cellcolor{gray!6}{3.603} & \cellcolor{gray!6}{0.021} & \cellcolor{gray!6}{\textbf{0.02}} & \cellcolor{gray!6}{3.358} & \cellcolor{gray!6}{3.580} & \cellcolor{gray!6}{0.025} & \cellcolor{gray!6}{\textbf{0.024}} & \cellcolor{gray!6}{3.367} & \cellcolor{gray!6}{2.692} & \cellcolor{gray!6}{0.032} & \cellcolor{gray!6}{\textbf{0.028}}\\
250 & 500 & 3.167 & 34.596 & 0.012 & \textbf{0.011} & 3.184 & 8.250 & 0.012 & \textbf{0.011} & 3.475 & 9.579 & 0.015 & \textbf{0.014} & 3.617 & 5.845 & 0.020 & \textbf{0.017}\\
\addlinespace
\cellcolor{gray!6}{500} & \cellcolor{gray!6}{10} & \cellcolor{gray!6}{2.122} & \cellcolor{gray!6}{3.507} & \cellcolor{gray!6}{0.317} & \cellcolor{gray!6}{\textbf{0.298}} & \cellcolor{gray!6}{2.117} & \cellcolor{gray!6}{2.742} & \cellcolor{gray!6}{0.389} & \cellcolor{gray!6}{\textbf{0.356}} & \cellcolor{gray!6}{2.346} & \cellcolor{gray!6}{2.476} & \cellcolor{gray!6}{0.5} & \cellcolor{gray!6}{\textbf{0.471}} & \cellcolor{gray!6}{2.393} & \cellcolor{gray!6}{2.335} & \cellcolor{gray!6}{0.607} & \cellcolor{gray!6}{\textbf{0.591}}\\
500 & 50 & 2.768 & 4.480 & 0.094 & \textbf{0.078} & 3.108 & 2.676 & 0.11 & \textbf{0.087} & 2.933 & 2.210 & 0.144 & \textbf{0.107} & 2.783 & 2.322 & 0.184 & \textbf{0.139}\\
\cellcolor{gray!6}{500} & \cellcolor{gray!6}{100} & \cellcolor{gray!6}{2.889} & \cellcolor{gray!6}{2.642} & \cellcolor{gray!6}{0.054} & \cellcolor{gray!6}{\textbf{0.044}} & \cellcolor{gray!6}{2.772} & \cellcolor{gray!6}{2.576} & \cellcolor{gray!6}{0.061} & \cellcolor{gray!6}{\textbf{0.05}} & \cellcolor{gray!6}{3.029} & \cellcolor{gray!6}{2.403} & \cellcolor{gray!6}{0.075} & \cellcolor{gray!6}{\textbf{0.058}} & \cellcolor{gray!6}{3.083} & \cellcolor{gray!6}{2.220} & \cellcolor{gray!6}{0.093} & \cellcolor{gray!6}{\textbf{0.07}}\\
500 & 250 & 2.781 & 15.625 & 0.029 & \textbf{0.023} & 2.970 & 4.052 & 0.03 & \textbf{0.025} & 3.262 & 2.361 & 0.036 & \textbf{0.03} & 3.649 & 2.237 & 0.043 & \textbf{0.035}\\
\cellcolor{gray!6}{500} & \cellcolor{gray!6}{500} & \cellcolor{gray!6}{2.359} & \cellcolor{gray!6}{5.530} & \cellcolor{gray!6}{0.017} & \cellcolor{gray!6}{\textbf{0.013}} & \cellcolor{gray!6}{3.333} & \cellcolor{gray!6}{6.459} & \cellcolor{gray!6}{0.021} & \cellcolor{gray!6}{\textbf{0.018}} & \cellcolor{gray!6}{3.675} & \cellcolor{gray!6}{3.509} & \cellcolor{gray!6}{0.022} & \cellcolor{gray!6}{\textbf{0.019}} & \cellcolor{gray!6}{3.823} & \cellcolor{gray!6}{2.741} & \cellcolor{gray!6}{0.024} & \cellcolor{gray!6}{\textbf{0.02}}\\
\addlinespace
1000 & 10 & 2.256 & 2.942 & 0.404 & \textbf{0.359} & 2.246 & 3.266 & 0.473 & \textbf{0.421} & 2.458 & 2.502 & 0.603 & \textbf{0.554} & 2.399 & 2.446 & 0.765 & \textbf{0.731}\\
\cellcolor{gray!6}{1000} & \cellcolor{gray!6}{50} & \cellcolor{gray!6}{2.339} & \cellcolor{gray!6}{4.734} & \cellcolor{gray!6}{0.107} & \cellcolor{gray!6}{\textbf{0.088}} & \cellcolor{gray!6}{2.860} & \cellcolor{gray!6}{2.932} & \cellcolor{gray!6}{0.14} & \cellcolor{gray!6}{\textbf{0.094}} & \cellcolor{gray!6}{2.866} & \cellcolor{gray!6}{2.389} & \cellcolor{gray!6}{0.165} & \cellcolor{gray!6}{\textbf{0.114}} & \cellcolor{gray!6}{2.841} & \cellcolor{gray!6}{2.397} & \cellcolor{gray!6}{0.210} & \cellcolor{gray!6}{\textbf{0.15}}\\
1000 & 100 & 2.788 & 2.947 & 0.069 & \textbf{0.048} & 3.375 & 2.729 & 0.08 & \textbf{0.055} & 3.239 & 2.353 & 0.087 & \textbf{0.062} & 3.306 & 2.167 & 0.112 & \textbf{0.076}\\
\cellcolor{gray!6}{1000} & \cellcolor{gray!6}{250} & \cellcolor{gray!6}{2.703} & \cellcolor{gray!6}{3.391} & \cellcolor{gray!6}{0.043} & \cellcolor{gray!6}{\textbf{0.023}} & \cellcolor{gray!6}{3.445} & \cellcolor{gray!6}{2.951} & \cellcolor{gray!6}{0.055} & \cellcolor{gray!6}{\textbf{0.031}} & \cellcolor{gray!6}{3.561} & \cellcolor{gray!6}{2.406} & \cellcolor{gray!6}{0.058} & \cellcolor{gray!6}{\textbf{0.034}} & \cellcolor{gray!6}{3.813} & \cellcolor{gray!6}{2.188} & \cellcolor{gray!6}{0.055} & \cellcolor{gray!6}{\textbf{0.036}}\\
1000 & 500 & 2.995 & 3.471 & 0.042 & \textbf{0.015} & 2.956 & 2.814 & 0.041 & \textbf{0.023} & 3.676 & 2.413 & 0.048 & \textbf{0.025} & 3.804 & 2.242 & 0.043 & \textbf{0.025}\\
\addlinespace
\cellcolor{gray!6}{2500} & \cellcolor{gray!6}{10} & \cellcolor{gray!6}{2.240} & \cellcolor{gray!6}{4.955} & \cellcolor{gray!6}{0.479} & \cellcolor{gray!6}{\textbf{0.399}} & \cellcolor{gray!6}{2.086} & \cellcolor{gray!6}{3.721} & \cellcolor{gray!6}{0.597} & \cellcolor{gray!6}{\textbf{0.5}} & \cellcolor{gray!6}{2.355} & \cellcolor{gray!6}{2.465} & \cellcolor{gray!6}{0.736} & \cellcolor{gray!6}{\textbf{0.66}} & \cellcolor{gray!6}{2.415} & \cellcolor{gray!6}{2.373} & \cellcolor{gray!6}{0.939} & \cellcolor{gray!6}{\textbf{0.88}}\\
2500 & 50 & 2.561 & 3.107 & 0.123 & \textbf{0.093} & 3.073 & 2.698 & 0.156 & \textbf{0.097} & 2.992 & 2.712 & 0.201 & \textbf{0.121} & 3.043 & 2.276 & 0.247 & \textbf{0.159}\\
\cellcolor{gray!6}{2500} & \cellcolor{gray!6}{100} & \cellcolor{gray!6}{2.686} & \cellcolor{gray!6}{4.588} & \cellcolor{gray!6}{0.086} & \cellcolor{gray!6}{\textbf{0.048}} & \cellcolor{gray!6}{3.061} & \cellcolor{gray!6}{2.720} & \cellcolor{gray!6}{0.091} & \cellcolor{gray!6}{\textbf{0.057}} & \cellcolor{gray!6}{3.221} & \cellcolor{gray!6}{2.383} & \cellcolor{gray!6}{0.108} & \cellcolor{gray!6}{\textbf{0.065}} & \cellcolor{gray!6}{3.204} & \cellcolor{gray!6}{2.267} & \cellcolor{gray!6}{0.129} & \cellcolor{gray!6}{\textbf{0.079}}\\
2500 & 250 & 2.631 & 8.042 & 0.07 & \textbf{0.022} & 3.459 & 2.797 & 0.098 & \textbf{0.032} & 3.589 & 2.432 & 0.087 & \textbf{0.037} & 3.986 & 2.280 & 0.085 & \textbf{0.038}\\
\cellcolor{gray!6}{2500} & \cellcolor{gray!6}{500} & \cellcolor{gray!6}{3.101} & \cellcolor{gray!6}{8.044} & \cellcolor{gray!6}{0.096} & \cellcolor{gray!6}{\textbf{0.013}} & \cellcolor{gray!6}{3.468} & \cellcolor{gray!6}{3.191} & \cellcolor{gray!6}{0.11} & \cellcolor{gray!6}{\textbf{0.029}} & \cellcolor{gray!6}{3.527} & \cellcolor{gray!6}{2.412} & \cellcolor{gray!6}{0.092} & \cellcolor{gray!6}{\textbf{0.026}} & \cellcolor{gray!6}{4.099} & \cellcolor{gray!6}{2.259} & \cellcolor{gray!6}{0.088} & \cellcolor{gray!6}{\textbf{0.028}}\\
\bottomrule
\textbf{\footnotesize Note:} &
\multicolumn{17}{l}{\footnotesize Lower is better; best method for each $N$,$K$,$d$, in bold. Values are Root Mean Square Error for estimated vs true latent factors, averaged over $d$ dimensions, }\\
& \multicolumn{17}{l}{\footnotesize $N$ units, and 50 simulations. For Bayesian models estimates are posterior means computed by averaging over 10000 posterior samples. }\\
& \multicolumn{17}{l}{\footnotesize  All results from Bayesian models are computed from 4000 posterior samples obtained from 4 parallel MCMC chains}\\
& \multicolumn{17}{l}{\footnotesize after 2000 burn-in iterations.}
\end{tabular}}
\end{table}

We see that our proposed methodology performs uniformly better than both PCA and the same model without $M$-matrices. That is evidence that including $M$-matrices in the model does indeed lead to superior estimation performance. Allowing for correlated factors also seems to lead to improvements in estimation error, demonstrating the usefulness of that addition to our modeling framework. We report MSE and 95\% Credible Interval coverage results for both the latent factors and loadings in Appendix B.

\begin{table}
\caption{ESS convergence for $\btheta$.}
\label{tab:thetaess}
\centering
\resizebox{0.98\textwidth}{!}{%
\begin{tabular}{rr|lll|lll|lll|lll}
\toprule
N & K &  \multicolumn{3}{c|}{d=2} & \multicolumn{3}{c|}{d=3} & 
\multicolumn{3}{c|}{d=5} & \multicolumn{3}{c}{d=8} \\
\midrule
&  & IRT & IRT-M & IRT-M & IRT & IRT-M  & IRT-M & IRT & IRT-M  & IRT-M  & IRT  & IRT-M  & IRT-M \\
\multicolumn{2}{c|}{Correlated $\btheta$?} & No & No & Yes & No & No & Yes & No & No & Yes & No & No & Yes \\
\midrule
\cellcolor{gray!6}{10} & \cellcolor{gray!6}{10} & \cellcolor{gray!6}{39} & \cellcolor{gray!6}{74} & \cellcolor{gray!6}{\textbf{77}} & \cellcolor{gray!6}{38} & \cellcolor{gray!6}{65} & \cellcolor{gray!6}{\textbf{68}} & \cellcolor{gray!6}{48} & \cellcolor{gray!6}{53} & \cellcolor{gray!6}{\textbf{58}} & \cellcolor{gray!6}{--} & \cellcolor{gray!6}{--} & \cellcolor{gray!6}{--}\\
10 & 50 & 21 & 30 & \textbf{32} & 16 & 29 & \textbf{30} & 13 & 22 & \textbf{24} & -- & -- & --\\
\cellcolor{gray!6}{10} & \cellcolor{gray!6}{100} & \cellcolor{gray!6}{18} & \cellcolor{gray!6}{23} & \cellcolor{gray!6}{\textbf{24}} & \cellcolor{gray!6}{14} & \cellcolor{gray!6}{22} & \cellcolor{gray!6}{\textbf{23}} & \cellcolor{gray!6}{10} & \cellcolor{gray!6}{18} & \cellcolor{gray!6}{\textbf{19}} & \cellcolor{gray!6}{--} & \cellcolor{gray!6}{--} & \cellcolor{gray!6}{--}\\
10 & 250 & 15 & 18 & \textbf{19} & 12 & 18 & \textbf{19} & 8 & 16 & \textbf{16} & -- & -- & --\\
\cellcolor{gray!6}{10} & \cellcolor{gray!6}{500} & \cellcolor{gray!6}{14} & \cellcolor{gray!6}{16} & \cellcolor{gray!6}{\textbf{17}} & \cellcolor{gray!6}{11} & \cellcolor{gray!6}{\textbf{17}} & \cellcolor{gray!6}{17} & \cellcolor{gray!6}{7} & \cellcolor{gray!6}{\textbf{15}} & \cellcolor{gray!6}{15} & \cellcolor{gray!6}{--} & \cellcolor{gray!6}{--} & \cellcolor{gray!6}{--}\\
\addlinespace
50 & 10 & 68 & \textbf{149} & 149 & 49 & 133 & \textbf{134} & 43 & 121 & \textbf{123} & 51 & 108 & \textbf{112}\\
\cellcolor{gray!6}{50} & \cellcolor{gray!6}{50} & \cellcolor{gray!6}{49} & \cellcolor{gray!6}{\textbf{83}} & \cellcolor{gray!6}{83} & \cellcolor{gray!6}{35} & \cellcolor{gray!6}{78} & \cellcolor{gray!6}{\textbf{78}} & \cellcolor{gray!6}{25} & \cellcolor{gray!6}{65} & \cellcolor{gray!6}{\textbf{65}} & \cellcolor{gray!6}{20} & \cellcolor{gray!6}{49} & \cellcolor{gray!6}{\textbf{49}}\\
50 & 100 & 42 & \textbf{63} & 61 & 32 & 62 & \textbf{62} & 23 & 55 & \textbf{55} & 17 & 43 & \textbf{43}\\
\cellcolor{gray!6}{50} & \cellcolor{gray!6}{250} & \cellcolor{gray!6}{30} & \cellcolor{gray!6}{\textbf{44}} & \cellcolor{gray!6}{44} & \cellcolor{gray!6}{26} & \cellcolor{gray!6}{49} & \cellcolor{gray!6}{\textbf{49}} & \cellcolor{gray!6}{21} & \cellcolor{gray!6}{47} & \cellcolor{gray!6}{\textbf{47}} & \cellcolor{gray!6}{16} & \cellcolor{gray!6}{\textbf{38}} & \cellcolor{gray!6}{38}\\
50 & 500 & 27 & \textbf{38} & 37 & 23 & 43 & \textbf{44} & 19 & \textbf{43} & 43 & 16 & \textbf{36} & 36\\
\addlinespace
\cellcolor{gray!6}{100} & \cellcolor{gray!6}{10} & \cellcolor{gray!6}{97} & \cellcolor{gray!6}{\textbf{180}} & \cellcolor{gray!6}{179} & \cellcolor{gray!6}{67} & \cellcolor{gray!6}{\textbf{168}} & \cellcolor{gray!6}{167} & \cellcolor{gray!6}{54} & \cellcolor{gray!6}{\textbf{148}} & \cellcolor{gray!6}{148} & \cellcolor{gray!6}{57} & \cellcolor{gray!6}{\textbf{140}} & \cellcolor{gray!6}{139}\\
100 & 50 & 72 & \textbf{117} & 117 & 55 & \textbf{110} & 110 & 37 & \textbf{93} & 92 & 28 & \textbf{71} & 70\\
\cellcolor{gray!6}{100} & \cellcolor{gray!6}{100} & \cellcolor{gray!6}{61} & \cellcolor{gray!6}{\textbf{93}} & \cellcolor{gray!6}{92} & \cellcolor{gray!6}{46} & \cellcolor{gray!6}{91} & \cellcolor{gray!6}{\textbf{92}} & \cellcolor{gray!6}{32} & \cellcolor{gray!6}{\textbf{80}} & \cellcolor{gray!6}{79} & \cellcolor{gray!6}{24} & \cellcolor{gray!6}{\textbf{63}} & \cellcolor{gray!6}{62}\\
100 & 250 & 45 & 67 & \textbf{68} & 38 & 72 & \textbf{72} & 30 & \textbf{68} & 68 & 23 & \textbf{57} & 56\\
\cellcolor{gray!6}{100} & \cellcolor{gray!6}{500} & \cellcolor{gray!6}{39} & \cellcolor{gray!6}{\textbf{58}} & \cellcolor{gray!6}{58} & \cellcolor{gray!6}{34} & \cellcolor{gray!6}{61} & \cellcolor{gray!6}{\textbf{62}} & \cellcolor{gray!6}{27} & \cellcolor{gray!6}{\textbf{63}} & \cellcolor{gray!6}{63} & \cellcolor{gray!6}{22} & \cellcolor{gray!6}{\textbf{54}} & \cellcolor{gray!6}{54}\\
\addlinespace
250 & 10 & 151 & \textbf{230} & 229 & 112 & \textbf{213} & 207 & 85 & \textbf{193} & 185 & 79 & \textbf{179} & 173\\
\cellcolor{gray!6}{250} & \cellcolor{gray!6}{50} & \cellcolor{gray!6}{122} & \cellcolor{gray!6}{165} & \cellcolor{gray!6}{\textbf{165}} & \cellcolor{gray!6}{95} & \cellcolor{gray!6}{\textbf{154}} & \cellcolor{gray!6}{153} & \cellcolor{gray!6}{67} & \cellcolor{gray!6}{\textbf{130}} & \cellcolor{gray!6}{127} & \cellcolor{gray!6}{50} & \cellcolor{gray!6}{\textbf{104}} & \cellcolor{gray!6}{100}\\
250 & 100 & 107 & \textbf{143} & 143 & 84 & 135 & \textbf{136} & 60 & \textbf{117} & 116 & 44 & \textbf{95} & 93\\
\cellcolor{gray!6}{250} & \cellcolor{gray!6}{250} & \cellcolor{gray!6}{82} & \cellcolor{gray!6}{111} & \cellcolor{gray!6}{\textbf{114}} & \cellcolor{gray!6}{68} & \cellcolor{gray!6}{109} & \cellcolor{gray!6}{\textbf{112}} & \cellcolor{gray!6}{48} & \cellcolor{gray!6}{99} & \cellcolor{gray!6}{\textbf{101}} & \cellcolor{gray!6}{37} & \cellcolor{gray!6}{\textbf{85}} & \cellcolor{gray!6}{84}\\
250 & 500 & 67 & 90 & \textbf{95} & 56 & 95 & \textbf{97} & 43 & 92 & \textbf{93} & 34 & \textbf{80} & 80\\
\addlinespace
\cellcolor{gray!6}{500} & \cellcolor{gray!6}{10} & \cellcolor{gray!6}{199} & \cellcolor{gray!6}{\textbf{253}} & \cellcolor{gray!6}{248} & \cellcolor{gray!6}{157} & \cellcolor{gray!6}{\textbf{242}} & \cellcolor{gray!6}{237} & \cellcolor{gray!6}{121} & \cellcolor{gray!6}{\textbf{229}} & \cellcolor{gray!6}{216} & \cellcolor{gray!6}{108} & \cellcolor{gray!6}{\textbf{206}} & \cellcolor{gray!6}{194}\\
500 & 50 & 162 & \textbf{194} & 193 & 133 & \textbf{178} & 176 & 102 & \textbf{157} & 152 & 76 & \textbf{133} & 126\\
\cellcolor{gray!6}{500} & \cellcolor{gray!6}{100} & \cellcolor{gray!6}{142} & \cellcolor{gray!6}{174} & \cellcolor{gray!6}{\textbf{176}} & \cellcolor{gray!6}{120} & \cellcolor{gray!6}{\textbf{160}} & \cellcolor{gray!6}{159} & \cellcolor{gray!6}{92} & \cellcolor{gray!6}{\textbf{143}} & \cellcolor{gray!6}{141} & \cellcolor{gray!6}{69} & \cellcolor{gray!6}{\textbf{115}} & \cellcolor{gray!6}{111}\\
500 & 250 & 122 & 148 & \textbf{152} & 102 & 139 & \textbf{145} & 80 & 125 & \textbf{126} & 59 & \textbf{105} & 104\\
\cellcolor{gray!6}{500} & \cellcolor{gray!6}{500} & \cellcolor{gray!6}{102} & \cellcolor{gray!6}{126} & \cellcolor{gray!6}{\textbf{133}} & \cellcolor{gray!6}{86} & \cellcolor{gray!6}{125} & \cellcolor{gray!6}{\textbf{130}} & \cellcolor{gray!6}{65} & \cellcolor{gray!6}{113} & \cellcolor{gray!6}{\textbf{117}} & \cellcolor{gray!6}{50} & \cellcolor{gray!6}{98} & \cellcolor{gray!6}{\textbf{99}}\\
\addlinespace
1000 & 10 & 248 & \textbf{277} & 269 & 203 & \textbf{265} & 255 & 165 & \textbf{240} & 232 & 142 & \textbf{232} & 216\\
\cellcolor{gray!6}{1000} & \cellcolor{gray!6}{50} & \cellcolor{gray!6}{188} & \cellcolor{gray!6}{\textbf{201}} & \cellcolor{gray!6}{199} & \cellcolor{gray!6}{174} & \cellcolor{gray!6}{\textbf{202}} & \cellcolor{gray!6}{198} & \cellcolor{gray!6}{139} & \cellcolor{gray!6}{\textbf{177}} & \cellcolor{gray!6}{167} & \cellcolor{gray!6}{108} & \cellcolor{gray!6}{\textbf{150}} & \cellcolor{gray!6}{138}\\
1000 & 100 & 180 & \textbf{194} & 192 & 156 & \textbf{181} & 176 & 126 & \textbf{159} & 151 & 100 & \textbf{134} & 127\\
\cellcolor{gray!6}{1000} & \cellcolor{gray!6}{250} & \cellcolor{gray!6}{157} & \cellcolor{gray!6}{171} & \cellcolor{gray!6}{\textbf{174}} & \cellcolor{gray!6}{142} & \cellcolor{gray!6}{\textbf{164}} & \cellcolor{gray!6}{163} & \cellcolor{gray!6}{111} & \cellcolor{gray!6}{\textbf{143}} & \cellcolor{gray!6}{140} & \cellcolor{gray!6}{89} & \cellcolor{gray!6}{\textbf{120}} & \cellcolor{gray!6}{116}\\
1000 & 500 & 148 & 159 & \textbf{165} & 123 & 151 & \textbf{151} & 100 & \textbf{135} & 133 & 77 & \textbf{112} & 111\\
\addlinespace
\cellcolor{gray!6}{2500} & \cellcolor{gray!6}{10} & \cellcolor{gray!6}{\textbf{289}} & \cellcolor{gray!6}{286} & \cellcolor{gray!6}{275} & \cellcolor{gray!6}{275} & \cellcolor{gray!6}{\textbf{294}} & \cellcolor{gray!6}{273} & \cellcolor{gray!6}{228} & \cellcolor{gray!6}{\textbf{274}} & \cellcolor{gray!6}{251} & \cellcolor{gray!6}{194} & \cellcolor{gray!6}{\textbf{248}} & \cellcolor{gray!6}{243}\\
2500 & 50 & 216 & \textbf{217} & 211 & \textbf{214} & 206 & 198 & 188 & \textbf{193} & 179 & 156 & \textbf{169} & 148\\
\cellcolor{gray!6}{2500} & \cellcolor{gray!6}{100} & \cellcolor{gray!6}{204} & \cellcolor{gray!6}{\textbf{206}} & \cellcolor{gray!6}{203} & \cellcolor{gray!6}{\textbf{196}} & \cellcolor{gray!6}{196} & \cellcolor{gray!6}{188} & \cellcolor{gray!6}{173} & \cellcolor{gray!6}{\textbf{176}} & \cellcolor{gray!6}{164} & \cellcolor{gray!6}{142} & \cellcolor{gray!6}{\textbf{150}} & \cellcolor{gray!6}{135}\\
2500 & 250 & 189 & \textbf{193} & 187 & 176 & \textbf{183} & 175 & 153 & \textbf{161} & 149 & 130 & \textbf{139} & 125\\
\cellcolor{gray!6}{2500} & \cellcolor{gray!6}{500} & \cellcolor{gray!6}{184} & \cellcolor{gray!6}{\textbf{187}} & \cellcolor{gray!6}{186} & \cellcolor{gray!6}{167} & \cellcolor{gray!6}{\textbf{173}} & \cellcolor{gray!6}{164} & \cellcolor{gray!6}{141} & \cellcolor{gray!6}{\textbf{151}} & \cellcolor{gray!6}{143} & \cellcolor{gray!6}{116} & \cellcolor{gray!6}{\textbf{128}} & \cellcolor{gray!6}{120}\\
\bottomrule
\textbf{\footnotesize Note:} & \multicolumn{13}{l}{\footnotesize Higher is better. Best method for each $N$,$K$,$d$, in bold. Values are ESS averaged over $N$ units, $d$ dimensions, }\\
&\multicolumn{13}{l}{\footnotesize and 50 simulations. ESS is a statistic that outputs the number of fully i.i.d. samples that have the same estimation }\\
&\multicolumn{13}{l}{\footnotesize power as the autocorrelated MCMC samples. Here ESS is computed over 10000 posterior samples. }\\
& \multicolumn{13}{l}{\footnotesize  All results from Bayesian models are computed from 4000 posterior samples obtained from 4 parallel MCMC chains}\\
& \multicolumn{13}{l}{\footnotesize after 2000 burn-in iterations.}
\end{tabular}}
\end{table}

We next compare our model (IRT-M) with the other Bayesian method (IRT), in terms of speed of convergence of the MCMC sampler for estimating the latent factors. To perform that comparison we use the version of the Effective Sample Size (ESS) statistic proposed in \cite{vehtari2021rank}. Results are shown in Table \ref{tab:thetaess}; larger values denote more information being captured in the Markov chain. We see that IRT-M displays similar or better convergence to simple IRT after the same number of iterations. This is generally true for both the version of our model with correlated factors and the version without; however, the uncorrelated model displays overall best convergence. This is expected, as exploring the posterior of correlated factors is generally harder than exploring that of uncorrelated factors. In addition, Tables \ref{tab:thetagew} and \ref{tab:thetarhat} in Appendix B show that our models achieve similar convergence performance according to other widely used MCMC convergence statistics. 

Those results indicate that our proposed approach can achieve performance superior to the standard methodology in terms of MSE without requiring the same strong assumptions and while displaying similar convergence properties. The last is notable because our model requires sampling from a truncated multivariate normal posterior when loadings are direction-constrained by $M$: that sampling step is nontrivial and requires MCMC tools like rejection or Gibbs sampling in itself \citep{wilhelm2010tmvtnorm}. Our convergence comparison shows that this harder sampling step does not affect speed and MCMC mixing performance. Note here that including the prior on variance hurts us in terms of relative convergence, but we still do well despite that.

\paragraph{Performance under misspecification of $M$-matrices} We have seen that our model performs well given a correctly specified set of $M$-matrices, but in real applications coding of the $M$-matrices is unlikely to be perfect. Thus, we also assess the robustness of our methodology to misspecification of the $M$-matrices. We do so in the same simulated-data setting as before, but this time our model is given progressively more misspecified $M$-matrices. Those misspecified $M$ matrices have two effects, and so misspecification affects the model in two ways. One, it alters elements in the diagonals of the matrices, which alters whether or not a certain item loads on a certain factor. Two, those altered loadings are used to generate anchor points, using the procedure described previously.

\begin{figure}[!htbp]
    \caption{MSE of the 3-dimensional IRT model at progressively more misspecified $M$-matrices. }
    \label{Fig:Misspecification}
    \includegraphics[width=\linewidth]{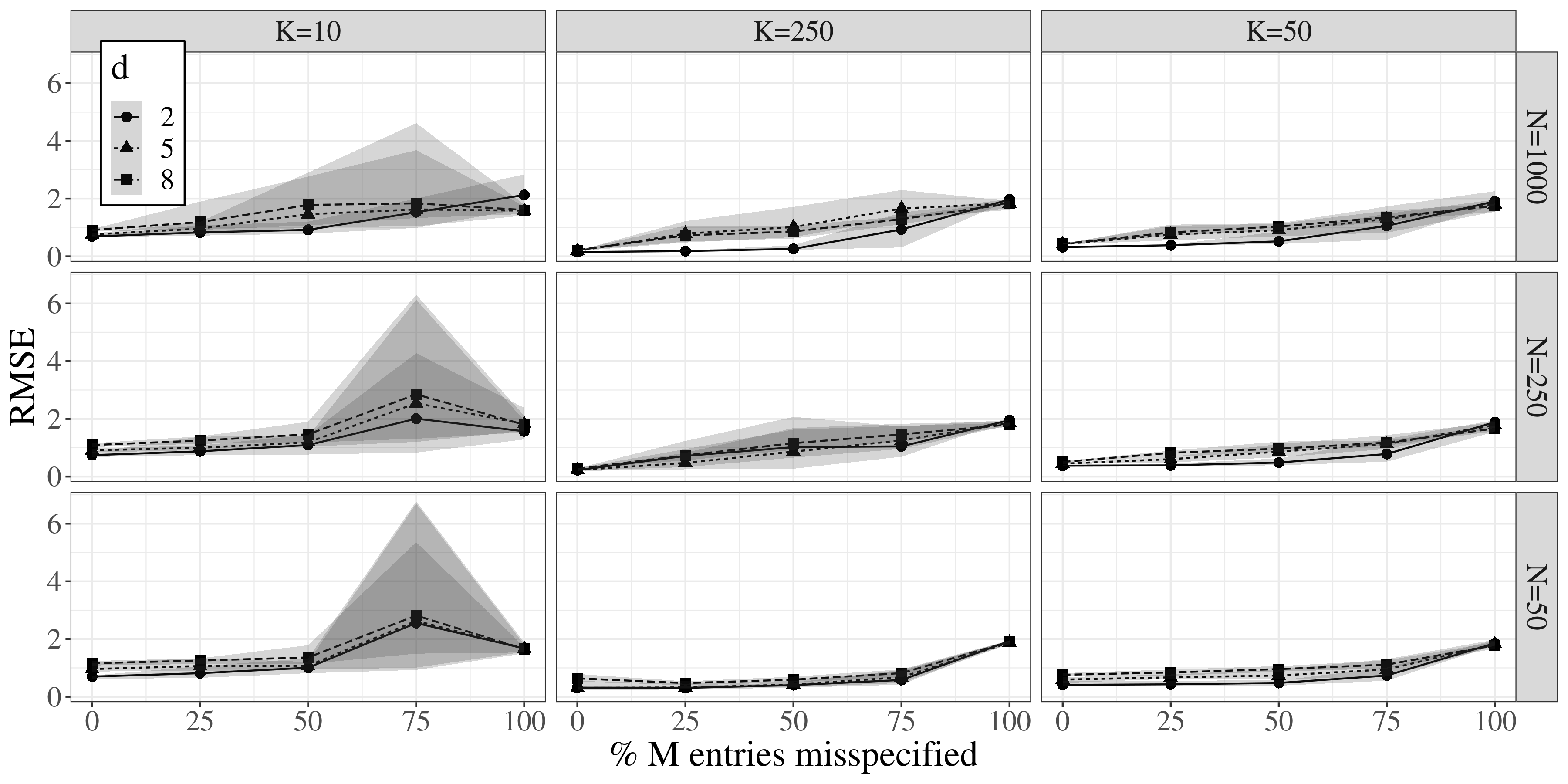}
  
    \footnotesize{\textbf{Note:} Values on the horizontal axis represent percentages of misspecification of $M$-matrix diagonals. The leftmost point corresponds to a model in which the $M$-matrices are completely correct, while the rightmost point corresponds to a model in which the $M$-matrices are completely misspecified. }
\end{figure}

Results are shown in Figure \ref{Fig:Misspecification}. The figure shows an expected trend: when the $M$ matrices are in large part misspecified, model performance is poor, while a correctly specified model displays very low error. The key point, however, is that, in general, even with only half of the correct specification for the $M$-matrices, our model still performs very well.

\vspace{-.1in}
\subsection*{Application to Roll Call Data}

Our model performs quite well according to reasonable benchmarks, as long as there exists ground truth to capture. Our motivating example shows that IRT-M also produces reasonable results in the absence of ground truth. However, in that example we had no existing measures with which to compare ours. Therefore, for our final exercise in model validation, we apply the same procedure detailed in our motivating example to Congressional roll call data from four sources: the 85$^{th}$ and 109$^{th}$ US House and Senate. Roll call data are commonly used as inputs to ideal-point estimation techniques \citep[e.g.,][]{poole1985, poole1991, clinton2004, treier2011, aldrich2014, tahk2018}. Further, those techniques use exogenous information about legislators' latent ideological positions to identify the models, providing a level of ground truth. In contrast, to apply IRT-M we will code bills---the items in this context---and make no assumptions regarding legislator positions. We can then compare latent positions derived from applying IRT-M with a two-dimensional latent space to existing ideal point measures such as DW-NOMINATE (DWN) scores, obtainable from \url{voteview.com}. That comparison not only reinforces the validity of IRT-M, it also helps us understand better the substantive consequences of employing correlated latent dimensions and varying what gets coded as part of each latent dimension.

The first step in applying IRT-M to roll call data is to specify a set of theoretically-informed latent dimensions. Our approach does not fix the number of latent dimensions; however, to compare to DWN we need to limit the number of latent dimensions to two. As there is no theoretically unique set of two latent dimensions given the range of substantive topics addressed in Congress, we opted for three different sets of coding rules. Coding rule A uses dimensions corresponding to 1) Economic/Redistribution and 2) Social/Cultural/Civil Rights/Equality. Coding rule B uses dimensions corresponding to 1) Economy/Distribution/Power and 2) Civil Rights. Coding rule C uses dimensions corresponding to 1) Economy/Public Distribution/Power and 2) Civil Rights/Redistribution. Bills not corresponding to any of those topics are coded $0$ for all latent dimensions. The major differences between coding rules are where Redistribution falls and what else gets lumped together with Civil Rights. Appendix C contains a full description of all three coding rules, and we make all coded bills and files to replicate figures available as well.

\begin{sidewaysfigure}
\centering

\begin{subfigure}[t]{.47\textwidth}
\centering
\includegraphics[width=\linewidth]{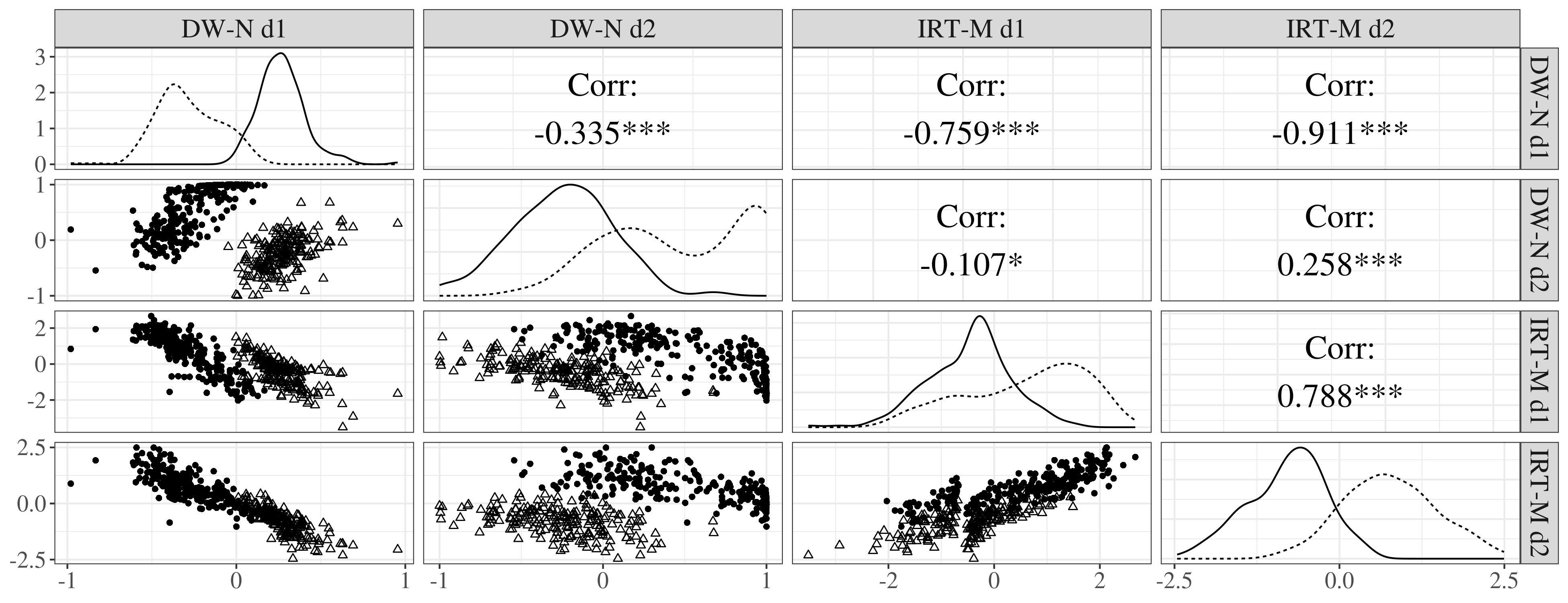}
        \caption{85$^{th}$ House, coding rule A}\label{figcod1:fig_a}
\end{subfigure}
\begin{subfigure}[t]{.47\textwidth}
\centering
\includegraphics[width=\linewidth]{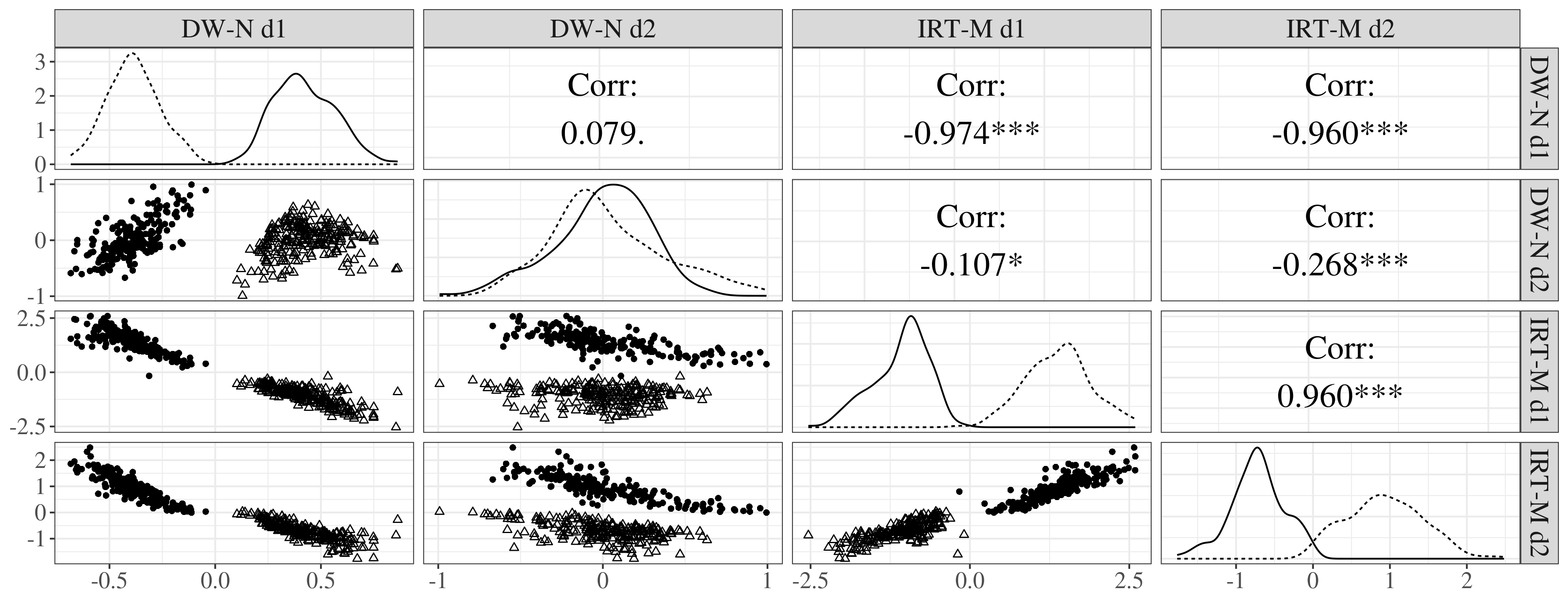}
\caption{109$^{th}$ House, coding rule A}\label{figcod1:fig_b}
\end{subfigure}

\medskip

\begin{subfigure}[t]{.47\textwidth}
\centering
\vspace{0pt}
\includegraphics[width=\linewidth]{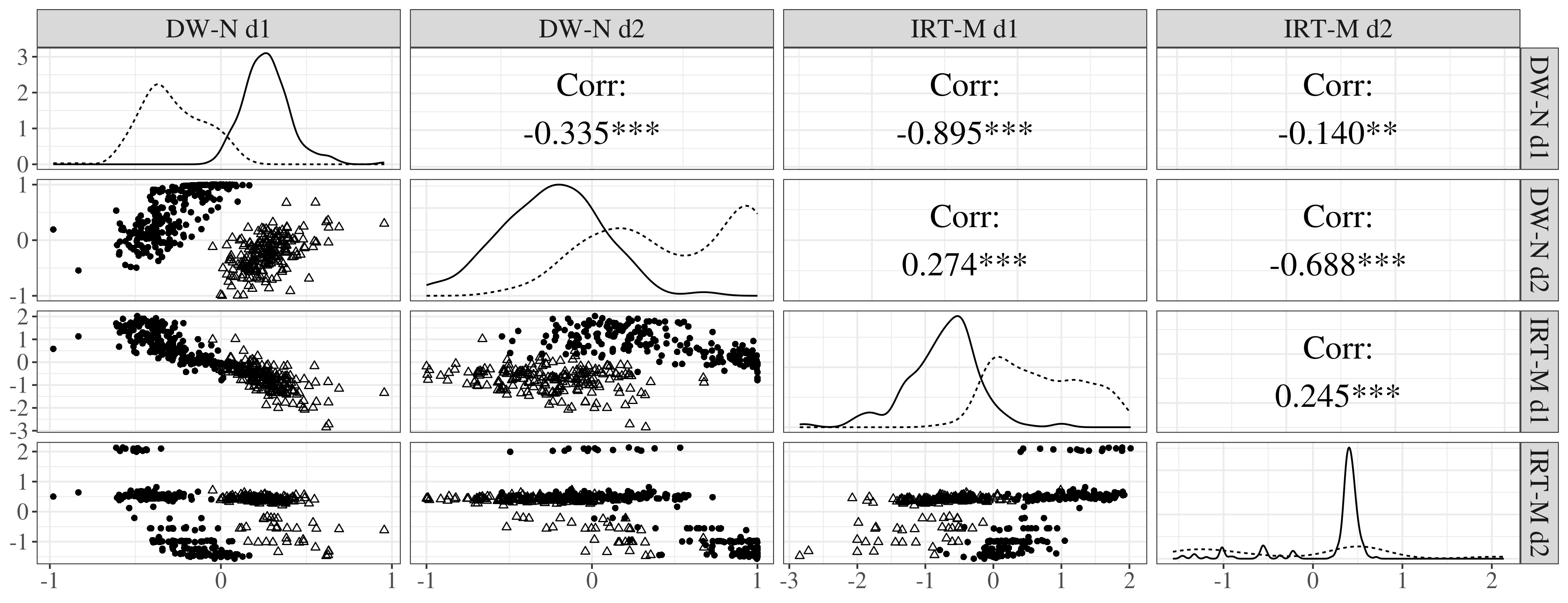}
\caption{85$^{th}$ House, coding rule B}\label{figcod1:fig_c}
\end{subfigure}
\begin{subfigure}[t]{.47\textwidth}
\centering
\vspace{0pt}
\includegraphics[width=\linewidth]{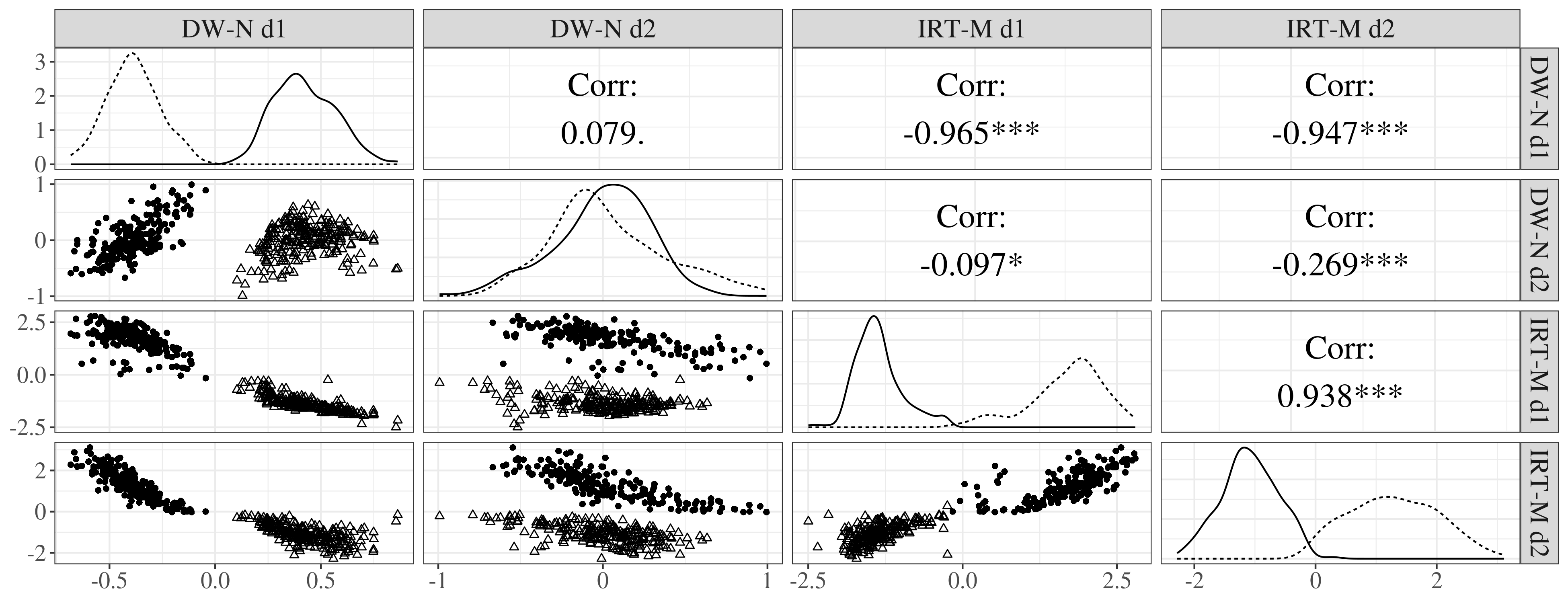}
\caption{109$^{th}$ House, coding rule B}\label{figcod1:fig_d}
\end{subfigure}

\medskip

\begin{subfigure}[t]{.47\textwidth}
\centering
\vspace{0pt}
\includegraphics[width=\linewidth]{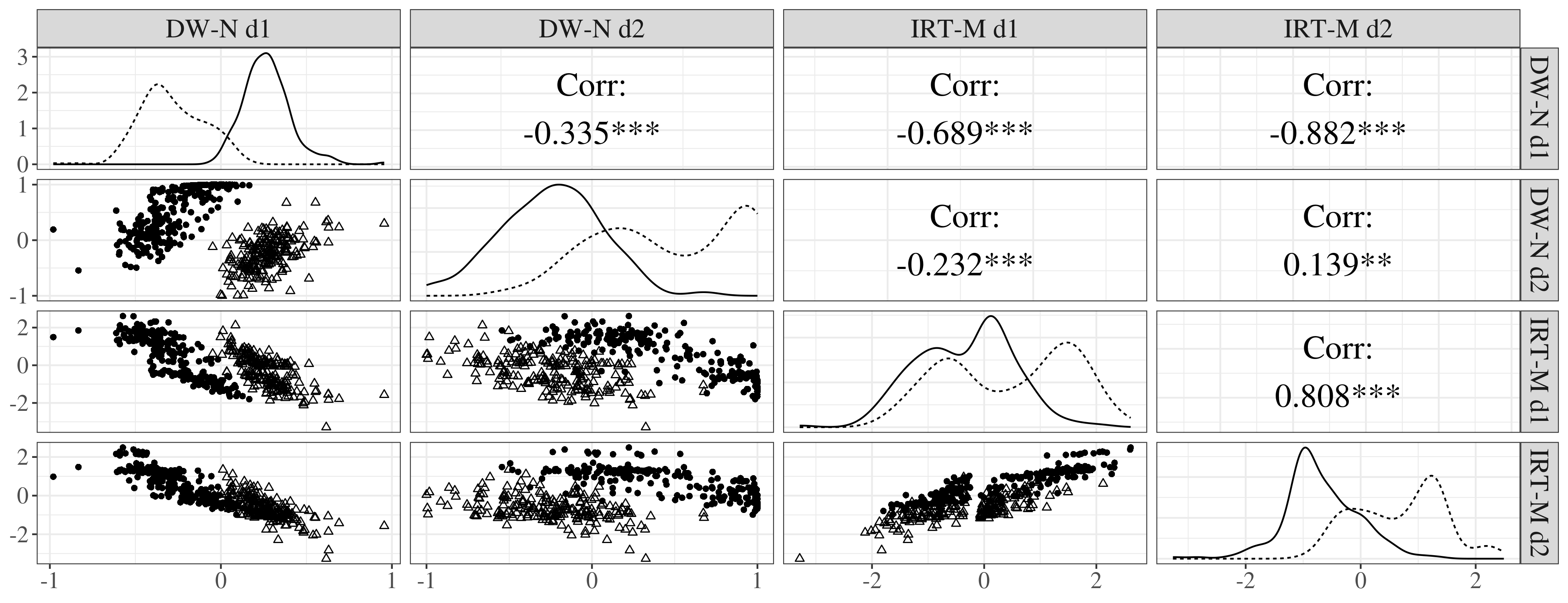}
\caption{85$^{th}$ House, coding rule C}\label{figcod1:fig_e}
\end{subfigure}
\begin{subfigure}[t]{.47\textwidth}
\centering
\vspace{0pt}
\includegraphics[width=\linewidth]{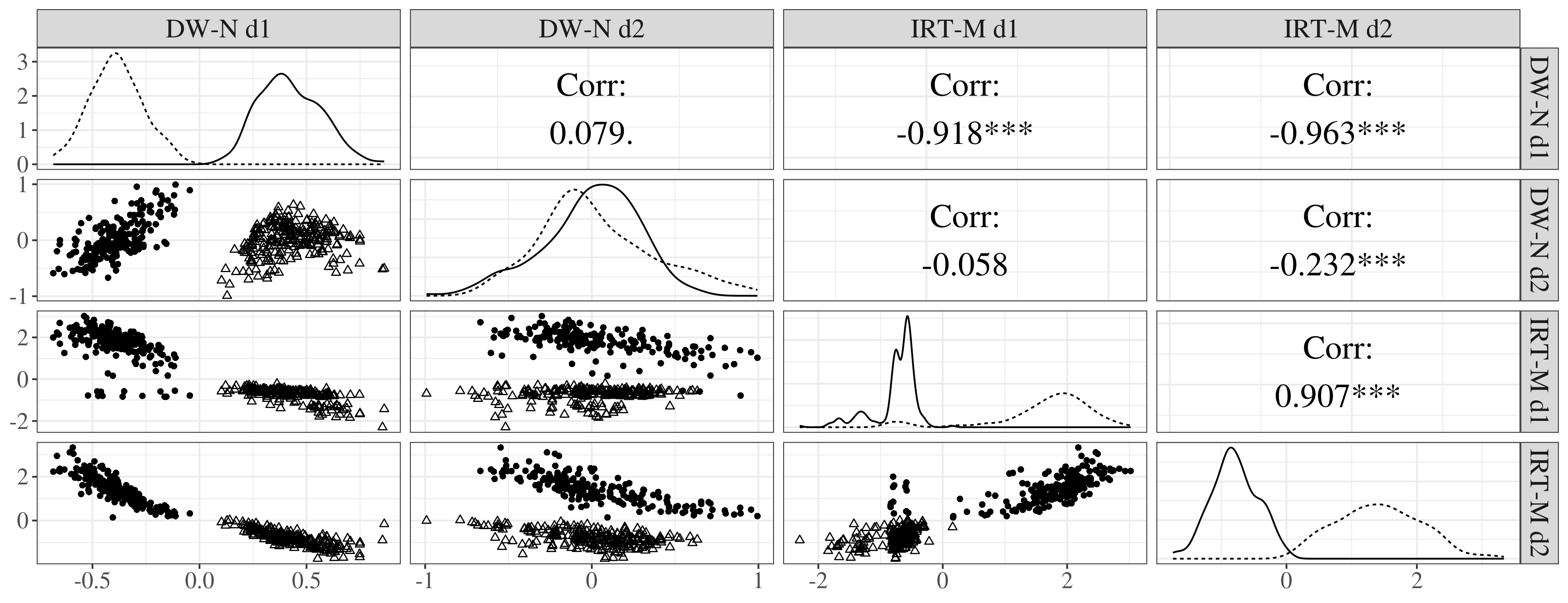}
\caption{109$^{th}$ House, coding rule C}\label{figcod1:fig_f}
\end{subfigure}

\footnotesize{\textbf{Note:} Each row/colum within each subfigure is one of the latent dimensions estimated either by Nominate or IRT-M. The bottom triangle of each subfigure displays scatterplots with each pair of dimensions on each axis. The diagonal contains density plots for each pair of dimensions. The top triangle contains Spearman correlation coefficients for each pair of dimensions.}

\caption{Correlations between IRT-M and DW-NOMINATE ideal points in the House. Solid line, Triangle=Democratic, Dashed line, filled circle=Republican}
\label{fig:nomcorrH}
\end{sidewaysfigure}

Figure \ref{fig:nomcorrH} expresses correlations between different latent dimensions for the 85$^{th}$ and 109$^{th}$ House of Representatives. A similar figure for the Senate is in Appendix D. There are six plots in the figure, one for each House and coding rule combination. Within each plot, the four rows and columns correspond to the first and second latent dimensions derived from DWN and IRT-M. Within each plot are sixteen subplots, one for each pair of those four dimensions. The subplots along the diagonal illustrate the distribution of that latent dimension for each of Democrats and Republicans. Above the diagonal, which will be our focus, are correlations between the latent dimension in the column and that in the row, both in the aggregate and broken down by political party. We will focus on those correlations in our brief discussion of model validity, though there is much more one could do with our analysis than we have space for here. High correlations between any two dimensions indicate that those dimensions predict voting behavior similarly.

We first highlight two features of Figure \ref{fig:nomcorrH} that speak to model validity. One, in all but coding rule B of the 85$^{th}$ House, our two dimensions are highly correlated. In contrast, DWN's two dimensions are mildly correlated in the 85$^{th}$ House and nearly uncorrelated in the 109$^{th}$. Given the breadth of bill topics covered in all dimensions we have coded, save for the second dimension in coding rule B which uniquely specifies civil rights, that suggests there may be an underlying factor, such as partisanship, driving voting behavior across a range of bills that may obscure more complex ideological preferences \citep{aldrich2014}. The increasing correlation of our two dimensions over time is consistent with that point. The difference observed in coding rule B of the 85$^{th}$ House is also consistent with that point, given the known schism in the Democratic party at that time over civil rights.

Two, our first latent dimension is consistently highly correlated with DWN's first dimension. That suggests that we are capturing a similar theoretical concept to that captured in DWN's first, economic, dimension. For all but coding rule B of the 85$^{th}$ House, which solely captures Civil Rights, our second dimension is also strongly correlated with DWN's first, for reasons we have noted. However, for coding rule B of the 85$^{th}$ House, our second dimension is much more highly correlated with DWN's second dimension. That provides further confidence in the validity of our measures, since that dimension in DWN is typically interpreted as being related to civil rights as well, unlike DWN's second dimension in the 109$^{th}$ House, which does not share that interpretation.

Together, those two features of Figure \ref{fig:nomcorrH} support our claim to the validity of our approach, and lend us more confidence that we are able to capture substantive meanings without making assumptions about legislators' preferences. They also illustrate the substantive effects of two aspects of IRT-M: correlated latent dimensions and coding rules. Modeling correlated latent dimensions lets us capture a scenario in which at least one plausible underlying factor drives more than one latent dimension. The underlying factor does not eliminate the meaning of the two latent dimensions---they still have the substantive meanings our coding rules constrained them to have---but it does suggest a richer causal story, of the type described in \cite{aldrich2014}. That richer story also suggests that, absent correlated dimensions and our coding rules, one dimension found by an unconstrained IRT model may capture a complex combination of theoretical concepts driven by an underlying factor, while other dimensions capture concepts not present in the theory, and potentially of less importance. It is possible that DWN applied to the 109$^{th}$ House is doing just that. Its first dimension, termed ``Economic/Redistributive'' on \url{voteview.com}, may be capturing the influence of partisanship, while its second dimension, left without a meaning and called ``Other Votes'' on \url{voteview.com}, may be capturing behavior that is not part of the underlying theory of two-dimensional ideological voting.

Using coding rules to capture the link between theory and measure has several implications. One is that, as long as the coding rules are adjusted for variation in meaning across time or space, the substantive meaning of the latent dimensions will remain constant. As an example, in coding rule B, the second dimension captures civil rights in both time periods. We can see that the split in the Democratic party during the 85$^{th}$ House is no longer appreciably present by the time of the 109$^{th}$ House as both dimensions are highly correlated in the 109$^{th}$ House. A second implication is that the items that go into what theoretical concepts each dimension captures fundamentally change what each dimension is measuring. As we saw, isolating civil rights as its own dimension leads to a very different measure than what we obtained under either set of coding rules in which it was combined with other related, but theoretically distinct, concepts.

\section*{Conclusion}

Measurement is the necessary bridge between theory and empirical test. In the social sciences, it's also the weak link: we traffic in complex concepts such as ideology, identity, and legitimacy, and deriving appropriate measures of those concepts is not trivial. Yet, without measurement that matches our theoretical constructs, our careful empirical studies may not truly be testing what they were intended to test. Further, in the absence of measurement that holds its meaning across time and place, it is difficult to build on each other's work. What captures an aspect of ideology or legitimacy in one context, for instance, may not carry into another.

Dimensional-reduction techniques, such as Item Response Theory (IRT) models, produce improvable measures of latent variables that are thought to underlie behavior, and their construction is transparent given knowledge of the data from which they draw. However, such unsupervised methods do not provide the latent dimensions they discover with intrinsic substantive interpretations. Prior approaches to assigning substantive meaning to latent dimensions either require additional information about units in the data---for example, the ideological positions of certain well-known legislators---or limit the number of latent dimensions to one. As a result, prior approaches cannot solve the problem of substantive interpretation when additional information is not available---as it is not, for example, in anonymous surveys---and the theory under investigation includes multiple, potentially correlated, latent dimensions.

We offer a novel solution to the problem of substantive interpretation: the IRT-M model. Applying the IRT-M model requires coding all responses by a set of units to a set of items---for example, responses by people to survey questions or votes by legislators on bills---according to whether and how that response could be predicted by each latent dimension in one's theory. That coding, used in conjunction with Bayesian Item Response Theory, allows IRT-M to produce posterior distributions over each unit's position on each of the latent dimensions, with each latent dimension having the substantive meaning specified in one's theory. We provide two worked examples of IRT-M's use: a motivating example applying it to survey data and latent dimensions of threats and attitudes, and an example applying it to roll call data in the U.S. Congress used as model validation. We also provide an R package that will allow analysts to apply the IRT-M model to their own data and theories.

In applying IRT-M, it is important to keep in mind the importance of theory to the approach. IRT-M is not designed to provide latent dimensions that would best predict responses to items by units in the data. Rather, it is designed to best measure theoretical concepts that may have driven behavior by data units. Thus, a theory as to how the latent dimensions are affecting the responses of the units is essential to the approach. That theory drives the coding of each item-dimension pair. As long as the theory is applied consistently in coding, one can apply IRT-M to disparate data sources across time and place. In all cases, IRT-M will return substantively meaningful latent dimensions, allowing one to make comparisons that previously required exogenous assumptions on individual units. For example, in the roll call application, ideal points located on a civil rights dimension maintain more or less the same meaning across time, as long as sufficient votes related to civil rights continue to be taken.

In the future, we aim to extend our model to additional forms of data, but even limited to dichotomous item data, our approach makes possible consistent, theoretically meaningful measurements that can combine data from numerous sources. We expect such measurements to be substantially more precise than simple indices, let alone single proxies, of the types of theoretical concepts social scientists consider. For instance, our motivating example---employing IRT-M to draw out latent concepts from survey data---illustrates how one can use existing survey data to test social-scientific theories in a straightforward fashion.

\paragraph{Acknowledgement} The authors affirm this research did not involve human subjects. The authors declare no ethical issues or conflicts of interest in this research. This research was funded by the National Science Foundation under grant no. SES-1727249. Research documentation and/or data that support the findings of this study are openly available in the APSR Dataverse at https://doi.org/10.7910/DVN/3G.


\bibliographystyle{plainnat}
\bibliography{biblio}

\newpage
\setcounter{page}{1}
\begin{center}\LARGE{\textbf{Supplemental Information}}\end{center}

\section{Appendix A: Detailed Model Specification and Implementation}\label{app:model}

As stated in the main paper, we adopt the following hierarchical model:
\begin{align}
    Y_{ik}|\mu_{ik} &= \ind_{[\mu_{ik} > 0]}&& i=1\dots, N,\,k = 1, \dots, K\label{Eq:Ymodel2}\\
    \mu_{ik}|\blambda_k, \btheta_i, b_k &\sim \mathcal{N}(\blambda_k^T\btheta_i - b_k, 1)&& i=1\dots, N,\,k = 1, \dots, K \label{Eq:Mumodel2}\\
    \lambda_{kj}|m_{kjj} &\sim 
    \begin{cases}
    \mathcal{N}_{[0, \infty]}(0, m_{kjj}^2) &\mbox{ if } m_{kjj} > 0\\
    \mathcal{N}_{[-\infty, 0]}(0, m_{kjj}^2) &\mbox{ if } m_{kjj} < 0\\
    \delta_{[\lambda_{kj} = 0]} &\mbox{ if } m_{kjj} = 0\\
    \mathcal{N}(0, 5) &\mbox{ if } m_{kjj} = \texttt{NA}\\
    \end{cases}&& k=1, \dots, K,\, j=1,\dots,d\label{Eq:lambdaPrior2}\\
    b_k &\sim \mathcal{N}(0, 1)&& k=1,\dots,K\label{Eq:bPrior2}\\
    \btheta_i|\Sigma &\sim \mathcal{N}_d(\boldsymbol{0}, \boldsymbol{\Sigma})&& i=1, \dots, N\label{Eq:thetaPrior2}\\
    \Sigma|\nu_0, \mathbf{S}_0 &\sim \mathcal{IW}_d(\nu_0, \mathbf{S}_0),&&\label{Eq:SigmaPrior2}
\end{align}

Note that the $d$ subscript on $\mathcal{N}_d$ and $\mathcal{IW}_d$ refers to their dimension throughout; it is not an index. The outcome model is a classical ogive (probit) IRT \citep{lord1953application}. Specifically, we assume: 
\begin{equation}Pr(Y_{ik} = 1) = \Phi(\blambda_k^T \btheta_i - b_k)\label{eq:probone}\end{equation}
where $\Phi$ is the standard normal CDF, $\blambda_k$ is the usual vector of $d$ loadings associated with item $k$, $\btheta_i$ is the vector of $d$ latent factor scores for unit $i$, and $b_k$ is the baseline ``discrimination'' parameter for item $k$. Note that, as stated in the Extensions subsection, that equation can be replaced with a logistic or multinomial-logistic specification without altering the main core of the model. The (standard) idea behind the model is to decompose the likelihood of an affirmative (``1'') answer to item $k$ by unit $i$ into $d$ dimensions, each of which could have either a positive, negative, or zero effect on that likelihood. Since direct estimation of Eq. \eqref{eq:probone} is often complex due to its nonlinearity, we adopt the classical Gibbs sampling scheme proposed first by \cite{albert1992bayesian}, and now widely used. That scheme proposes the creation of the normally-distributed latent-variable $\mu_{ik}$ for every outcome (Eq. \eqref{Eq:Mumodel2}, and then conditional posterior updates for the other parameters in the model can be derived conditional on that variable. Notably, if we condition on $\mu_{ik}$, we see that $Y_{ik}$ becomes: $$Y_{ik}|\mu_{ik} = \ind_{[\mu_{ij} > 0]} = \begin{cases}1 &\mbox{ if } \mu_{ik} > 0 \\ 0 & \mbox{otherwise},\end{cases}$$
where we use $\ind_{{[\cdot]}}$ to represent the indicator function.

While so far our modeling choices have been standard, we now introduce a new model for the loadings $\blambda$ that implements our main theoretical contribution. For each item $k$, we model $\lambda_k$ conditionally on a $d\times d$ diagonal matrix specific to that item. The diagonal of such a matrix is allowed to contain either real values, or the special $\texttt{NA}$ code to denote missingness. We then assign a prior distribution to $\blambda_k$ conditional on the values of the diagonal of $M_k$: for each $j=1,\dots,d$, we independently draw $\lambda_{kj}$ from a positive-truncated normal distribution, if $m_{kjj} > 0$, or a negative-truncated normal distribution if $m_{kjj} < 0$. In essence, this constrains item $k$ to only either load positively or negatively on factor $j$, as specified by the user. Additionally, positive or negative entries of $m_{kjj}$ can also be given an absolute value of the user's response: this should encode any prior on the extent to which the item loads on the factor. If the user believes that item $k$ is strongly determined by factor $j$, then $m_{kjj}$ can have a large absolute value, whereas if the user believes that factor $j$ only contributes little to item $k$, then $m_{kjj}$ can have a small absolute value. This is then reflected in the model by using $m_{kjj}^2$ as variance for $\lambda_{kj}$. The user can also specify that item $k$ does not load on factor $j$ at all, i.e., a respondent's level of factor $j$ has no impact on whether that respondent will answer positively to item $k$; the user can do so by setting $m_{kjj} = 0$, in which case $\lambda_{kj}$ is drawn from a distribution that puts density 1 at $0$ and 0 everywhere else, which is represented by Dirac's delta function $\delta_{[\lambda_{kj} = 0]}$ in Eq. \eqref{Eq:lambdaPrior}. In this case, $\lambda_{kj} = 0$ w.p. 1. Finally, a user might not know whether item $k$ is expected to load positively, negatively, or not at all on factor $j$: in this case the user should set $m_{kjj} = \texttt{NA}$, and $\lambda_{kj}$ will be drawn from a standard normal distribution. The classical ``discrimination'' parameter is represented by $b_k$ in our modeling framework: we do not constrain sampling of this parameter in any way dependent on the $M$-matrix. The latent factors $\btheta_i$ are sampled independently from a $d$-dimensional multivariate normal distribution for each respondent, $i$ (Eq. \eqref{Eq:thetaPrior2}). Notably, the restrictions imposed on the loadings by our framework allow a covariance matrix $\Sigma$ to be learned for the factors: we accomplish this by placing a $d$-dimensional Inverse-Wishart conjugate prior on $\Sigma$  (Eq. \eqref{Eq:SigmaPrior2}). With this full specification, our model has only three hyperparameters: the $M$-matrix, $\nu_0$, and $\mathbf{S}_0$, the latter being the degrees of freedom and co-variance parameters for the Inverse-Wishart prior on $\Sigma$. 

\vspace{-.1in}
\subsection{Gibbs sampler}
Posterior sampling for the model introduced in the previous section is implemented via standard MCMC techniques. Specifically, we derived and implemented a Gibbs sampler for the model based on fully conjugate conditional posterior updates for all the model parameters. We reproduce the steps of the sampler here as pseudocode. Throughout the following we will denote vectors and matrices containing multiple parameters by omitting the relevant indices from the respective symbols, and by displaying those symbols in bold. For example, if $\mu_{ik}$ denotes the value of $\mu$ for data unit $i$ and item $k$, then $\bmu_i = [\mu_{i1}, \dots, \mu_{iK}]^T$ denotes a $K$-dimensional column vector containing all the values of $\mu$ associated with data unit $i$, and $\bmu$ a $N\times K$ matrix of values of $\mu$ for all $N$ data units and $K$ items in which the $i^{th}$ row is the (transposed) vector $\bmu_i$ and the $k^{th}$ column is a $N$-dimensional vector of all $\mu$ values associated with item $k$. 

All the full conditional posterior updates for each of the parameters in our sampler are derived using the standard formulas for normal-normal, and normal-inverse Wishart conjugate updates given, for example, in \cite{hoff2009first}, and as such we omit a full derivation here.

Our Gibbs sampler is as follows:

\noindent 1.  Initialize $\bmu \in \mathbb{R}^{N\times K}$, $\bb \in \mathbb{R}^{K}$, $\btheta \in \mathbb{R}^{N \times P}, \blambda \in \mathbb{R}^{K \times d}, \Sigma \in \mathbb{R}^{d\times d}$ by drawing once from their prior distributions as specified in Eqns. \eqref{Eq:Mumodel2}-\eqref{Eq:SigmaPrior2}.\\ 
\noindent 2. For $s = 1,\dots S$ sampling iterations:
    \begin{enumerate}
        \item For $i=1,\dots, N$ sample $\btheta_i \sim  \mathcal{N}_d(B_{i}, V_{i}^{-1})$, where $V_i = \blambda^T \blambda + \Sigma^{-1}$, and $B_i = V_i^{-1}\blambda^T(\bb + \bmu_i)$.
        \item For $i=1, \dots, N$, $k=1,\dots, K$ sample: $$\mu_{ik} \sim \begin{cases}
         \mathcal{N}_{[0, \infty]}(B_{ik}, 1) &\mbox{ if } Y_{ik} = 1\\
         \mathcal{N}_{[-\infty, 0]}(B_{ik}, 1) &\mbox{ if } Y_{ik} = 0\\
         \mathcal{N}(B_{ik}, 1) &\mbox{ if } Y_{ik} = \text{\texttt{NA}},
        \end{cases}$$
        where $B_{ik} = \blambda_k^{T} \btheta_i + b_k$. 
        \item Sample $\Sigma^* \sim \mathcal{IW}_d(N + \nu_0, \btheta^T\btheta + S_0)$. Initialize $d\times d$ diagonal matrix $\Sigma$, and for $i,j=1,\dots, d, j \neq i$, set $\Sigma_{ij} = \Sigma^*_{ij}/\Sigma^*_{ii}$. 
        \item For $k=1, \dots, K$, sample $b_k \sim \mathcal{N}(B_k V_k)$, where $V_k = \frac{1}{N + 1}$, and $B_k = V_k\sum_{i=1}^N \btheta_i \blambda_k^T - \mu_{ik}$
        \item For $k=1,\dots,K$, let $\bL_{k}$, and $\bU_k$ be $d$-dimensional vectors of lower and upper bounds for item $k$, where, for $j = 1,\dots,d$: 
        \begin{align*}
            \bL_{kj} = \begin{cases} 
            - \infty &\mbox{ if } m_{kjj} < 0\\
            - \infty &\mbox{ if } m_{kjj} = \texttt{NA}\\
            0 &\mbox{ if } m_{kjj} \geq 0
            \end{cases}, \quad
            \bU_{kj} = \begin{cases} 
            \infty &\mbox{ if } m_{kjj} > 0\\
            \infty &\mbox{ if } m_{kjj} = \texttt{NA}\\
            0 &\mbox{ if } m_{kjj} \leq 0.
            \end{cases}
        \end{align*}
        Additionally define the $d\times d$ diagonal matrix $\Omega_k$, such that the $j^{th}$ element of the diagonal is defined as: $\Omega_{kjj} = |m_{kjj}|^2$. Finally, sample: $\blambda_k \sim \mathcal{N}_{d, [\bL_k, \bU_k]}(B_k, V_k^{-1})$, where: $V_k = \btheta^t\btheta + \Omega_k^{-1}$, and $B_k = V_k^{-1}\btheta^t(b_k + \bmu_{k})$, and $\mathcal{N}_{[d, \bL_k, \bU_k]}$ is the truncated multivariate normal distribution of dimension $d$, where each dimension is truncated between the bounds defined in the respective dimensions of the vectors $\bL_k$ and $\bU_k$, and such that if $L_{kj} = U_{kj}$, then $\lambda_{kj} = L_{kj}$ when sampled from this distribution. 
        \item Store the values of $\bmu$, $\btheta$, $\blambda$, $\bb$, $\Sigma$ sampled at this iteration. 
    \end{enumerate}

\subsection*{Learning independent factors with correlated loadings}
Our model can be modified to allow for learning correlated loadings but independent factors. This is useful in case the analyst is interested in explicitly independent latent dimensions, and maintains the rotation invariance requirements needed for model identification. 
In order to implement this modification it is sufficient to first replace the prior on $\btheta_i$ defined in Eq. (\ref{Eq:thetaPrior2}) with $\btheta_i \sim \mathcal{N}_d(\mathbf{0}, \mathbf{I})$, where $\mathbf{I}$ is the $d\times d$ identity matrix. Second, the prior on $\lambda_{kj}$ in Equation \eqref{Eq:lambdaPrior2} should be modified to $\blambda_k \sim \mathcal{N}_{[d, \bL_k, \bU_k]}(\mathbf{0}, \Omega_k)$, where $\bL_k$ and $\bU_k$ are lower and upper bounds on each of the $d$-dimensions of the normal distribution and they are defined at step (e) of the Gibbs sampler introduced earlier. Finally, the $d$-dimensional matrix $\Omega_k$ should be given the prior induced by sampling $\Omega_k^* \sim \mathcal{IW}_d(\nu_0, \mathbf{S}_0)$, and then setting each diagonal element of $\Omega_k$ to: $$\Omega_{kjj} = \begin{cases}\Omega^*_{kjj} &\mbox{ if } m_{kjj} \neq 0\\ 0 &\mbox{ if } m_{kjj} = 0\end{cases},$$ 
and each off-diagonal element of $\Omega_{kj\ell}$ to:
$$\Omega_{kj\ell} = \begin{cases}\Omega^*_{kj\ell}/\Omega_{kjj} & \mbox{ if } m_{kjj} \neq 0\\ 
0 & \mbox{ if } m_{kjj} = 0\end{cases}.$$
Sampling from this model is possible by adapting the Gibbs sampling scheme used for the model with correlated factors.

\subsection*{Over-Identification}
Our approach provides a method for researchers to link their theoretical expectations about latent constructs to models that estimate those constructs from data. That implies that, in some settings, researchers' theoretical expectations might result in more than the $d(d-1)$ model constraints required for model identification. In this case it is said that the model is \textit{over-identified}: an identified (but not over-identified) model will learn the unique value of the latent dimensions that best fits the data according to some numerical measure of fit (this can be shown to approximately equal a penalized L2 distance between observed data and factor-loading combinations in the case of our model). Conversely, an over-identified model may not return latent dimensions that best maximize model fit, since the additional constraints imposed on the model may be ruling out precisely those values of the latent parameters that maximize fit. 

Depending on the researcher's needs, over-identification may or may not be desirable: a researcher that has no strong theoretical priors over the \textit{meaning} of their dimensions may be fine with latent quantities that are learned to maximize some numerical measure of fit, while researchers who specifically want to target the estimation of latent dimensions that conform to some theoretical expectation may still want to impose constraints on such latent quantities, even if this means that their model will not be the one that best fits the data. If the latent dimensions found by a model that is not over-identified are very different from those found by an over-identified model, then that should be a sign for the researcher that the data may not support the theoretical priors they hold about their latent dimensions. In that case, either the theory needs revision or the data are a poor match to the theory. Because of this, a practical suggestion for researchers is to first fit a model with the bare minimum number of constraints required for identification, and then fit a model with all the constraints that they want to impose. Comparing the dimensions returned by each model will be informative: if the latent quantities are very different, then either the data are a poor fit for the theory or the theory behind the over-identified model may need to be revised.  

Finally, we remark that our Simulation depicted in Figure \ref{Fig:Misspecification} shows that even in the event that a model is over-identified with constraints that are misspecified, according to some population model, latent dimensions output by IRT-M are still largely similar to their target values as long as not too many of the constraints are misspecified. 

\section{Appendix B: Additional Simulation Information}
\subsection{General Simulation Setup}
All simulations were run for each $N, K, d$-triple a total of 50 times. At each iteration data was generated from the model detailed in the paper as follows: 
\begin{align*}
    \rho_{j, \ell} &\sim \text{Uniform}(-1, 1) & j, \ell = 1, \dots, d\\
    \Sigma &= \begin{bmatrix}
                1, & \dots, &\rho_{1, d}\\
                \rho_{1,2}, &\dots, & \rho_{2, d}\\ 
                 &   \ddots & \\
                 \rho_{1, d},& \dots, &1 \\
              \end{bmatrix}\\
    \btheta_i &\sim \mathcal{N}_d(\mathbf{0}, \Sigma) & i = 1,\dots, N\\
    u_{kj} &\sim \text{Uniform}(0, 1), &k = 1, \dots, K,\; j = 1,\dots, d\\
    \lambda_{k, j} & \sim \begin{cases} \mathcal{N}(0, 1) &\mbox{ if } u_{ij} > 0.25\\
    0 &\mbox{ otherwise. }\end{cases}&k = 1, \dots, K,\; j = 1,\dots, d\\
    b_k &\sim \mathcal{N}(0, 1)&k = 1, \dots, K,\\
    \mu_{ik} &\sim \mathcal{N}(\blambda_k^T\btheta_i - b_k, 1)&i = 1, \dots, N,\; k = 1,\dots, K\\
    Y_{ik} & = \ind_{[\mu_{ik} > 0]}. &i = 1, \dots, N,\; k = 1,\dots, K
\end{align*}
We keep the proportion of loadings that are 0 fixed at 25\% in all our simulations. $M$-matrices for our models are then generated according to the sign of the generated $\lambda_{k,j}$; i.e., we generate, for $k =1, \dots K$:
\begin{align*}
    \bM_k = \begin{bmatrix}
        \text{sign}(\lambda_{kj}), &\dots, & 0\\
        & \ddots & \\
        0, &\dots, & \text{sign}(\lambda_{kj})\\
    \end{bmatrix},\; \text{ with }\; \text{sign}(\lambda_{kj}) = \begin{cases}1 &\mbox{ if }\lambda_{kj} > 0 \\ -1 &\mbox{ if }\lambda_{kj} < 0 \\ 0 &\mbox{ if }\lambda_{kj} = 0 \\ \end{cases}.
\end{align*}

\subsection{Additional Simulation Results}
We report additional results from the simulations conducted and introduced in the main paper.

Tables \ref{tab:lambdamse} and \ref{tab:lambdacov} respectively show MSE and 95\% credible interval coverage for learning the loadings, $\blambda$. We see that IRT-M generally performs well in learning loadings, and largely outperforms PCA and traditional IRT. Adding factor correlation does seem to give the model a small boost in performance for this task, especially as the amount of respondents ($N$) grows. In terms of coverage all models seem to undercover the true parameter values: this is somewhat expected as learning individualized parameters with good uncertainty estimation is a generally hard problem. Nonetheless, IRT-M presents a substantial improvement in terms of coverage for both $\blambda$ and $\btheta$ over the traditional methodologies. 

\begin{table}
\caption{MSE for $\blambda$.}
\label{tab:lambdamse}
\centering
\resizebox{0.98\textwidth}{!}{%
\begin{tabular}{rr|llll|llll|llll|llll}
\toprule
N & K &  \multicolumn{4}{c|}{d=2} & \multicolumn{4}{c|}{d=3} & 
\multicolumn{4}{c|}{d=5} & \multicolumn{4}{c}{d=8} \\
\midrule
&  & IRT & IRT-M & IRT-M & IRT & IRT-M  & IRT-M & IRT & IRT-M  & IRT-M  & IRT  & IRT-M  & IRT-M \\
\multicolumn{2}{c|}{Correlated $\btheta$?} & No & No & Yes & No & No & Yes & No & No & Yes & No & No & Yes \\
\midrule
\cellcolor{gray!6}{10} & \cellcolor{gray!6}{10} & \cellcolor{gray!6}{2.077} & \cellcolor{gray!6}{3.811} & \cellcolor{gray!6}{\textbf{0.418}} & \cellcolor{gray!6}{0.421} & \cellcolor{gray!6}{2.446} & \cellcolor{gray!6}{3.147} & \cellcolor{gray!6}{\textbf{0.468}} & \cellcolor{gray!6}{0.481} & \cellcolor{gray!6}{2.429} & \cellcolor{gray!6}{2.947} & \cellcolor{gray!6}{\textbf{0.592}} & \cellcolor{gray!6}{0.592} & \cellcolor{gray!6}{--} & \cellcolor{gray!6}{--} & \cellcolor{gray!6}{--} & \cellcolor{gray!6}{--}\\
10 & 50 & 1.788 & 3.341 & 0.301 & \textbf{0.296} & 1.985 & 2.904 & 0.339 & \textbf{0.338} & 2.079 & 2.799 & \textbf{0.359} & 0.362 & -- & -- & -- & --\\
\cellcolor{gray!6}{10} & \cellcolor{gray!6}{100} & \cellcolor{gray!6}{2.146} & \cellcolor{gray!6}{6.458} & \cellcolor{gray!6}{\textbf{0.295}} & \cellcolor{gray!6}{0.299} & \cellcolor{gray!6}{1.970} & \cellcolor{gray!6}{4.284} & \cellcolor{gray!6}{\textbf{0.302}} & \cellcolor{gray!6}{0.302} & \cellcolor{gray!6}{1.976} & \cellcolor{gray!6}{3.087} & \cellcolor{gray!6}{0.334} & \cellcolor{gray!6}{\textbf{0.333}} & \cellcolor{gray!6}{--} & \cellcolor{gray!6}{--} & \cellcolor{gray!6}{--} & \cellcolor{gray!6}{--}\\
10 & 250 & 2.036 & 7.229 & 0.283 & \textbf{0.282} & 1.931 & 4.621 & 0.301 & \textbf{0.3} & 2.036 & 3.247 & 0.324 & \textbf{0.322} & -- & -- & -- & --\\
\cellcolor{gray!6}{10} & \cellcolor{gray!6}{500} & \cellcolor{gray!6}{1.925} & \cellcolor{gray!6}{9.382} & \cellcolor{gray!6}{\textbf{0.285}} & \cellcolor{gray!6}{0.285} & \cellcolor{gray!6}{1.933} & \cellcolor{gray!6}{6.991} & \cellcolor{gray!6}{0.287} & \cellcolor{gray!6}{\textbf{0.286}} & \cellcolor{gray!6}{1.955} & \cellcolor{gray!6}{3.099} & \cellcolor{gray!6}{\textbf{0.312}} & \cellcolor{gray!6}{0.312} & \cellcolor{gray!6}{--} & \cellcolor{gray!6}{--} & \cellcolor{gray!6}{--} & \cellcolor{gray!6}{--}\\
\addlinespace
50 & 10 & 2.245 & 3.477 & 0.31 & \textbf{0.299} & 2.204 & 3.320 & 0.393 & \textbf{0.391} & 2.369 & 3.365 & 0.521 & \textbf{0.515} & 2.507 & 2.981 & \textbf{0.575} & 0.575\\
\cellcolor{gray!6}{50} & \cellcolor{gray!6}{50} & \cellcolor{gray!6}{2.017} & \cellcolor{gray!6}{3.266} & \cellcolor{gray!6}{\textbf{0.158}} & \cellcolor{gray!6}{0.158} & \cellcolor{gray!6}{1.990} & \cellcolor{gray!6}{3.246} & \cellcolor{gray!6}{\textbf{0.178}} & \cellcolor{gray!6}{0.179} & \cellcolor{gray!6}{2.040} & \cellcolor{gray!6}{2.666} & \cellcolor{gray!6}{0.207} & \cellcolor{gray!6}{\textbf{0.206}} & \cellcolor{gray!6}{2.067} & \cellcolor{gray!6}{2.511} & \cellcolor{gray!6}{0.257} & \cellcolor{gray!6}{\textbf{0.253}}\\
50 & 100 & 2.394 & 6.855 & 0.151 & \textbf{0.149} & 1.917 & 4.472 & \textbf{0.157} & 0.157 & 2.062 & 2.917 & 0.175 & \textbf{0.174} & 2.027 & 2.495 & 0.219 & \textbf{0.216}\\
\cellcolor{gray!6}{50} & \cellcolor{gray!6}{250} & \cellcolor{gray!6}{2.002} & \cellcolor{gray!6}{15.945} & \cellcolor{gray!6}{\textbf{0.139}} & \cellcolor{gray!6}{0.139} & \cellcolor{gray!6}{2.074} & \cellcolor{gray!6}{9.536} & \cellcolor{gray!6}{\textbf{0.14}} & \cellcolor{gray!6}{0.141} & \cellcolor{gray!6}{2.083} & \cellcolor{gray!6}{4.385} & \cellcolor{gray!6}{0.16} & \cellcolor{gray!6}{\textbf{0.158}} & \cellcolor{gray!6}{1.949} & \cellcolor{gray!6}{2.847} & \cellcolor{gray!6}{\textbf{0.186}} & \cellcolor{gray!6}{0.186}\\
50 & 500 & 2.052 & 16.590 & \textbf{0.122} & 0.123 & 2.049 & 17.645 & \textbf{0.132} & 0.132 & 2.028 & 4.888 & \textbf{0.15} & 0.15 & 2.009 & 3.168 & 0.185 & \textbf{0.184}\\
\addlinespace
\cellcolor{gray!6}{100} & \cellcolor{gray!6}{10} & \cellcolor{gray!6}{2.319} & \cellcolor{gray!6}{3.616} & \cellcolor{gray!6}{0.285} & \cellcolor{gray!6}{\textbf{0.273}} & \cellcolor{gray!6}{2.286} & \cellcolor{gray!6}{3.635} & \cellcolor{gray!6}{0.402} & \cellcolor{gray!6}{\textbf{0.386}} & \cellcolor{gray!6}{2.327} & \cellcolor{gray!6}{3.250} & \cellcolor{gray!6}{0.478} & \cellcolor{gray!6}{\textbf{0.463}} & \cellcolor{gray!6}{2.403} & \cellcolor{gray!6}{3.031} & \cellcolor{gray!6}{0.609} & \cellcolor{gray!6}{\textbf{0.582}}\\
100 & 50 & 2.080 & 3.419 & 0.139 & \textbf{0.138} & 2.145 & 2.877 & 0.139 & \textbf{0.137} & 2.094 & 2.821 & 0.17 & \textbf{0.165} & 2.060 & 2.473 & 0.21 & \textbf{0.203}\\
\cellcolor{gray!6}{100} & \cellcolor{gray!6}{100} & \cellcolor{gray!6}{1.763} & \cellcolor{gray!6}{6.281} & \cellcolor{gray!6}{\textbf{0.104}} & \cellcolor{gray!6}{0.104} & \cellcolor{gray!6}{1.984} & \cellcolor{gray!6}{3.265} & \cellcolor{gray!6}{0.109} & \cellcolor{gray!6}{\textbf{0.108}} & \cellcolor{gray!6}{2.104} & \cellcolor{gray!6}{2.781} & \cellcolor{gray!6}{0.134} & \cellcolor{gray!6}{\textbf{0.133}} & \cellcolor{gray!6}{2.014} & \cellcolor{gray!6}{2.454} & \cellcolor{gray!6}{0.155} & \cellcolor{gray!6}{\textbf{0.153}}\\
100 & 250 & 1.967 & 7.106 & \textbf{0.082} & 0.082 & 1.870 & 5.350 & 0.098 & \textbf{0.097} & 2.049 & 3.724 & \textbf{0.11} & 0.111 & 2.007 & 2.982 & 0.127 & \textbf{0.126}\\
\cellcolor{gray!6}{100} & \cellcolor{gray!6}{500} & \cellcolor{gray!6}{2.013} & \cellcolor{gray!6}{27.811} & \cellcolor{gray!6}{\textbf{0.074}} & \cellcolor{gray!6}{0.074} & \cellcolor{gray!6}{1.878} & \cellcolor{gray!6}{11.669} & \cellcolor{gray!6}{0.095} & \cellcolor{gray!6}{\textbf{0.094}} & \cellcolor{gray!6}{2.054} & \cellcolor{gray!6}{4.519} & \cellcolor{gray!6}{\textbf{0.101}} & \cellcolor{gray!6}{0.101} & \cellcolor{gray!6}{2.019} & \cellcolor{gray!6}{3.136} & \cellcolor{gray!6}{\textbf{0.115}} & \cellcolor{gray!6}{0.115}\\
\addlinespace
250 & 10 & 2.443 & 4.047 & 0.289 & \textbf{0.254} & 2.433 & 3.888 & 0.437 & \textbf{0.411} & 2.463 & 3.603 & 0.556 & \textbf{0.545} & 2.494 & 3.304 & 0.629 & \textbf{0.603}\\
\cellcolor{gray!6}{250} & \cellcolor{gray!6}{50} & \cellcolor{gray!6}{2.191} & \cellcolor{gray!6}{4.766} & \cellcolor{gray!6}{0.118} & \cellcolor{gray!6}{\textbf{0.103}} & \cellcolor{gray!6}{2.040} & \cellcolor{gray!6}{3.027} & \cellcolor{gray!6}{0.132} & \cellcolor{gray!6}{\textbf{0.116}} & \cellcolor{gray!6}{2.060} & \cellcolor{gray!6}{2.898} & \cellcolor{gray!6}{0.141} & \cellcolor{gray!6}{\textbf{0.126}} & \cellcolor{gray!6}{2.084} & \cellcolor{gray!6}{2.545} & \cellcolor{gray!6}{0.195} & \cellcolor{gray!6}{\textbf{0.172}}\\
250 & 100 & 1.753 & 4.496 & 0.065 & \textbf{0.06} & 2.018 & 3.728 & 0.093 & \textbf{0.087} & 2.084 & 3.079 & 0.118 & \textbf{0.11} & 2.002 & 2.376 & 0.127 & \textbf{0.117}\\
\cellcolor{gray!6}{250} & \cellcolor{gray!6}{250} & \cellcolor{gray!6}{1.982} & \cellcolor{gray!6}{5.132} & \cellcolor{gray!6}{0.04} & \cellcolor{gray!6}{\textbf{0.039}} & \cellcolor{gray!6}{1.998} & \cellcolor{gray!6}{3.818} & \cellcolor{gray!6}{\textbf{0.068}} & \cellcolor{gray!6}{0.069} & \cellcolor{gray!6}{2.066} & \cellcolor{gray!6}{3.386} & \cellcolor{gray!6}{\textbf{0.073}} & \cellcolor{gray!6}{0.073} & \cellcolor{gray!6}{1.971} & \cellcolor{gray!6}{2.648} & \cellcolor{gray!6}{\textbf{0.079}} & \cellcolor{gray!6}{0.079}\\
250 & 500 & 1.992 & 15.136 & 0.06 & \textbf{0.059} & 2.069 & 4.849 & \textbf{0.053} & 0.053 & 1.994 & 4.257 & \textbf{0.064} & 0.065 & 2.027 & 3.005 & 0.072 & \textbf{0.071}\\
\addlinespace
\cellcolor{gray!6}{500} & \cellcolor{gray!6}{10} & \cellcolor{gray!6}{2.357} & \cellcolor{gray!6}{5.684} & \cellcolor{gray!6}{0.185} & \cellcolor{gray!6}{\textbf{0.161}} & \cellcolor{gray!6}{2.156} & \cellcolor{gray!6}{3.039} & \cellcolor{gray!6}{0.398} & \cellcolor{gray!6}{\textbf{0.354}} & \cellcolor{gray!6}{2.387} & \cellcolor{gray!6}{3.006} & \cellcolor{gray!6}{0.66} & \cellcolor{gray!6}{\textbf{0.562}} & \cellcolor{gray!6}{2.468} & \cellcolor{gray!6}{3.139} & \cellcolor{gray!6}{0.724} & \cellcolor{gray!6}{\textbf{0.688}}\\
500 & 50 & 2.066 & 4.473 & 0.106 & \textbf{0.074} & 2.180 & 2.962 & 0.127 & \textbf{0.094} & 2.145 & 2.564 & 0.18 & \textbf{0.133} & 2.046 & 2.491 & 0.204 & \textbf{0.161}\\
\cellcolor{gray!6}{500} & \cellcolor{gray!6}{100} & \cellcolor{gray!6}{2.039} & \cellcolor{gray!6}{5.515} & \cellcolor{gray!6}{0.093} & \cellcolor{gray!6}{\textbf{0.063}} & \cellcolor{gray!6}{1.948} & \cellcolor{gray!6}{3.081} & \cellcolor{gray!6}{0.093} & \cellcolor{gray!6}{\textbf{0.074}} & \cellcolor{gray!6}{2.041} & \cellcolor{gray!6}{2.630} & \cellcolor{gray!6}{0.114} & \cellcolor{gray!6}{\textbf{0.088}} & \cellcolor{gray!6}{1.983} & \cellcolor{gray!6}{2.430} & \cellcolor{gray!6}{0.128} & \cellcolor{gray!6}{\textbf{0.102}}\\
500 & 250 & 1.830 & 7.330 & 0.067 & \textbf{0.05} & 1.857 & 5.287 & 0.063 & \textbf{0.054} & 1.965 & 2.499 & 0.079 & \textbf{0.067} & 1.965 & 2.405 & 0.09 & \textbf{0.08}\\
\cellcolor{gray!6}{500} & \cellcolor{gray!6}{500} & \cellcolor{gray!6}{1.785} & \cellcolor{gray!6}{4.906} & \cellcolor{gray!6}{0.042} & \cellcolor{gray!6}{\textbf{0.035}} & \cellcolor{gray!6}{1.997} & \cellcolor{gray!6}{6.309} & \cellcolor{gray!6}{0.063} & \cellcolor{gray!6}{\textbf{0.059}} & \cellcolor{gray!6}{2.088} & \cellcolor{gray!6}{3.419} & \cellcolor{gray!6}{0.056} & \cellcolor{gray!6}{\textbf{0.055}} & \cellcolor{gray!6}{2.043} & \cellcolor{gray!6}{2.687} & \cellcolor{gray!6}{0.057} & \cellcolor{gray!6}{\textbf{0.054}}\\
\addlinespace
1000 & 10 & 2.379 & 3.173 & 0.177 & \textbf{0.162} & 2.255 & 3.427 & 0.251 & \textbf{0.222} & 2.478 & 3.250 & 0.691 & \textbf{0.586} & 2.406 & 3.255 & 0.85 & \textbf{0.763}\\
\cellcolor{gray!6}{1000} & \cellcolor{gray!6}{50} & \cellcolor{gray!6}{1.990} & \cellcolor{gray!6}{5.189} & \cellcolor{gray!6}{0.064} & \cellcolor{gray!6}{\textbf{0.04}} & \cellcolor{gray!6}{2.089} & \cellcolor{gray!6}{4.736} & \cellcolor{gray!6}{0.228} & \cellcolor{gray!6}{\textbf{0.122}} & \cellcolor{gray!6}{2.029} & \cellcolor{gray!6}{2.706} & \cellcolor{gray!6}{0.19} & \cellcolor{gray!6}{\textbf{0.128}} & \cellcolor{gray!6}{1.998} & \cellcolor{gray!6}{2.734} & \cellcolor{gray!6}{0.196} & \cellcolor{gray!6}{\textbf{0.143}}\\
1000 & 100 & 1.966 & 3.733 & 0.129 & \textbf{0.077} & 2.145 & 3.403 & 0.164 & \textbf{0.096} & 2.102 & 2.691 & 0.142 & \textbf{0.099} & 2.013 & 2.410 & 0.147 & \textbf{0.09}\\
\cellcolor{gray!6}{1000} & \cellcolor{gray!6}{250} & \cellcolor{gray!6}{1.796} & \cellcolor{gray!6}{3.445} & \cellcolor{gray!6}{0.064} & \cellcolor{gray!6}{\textbf{0.028}} & \cellcolor{gray!6}{2.073} & \cellcolor{gray!6}{3.943} & \cellcolor{gray!6}{0.126} & \cellcolor{gray!6}{\textbf{0.09}} & \cellcolor{gray!6}{2.010} & \cellcolor{gray!6}{2.647} & \cellcolor{gray!6}{0.118} & \cellcolor{gray!6}{\textbf{0.091}} & \cellcolor{gray!6}{2.082} & \cellcolor{gray!6}{2.397} & \cellcolor{gray!6}{0.098} & \cellcolor{gray!6}{\textbf{0.068}}\\
1000 & 500 & 2.054 & 3.820 & 0.072 & \textbf{0.027} & 1.950 & 2.967 & 0.074 & \textbf{0.057} & 2.025 & 2.635 & 0.087 & \textbf{0.069} & 1.989 & 2.367 & 0.081 & \textbf{0.067}\\
\addlinespace
\cellcolor{gray!6}{2500} & \cellcolor{gray!6}{10} & \cellcolor{gray!6}{2.389} & \cellcolor{gray!6}{6.342} & \cellcolor{gray!6}{0.34} & \cellcolor{gray!6}{\textbf{0.281}} & \cellcolor{gray!6}{2.377} & \cellcolor{gray!6}{5.862} & \cellcolor{gray!6}{0.339} & \cellcolor{gray!6}{\textbf{0.315}} & \cellcolor{gray!6}{2.421} & \cellcolor{gray!6}{3.248} & \cellcolor{gray!6}{0.838} & \cellcolor{gray!6}{\textbf{0.635}} & \cellcolor{gray!6}{2.420} & \cellcolor{gray!6}{3.033} & \cellcolor{gray!6}{0.8} & \cellcolor{gray!6}{\textbf{0.688}}\\
2500 & 50 & 2.093 & 3.569 & 0.141 & \textbf{0.042} & 2.018 & 3.427 & 0.206 & \textbf{0.154} & 2.040 & 3.081 & 0.238 & \textbf{0.112} & 2.086 & 2.622 & 0.243 & \textbf{0.161}\\
\cellcolor{gray!6}{2500} & \cellcolor{gray!6}{100} & \cellcolor{gray!6}{2.013} & \cellcolor{gray!6}{5.508} & \cellcolor{gray!6}{0.15} & \cellcolor{gray!6}{\textbf{0.063}} & \cellcolor{gray!6}{1.962} & \cellcolor{gray!6}{3.364} & \cellcolor{gray!6}{\textbf{0.146}} & \cellcolor{gray!6}{0.154} & \cellcolor{gray!6}{2.023} & \cellcolor{gray!6}{2.830} & \cellcolor{gray!6}{0.193} & \cellcolor{gray!6}{\textbf{0.095}} & \cellcolor{gray!6}{2.024} & \cellcolor{gray!6}{2.536} & \cellcolor{gray!6}{0.183} & \cellcolor{gray!6}{\textbf{0.103}}\\
2500 & 250 & 1.852 & 9.136 & 0.135 & \textbf{0.098} & 2.067 & 4.997 & \textbf{0.24} & 0.252 & 2.066 & 3.071 & 0.214 & \textbf{0.134} & 2.046 & 2.535 & 0.173 & \textbf{0.129}\\
\cellcolor{gray!6}{2500} & \cellcolor{gray!6}{500} & \cellcolor{gray!6}{2.023} & \cellcolor{gray!6}{8.175} & \cellcolor{gray!6}{0.154} & \cellcolor{gray!6}{\textbf{0.031}} & \cellcolor{gray!6}{2.021} & \cellcolor{gray!6}{5.870} & \cellcolor{gray!6}{0.205} & \cellcolor{gray!6}{\textbf{0.177}} & \cellcolor{gray!6}{1.952} & \cellcolor{gray!6}{2.682} & \cellcolor{gray!6}{0.165} & \cellcolor{gray!6}{\textbf{0.119}} & \cellcolor{gray!6}{2.063} & \cellcolor{gray!6}{2.494} & \cellcolor{gray!6}{0.164} & \cellcolor{gray!6}{\textbf{0.144}}\\
\bottomrule
\textbf{\footnotesize Note:} &
\multicolumn{17}{l}{\footnotesize Lower is better; best method for each $N$,$K$,$d$, in bold. Values are Root Mean Square Error for estimated vs true loadings, averaged over $d$ dimensions, }\\
& \multicolumn{17}{l}{\footnotesize $N$ units, and 50 simulations. For Bayesian models estimates are posterior means computed by averaging over 10000 posterior samples. }\\
& \multicolumn{17}{l}{\footnotesize All results from Bayesian models are computed from 4000 posterior samples obtained from 4 parallel MCMC chains after 2000 burn-in iterations.}
\end{tabular}}
\end{table}

\begin{table}
\caption{95\% Credible interval coverage for $\blambda$.}
\label{tab:lambdacov}
\centering
\resizebox{0.98\textwidth}{!}{%
\begin{tabular}{rr|lll|lll|lll|lll}
\toprule
N & K &  \multicolumn{3}{c|}{d=2} & \multicolumn{3}{c|}{d=3} & 
\multicolumn{3}{c|}{d=5} & \multicolumn{3}{c}{d=8} \\
\midrule
&  & IRT & IRT-M & IRT-M & IRT & IRT-M  & IRT-M & IRT & IRT-M  & IRT-M  & IRT  & IRT-M  & IRT-M \\
\multicolumn{2}{c|}{Correlated $\btheta$?} & No & No & Yes & No & No & Yes & No & No & Yes & No & No & Yes \\
\midrule
\cellcolor{gray!6}{10} & \cellcolor{gray!6}{10} & \cellcolor{gray!6}{\textbf{0.676}} & \cellcolor{gray!6}{0.596} & \cellcolor{gray!6}{0.598} & \cellcolor{gray!6}{\textbf{0.685}} & \cellcolor{gray!6}{0.581} & \cellcolor{gray!6}{0.585} & \cellcolor{gray!6}{\textbf{0.684}} & \cellcolor{gray!6}{0.587} & \cellcolor{gray!6}{0.589} & \cellcolor{gray!6}{--} & \cellcolor{gray!6}{--} & \cellcolor{gray!6}{--}\\
10 & 50 & 0.56 & \textbf{0.602} & 0.6 & 0.586 & 0.585 & \textbf{0.588} & \textbf{0.62} & 0.587 & 0.584 & -- & -- & --\\
\cellcolor{gray!6}{10} & \cellcolor{gray!6}{100} & \cellcolor{gray!6}{0.516} & \cellcolor{gray!6}{\textbf{0.602}} & \cellcolor{gray!6}{0.599} & \cellcolor{gray!6}{0.526} & \cellcolor{gray!6}{0.591} & \cellcolor{gray!6}{\textbf{0.593}} & \cellcolor{gray!6}{0.548} & \cellcolor{gray!6}{0.584} & \cellcolor{gray!6}{\textbf{0.586}} & \cellcolor{gray!6}{--} & \cellcolor{gray!6}{--} & \cellcolor{gray!6}{--}\\
10 & 250 & 0.478 & \textbf{0.596} & 0.595 & 0.488 & \textbf{0.593} & 0.593 & 0.497 & 0.586 & \textbf{0.588} & -- & -- & --\\
\cellcolor{gray!6}{10} & \cellcolor{gray!6}{500} & \cellcolor{gray!6}{0.479} & \cellcolor{gray!6}{\textbf{0.594}} & \cellcolor{gray!6}{0.594} & \cellcolor{gray!6}{0.469} & \cellcolor{gray!6}{0.596} & \cellcolor{gray!6}{\textbf{0.598}} & \cellcolor{gray!6}{0.48} & \cellcolor{gray!6}{0.587} & \cellcolor{gray!6}{\textbf{0.588}} & \cellcolor{gray!6}{--} & \cellcolor{gray!6}{--} & \cellcolor{gray!6}{--}\\
\addlinespace
50 & 10 & 0.54 & \textbf{0.598} & 0.589 & 0.57 & 0.587 & \textbf{0.589} & \textbf{0.626} & 0.568 & 0.575 & \textbf{0.66} & 0.585 & 0.583\\
\cellcolor{gray!6}{50} & \cellcolor{gray!6}{50} & \cellcolor{gray!6}{0.368} & \cellcolor{gray!6}{0.609} & \cellcolor{gray!6}{\textbf{0.611}} & \cellcolor{gray!6}{0.408} & \cellcolor{gray!6}{\textbf{0.614}} & \cellcolor{gray!6}{0.612} & \cellcolor{gray!6}{0.472} & \cellcolor{gray!6}{0.596} & \cellcolor{gray!6}{\textbf{0.598}} & \cellcolor{gray!6}{0.518} & \cellcolor{gray!6}{0.592} & \cellcolor{gray!6}{\textbf{0.596}}\\
50 & 100 & 0.369 & 0.613 & \textbf{0.615} & 0.377 & \textbf{0.621} & 0.621 & 0.412 & 0.608 & \textbf{0.611} & 0.453 & 0.59 & \textbf{0.592}\\
\cellcolor{gray!6}{50} & \cellcolor{gray!6}{250} & \cellcolor{gray!6}{0.38} & \cellcolor{gray!6}{0.626} & \cellcolor{gray!6}{\textbf{0.627}} & \cellcolor{gray!6}{0.368} & \cellcolor{gray!6}{0.623} & \cellcolor{gray!6}{\textbf{0.624}} & \cellcolor{gray!6}{0.369} & \cellcolor{gray!6}{0.617} & \cellcolor{gray!6}{\textbf{0.619}} & \cellcolor{gray!6}{0.399} & \cellcolor{gray!6}{\textbf{0.601}} & \cellcolor{gray!6}{0.601}\\
50 & 500 & 0.365 & \textbf{0.635} & 0.634 & 0.361 & \textbf{0.63} & 0.63 & 0.373 & \textbf{0.623} & 0.623 & 0.39 & 0.602 & \textbf{0.603}\\
\addlinespace
\cellcolor{gray!6}{100} & \cellcolor{gray!6}{10} & \cellcolor{gray!6}{0.396} & \cellcolor{gray!6}{0.568} & \cellcolor{gray!6}{\textbf{0.573}} & \cellcolor{gray!6}{0.457} & \cellcolor{gray!6}{0.529} & \cellcolor{gray!6}{\textbf{0.553}} & \cellcolor{gray!6}{0.527} & \cellcolor{gray!6}{0.546} & \cellcolor{gray!6}{\textbf{0.558}} & \cellcolor{gray!6}{\textbf{0.579}} & \cellcolor{gray!6}{0.552} & \cellcolor{gray!6}{0.558}\\
100 & 50 & 0.298 & 0.598 & \textbf{0.601} & 0.327 & 0.602 & \textbf{0.605} & 0.368 & 0.581 & \textbf{0.585} & 0.419 & 0.574 & \textbf{0.579}\\
\cellcolor{gray!6}{100} & \cellcolor{gray!6}{100} & \cellcolor{gray!6}{0.322} & \cellcolor{gray!6}{\textbf{0.623}} & \cellcolor{gray!6}{0.616} & \cellcolor{gray!6}{0.303} & \cellcolor{gray!6}{\textbf{0.617}} & \cellcolor{gray!6}{0.616} & \cellcolor{gray!6}{0.336} & \cellcolor{gray!6}{\textbf{0.602}} & \cellcolor{gray!6}{0.601} & \cellcolor{gray!6}{0.366} & \cellcolor{gray!6}{0.595} & \cellcolor{gray!6}{\textbf{0.599}}\\
100 & 250 & 0.308 & \textbf{0.637} & 0.637 & 0.291 & \textbf{0.624} & 0.623 & 0.308 & 0.619 & \textbf{0.621} & 0.327 & 0.61 & \textbf{0.611}\\
\cellcolor{gray!6}{100} & \cellcolor{gray!6}{500} & \cellcolor{gray!6}{0.304} & \cellcolor{gray!6}{\textbf{0.644}} & \cellcolor{gray!6}{0.644} & \cellcolor{gray!6}{0.3} & \cellcolor{gray!6}{\textbf{0.631}} & \cellcolor{gray!6}{0.63} & \cellcolor{gray!6}{0.306} & \cellcolor{gray!6}{\textbf{0.628}} & \cellcolor{gray!6}{0.628} & \cellcolor{gray!6}{0.306} & \cellcolor{gray!6}{0.617} & \cellcolor{gray!6}{\textbf{0.618}}\\
\addlinespace
250 & 10 & 0.318 & 0.581 & \textbf{0.586} & 0.373 & 0.504 & \textbf{0.512} & 0.379 & \textbf{0.52} & 0.519 & 0.449 & 0.505 & \textbf{0.512}\\
\cellcolor{gray!6}{250} & \cellcolor{gray!6}{50} & \cellcolor{gray!6}{0.226} & \cellcolor{gray!6}{\textbf{0.592}} & \cellcolor{gray!6}{0.592} & \cellcolor{gray!6}{0.222} & \cellcolor{gray!6}{0.545} & \cellcolor{gray!6}{\textbf{0.559}} & \cellcolor{gray!6}{0.251} & \cellcolor{gray!6}{0.537} & \cellcolor{gray!6}{\textbf{0.557}} & \cellcolor{gray!6}{0.287} & \cellcolor{gray!6}{0.515} & \cellcolor{gray!6}{\textbf{0.534}}\\
250 & 100 & 0.208 & \textbf{0.618} & 0.615 & 0.21 & 0.583 & \textbf{0.584} & 0.228 & 0.548 & \textbf{0.554} & 0.243 & 0.546 & \textbf{0.557}\\
\cellcolor{gray!6}{250} & \cellcolor{gray!6}{250} & \cellcolor{gray!6}{0.223} & \cellcolor{gray!6}{\textbf{0.635}} & \cellcolor{gray!6}{0.632} & \cellcolor{gray!6}{0.216} & \cellcolor{gray!6}{\textbf{0.612}} & \cellcolor{gray!6}{0.604} & \cellcolor{gray!6}{0.203} & \cellcolor{gray!6}{\textbf{0.602}} & \cellcolor{gray!6}{0.594} & \cellcolor{gray!6}{0.223} & \cellcolor{gray!6}{\textbf{0.595}} & \cellcolor{gray!6}{0.593}\\
250 & 500 & 0.267 & \textbf{0.629} & 0.625 & 0.207 & \textbf{0.633} & 0.629 & 0.218 & \textbf{0.622} & 0.614 & 0.228 & \textbf{0.613} & 0.609\\
\addlinespace
\cellcolor{gray!6}{500} & \cellcolor{gray!6}{10} & \cellcolor{gray!6}{0.225} & \cellcolor{gray!6}{0.567} & \cellcolor{gray!6}{\textbf{0.576}} & \cellcolor{gray!6}{0.252} & \cellcolor{gray!6}{0.496} & \cellcolor{gray!6}{\textbf{0.509}} & \cellcolor{gray!6}{0.317} & \cellcolor{gray!6}{0.454} & \cellcolor{gray!6}{\textbf{0.477}} & \cellcolor{gray!6}{0.343} & \cellcolor{gray!6}{0.473} & \cellcolor{gray!6}{\textbf{0.491}}\\
500 & 50 & 0.193 & 0.542 & \textbf{0.556} & 0.183 & 0.499 & \textbf{0.531} & 0.182 & 0.475 & \textbf{0.51} & 0.209 & 0.461 & \textbf{0.494}\\
\cellcolor{gray!6}{500} & \cellcolor{gray!6}{100} & \cellcolor{gray!6}{0.173} & \cellcolor{gray!6}{0.566} & \cellcolor{gray!6}{\textbf{0.574}} & \cellcolor{gray!6}{0.152} & \cellcolor{gray!6}{0.525} & \cellcolor{gray!6}{\textbf{0.544}} & \cellcolor{gray!6}{0.157} & \cellcolor{gray!6}{0.492} & \cellcolor{gray!6}{\textbf{0.519}} & \cellcolor{gray!6}{0.173} & \cellcolor{gray!6}{0.477} & \cellcolor{gray!6}{\textbf{0.511}}\\
500 & 250 & 0.177 & \textbf{0.576} & 0.565 & 0.148 & \textbf{0.567} & 0.565 & 0.134 & 0.541 & \textbf{0.544} & 0.151 & 0.525 & \textbf{0.533}\\
\cellcolor{gray!6}{500} & \cellcolor{gray!6}{500} & \cellcolor{gray!6}{0.183} & \cellcolor{gray!6}{\textbf{0.614}} & \cellcolor{gray!6}{0.607} & \cellcolor{gray!6}{0.159} & \cellcolor{gray!6}{\textbf{0.585}} & \cellcolor{gray!6}{0.575} & \cellcolor{gray!6}{0.154} & \cellcolor{gray!6}{\textbf{0.584}} & \cellcolor{gray!6}{0.574} & \cellcolor{gray!6}{0.152} & \cellcolor{gray!6}{\textbf{0.576}} & \cellcolor{gray!6}{0.57}\\
\addlinespace
1000 & 10 & 0.194 & \textbf{0.534} & 0.534 & 0.2 & 0.445 & \textbf{0.486} & 0.224 & 0.392 & \textbf{0.437} & 0.257 & 0.416 & \textbf{0.436}\\
\cellcolor{gray!6}{1000} & \cellcolor{gray!6}{50} & \cellcolor{gray!6}{0.15} & \cellcolor{gray!6}{0.547} & \cellcolor{gray!6}{\textbf{0.582}} & \cellcolor{gray!6}{0.136} & \cellcolor{gray!6}{0.443} & \cellcolor{gray!6}{\textbf{0.502}} & \cellcolor{gray!6}{0.129} & \cellcolor{gray!6}{0.411} & \cellcolor{gray!6}{\textbf{0.471}} & \cellcolor{gray!6}{0.151} & \cellcolor{gray!6}{0.41} & \cellcolor{gray!6}{\textbf{0.461}}\\
1000 & 100 & 0.144 & 0.51 & \textbf{0.545} & 0.12 & 0.466 & \textbf{0.512} & 0.111 & 0.443 & \textbf{0.502} & 0.122 & 0.411 & \textbf{0.477}\\
\cellcolor{gray!6}{1000} & \cellcolor{gray!6}{250} & \cellcolor{gray!6}{0.136} & \cellcolor{gray!6}{0.564} & \cellcolor{gray!6}{\textbf{0.575}} & \cellcolor{gray!6}{0.11} & \cellcolor{gray!6}{0.494} & \cellcolor{gray!6}{\textbf{0.507}} & \cellcolor{gray!6}{0.094} & \cellcolor{gray!6}{0.461} & \cellcolor{gray!6}{\textbf{0.492}} & \cellcolor{gray!6}{0.099} & \cellcolor{gray!6}{0.455} & \cellcolor{gray!6}{\textbf{0.495}}\\
1000 & 500 & 0.14 & \textbf{0.549} & 0.547 & 0.106 & 0.545 & \textbf{0.557} & 0.093 & 0.496 & \textbf{0.504} & 0.093 & 0.491 & \textbf{0.504}\\
\addlinespace
\cellcolor{gray!6}{2500} & \cellcolor{gray!6}{10} & \cellcolor{gray!6}{0.154} & \cellcolor{gray!6}{0.473} & \cellcolor{gray!6}{\textbf{0.525}} & \cellcolor{gray!6}{0.13} & \cellcolor{gray!6}{0.382} & \cellcolor{gray!6}{\textbf{0.432}} & \cellcolor{gray!6}{0.152} & \cellcolor{gray!6}{0.37} & \cellcolor{gray!6}{\textbf{0.395}} & \cellcolor{gray!6}{0.189} & \cellcolor{gray!6}{0.367} & \cellcolor{gray!6}{\textbf{0.398}}\\
2500 & 50 & 0.11 & 0.524 & \textbf{0.579} & 0.091 & 0.386 & \textbf{0.468} & 0.088 & 0.354 & \textbf{0.436} & 0.103 & 0.346 & \textbf{0.417}\\
\cellcolor{gray!6}{2500} & \cellcolor{gray!6}{100} & \cellcolor{gray!6}{0.12} & \cellcolor{gray!6}{0.464} & \cellcolor{gray!6}{\textbf{0.539}} & \cellcolor{gray!6}{0.078} & \cellcolor{gray!6}{0.421} & \cellcolor{gray!6}{\textbf{0.498}} & \cellcolor{gray!6}{0.071} & \cellcolor{gray!6}{0.365} & \cellcolor{gray!6}{\textbf{0.453}} & \cellcolor{gray!6}{0.078} & \cellcolor{gray!6}{0.348} & \cellcolor{gray!6}{\textbf{0.438}}\\
2500 & 250 & 0.122 & 0.513 & \textbf{0.555} & 0.074 & 0.425 & \textbf{0.486} & 0.061 & 0.385 & \textbf{0.456} & 0.062 & 0.354 & \textbf{0.435}\\
\cellcolor{gray!6}{2500} & \cellcolor{gray!6}{500} & \cellcolor{gray!6}{0.11} & \cellcolor{gray!6}{0.506} & \cellcolor{gray!6}{\textbf{0.535}} & \cellcolor{gray!6}{0.076} & \cellcolor{gray!6}{0.453} & \cellcolor{gray!6}{\textbf{0.503}} & \cellcolor{gray!6}{0.058} & \cellcolor{gray!6}{0.414} & \cellcolor{gray!6}{\textbf{0.471}} & \cellcolor{gray!6}{0.057} & \cellcolor{gray!6}{0.383} & \cellcolor{gray!6}{\textbf{0.449}}\\
\bottomrule
\textbf{\footnotesize Note:} & \multicolumn{13}{l}{\footnotesize Higher is better. Best method for each $N$,$K$,$d$, in bold. Values are proportion of times that the true value of $lambda_{kj}$ falls within}\\
& \multicolumn{13}{l}{ \footnotesize the 95\% Credible Interval generated by the posterior draws of each estimated latent loading. Proportions are computed over $K$}\\
& \multicolumn{13}{l}{\footnotesize items separately for each dimension of $\blambda$, resulting values are averaged across dimensions and 50 simulations.}\\
& \multicolumn{13}{l}{\footnotesize All results from Bayesian models are computed from 4000 posterior samples obtained from 4 parallel MCMC chains}\\
& \multicolumn{13}{l}{\footnotesize after 2000 burn-in iterations.}
\end{tabular}}
\end{table}

Table \ref{tab:thetacov} displays 95\% credible interval coverage for the factors, $\btheta$. Here we see that IRT-M performs well above standard IRT in terms of coverage. This gain is likely due to the fact that IRT-M can estimate more precise posteriors, thanks to the additional information provided to it by the $M$-matrices. Together with gains in estimation error as measured by MSE (Table \ref{tab:thetamse}) these gains in coverage are substantial enough to justify use of IRT-M instead of traditional IRT. 

\begin{table}
\caption{95\% Credible interval coverage for $\btheta$.}
\label{tab:thetacov}
\centering
\resizebox{0.98\textwidth}{!}{%
\begin{tabular}{rr|lll|lll|lll|lll}
\toprule
N & K &  \multicolumn{3}{c|}{d=2} & \multicolumn{3}{c|}{d=3} & 
\multicolumn{3}{c|}{d=5} & \multicolumn{3}{c}{d=8} \\
\midrule
&  & IRT & IRT-M & IRT-M & IRT & IRT-M  & IRT-M & IRT & IRT-M  & IRT-M  & IRT  & IRT-M  & IRT-M \\
\multicolumn{2}{c|}{Correlated $\btheta$?} & No & No & Yes & No & No & Yes & No & No & Yes & No & No & Yes \\
\midrule
\cellcolor{gray!6}{10} & \cellcolor{gray!6}{10} & \cellcolor{gray!6}{0.577} & \cellcolor{gray!6}{\textbf{0.68}} & \cellcolor{gray!6}{0.668} & \cellcolor{gray!6}{0.583} & \cellcolor{gray!6}{\textbf{0.718}} & \cellcolor{gray!6}{0.715} & \cellcolor{gray!6}{0.622} & \cellcolor{gray!6}{\textbf{0.775}} & \cellcolor{gray!6}{0.769} & \cellcolor{gray!6}{--} & \cellcolor{gray!6}{--} & \cellcolor{gray!6}{--}\\
10 & 50 & 0.434 & \textbf{0.66} & 0.658 & 0.450 & 0.669 & \textbf{0.67} & 0.501 & \textbf{0.711} & 0.704 & -- & -- & --\\
\cellcolor{gray!6}{10} & \cellcolor{gray!6}{100} & \cellcolor{gray!6}{0.296} & \cellcolor{gray!6}{\textbf{0.629}} & \cellcolor{gray!6}{0.613} & \cellcolor{gray!6}{0.351} & \cellcolor{gray!6}{\textbf{0.66}} & \cellcolor{gray!6}{0.656} & \cellcolor{gray!6}{0.391} & \cellcolor{gray!6}{0.685} & \cellcolor{gray!6}{\textbf{0.687}} & \cellcolor{gray!6}{--} & \cellcolor{gray!6}{--} & \cellcolor{gray!6}{--}\\
10 & 250 & 0.219 & \textbf{0.576} & 0.574 & 0.226 & 0.597 & \textbf{0.6} & 0.268 & 0.649 & \textbf{0.653} & -- & -- & --\\
\cellcolor{gray!6}{10} & \cellcolor{gray!6}{500} & \cellcolor{gray!6}{0.149} & \cellcolor{gray!6}{\textbf{0.534}} & \cellcolor{gray!6}{0.534} & \cellcolor{gray!6}{0.173} & \cellcolor{gray!6}{\textbf{0.58}} & \cellcolor{gray!6}{0.573} & \cellcolor{gray!6}{0.198} & \cellcolor{gray!6}{0.613} & \cellcolor{gray!6}{\textbf{0.618}} & \cellcolor{gray!6}{--} & \cellcolor{gray!6}{--} & \cellcolor{gray!6}{--}\\
\addlinespace
50 & 10 & 0.460 & \textbf{0.607} & 0.601 & 0.497 & \textbf{0.616} & 0.615 & 0.538 & 0.633 & \textbf{0.635} & 0.566 & \textbf{0.667} & 0.665\\
\cellcolor{gray!6}{50} & \cellcolor{gray!6}{50} & \cellcolor{gray!6}{0.254} & \cellcolor{gray!6}{0.599} & \cellcolor{gray!6}{\textbf{0.604}} & \cellcolor{gray!6}{0.304} & \cellcolor{gray!6}{0.591} & \cellcolor{gray!6}{\textbf{0.6}} & \cellcolor{gray!6}{0.355} & \cellcolor{gray!6}{0.604} & \cellcolor{gray!6}{\textbf{0.611}} & \cellcolor{gray!6}{0.406} & \cellcolor{gray!6}{0.623} & \cellcolor{gray!6}{\textbf{0.626}}\\
50 & 100 & 0.215 & 0.556 & \textbf{0.568} & 0.217 & 0.565 & \textbf{0.576} & 0.257 & 0.561 & \textbf{0.566} & 0.312 & 0.574 & \textbf{0.579}\\
\cellcolor{gray!6}{50} & \cellcolor{gray!6}{250} & \cellcolor{gray!6}{0.139} & \cellcolor{gray!6}{0.516} & \cellcolor{gray!6}{\textbf{0.519}} & \cellcolor{gray!6}{0.141} & \cellcolor{gray!6}{0.53} & \cellcolor{gray!6}{\textbf{0.531}} & \cellcolor{gray!6}{0.159} & \cellcolor{gray!6}{0.514} & \cellcolor{gray!6}{\textbf{0.519}} & \cellcolor{gray!6}{0.199} & \cellcolor{gray!6}{0.49} & \cellcolor{gray!6}{\textbf{0.495}}\\
50 & 500 & 0.097 & \textbf{0.484} & 0.484 & 0.103 & 0.489 & \textbf{0.503} & 0.121 & 0.471 & \textbf{0.481} & 0.142 & 0.438 & \textbf{0.444}\\
\addlinespace
\cellcolor{gray!6}{100} & \cellcolor{gray!6}{10} & \cellcolor{gray!6}{0.440} & \cellcolor{gray!6}{\textbf{0.599}} & \cellcolor{gray!6}{0.597} & \cellcolor{gray!6}{0.480} & \cellcolor{gray!6}{0.604} & \cellcolor{gray!6}{\textbf{0.607}} & \cellcolor{gray!6}{0.492} & \cellcolor{gray!6}{0.604} & \cellcolor{gray!6}{\textbf{0.607}} & \cellcolor{gray!6}{0.540} & \cellcolor{gray!6}{0.63} & \cellcolor{gray!6}{\textbf{0.631}}\\
100 & 50 & 0.263 & 0.589 & \textbf{0.6} & 0.294 & 0.587 & \textbf{0.6} & 0.325 & 0.588 & \textbf{0.602} & 0.369 & 0.594 & \textbf{0.605}\\
\cellcolor{gray!6}{100} & \cellcolor{gray!6}{100} & \cellcolor{gray!6}{0.194} & \cellcolor{gray!6}{0.572} & \cellcolor{gray!6}{\textbf{0.585}} & \cellcolor{gray!6}{0.211} & \cellcolor{gray!6}{0.564} & \cellcolor{gray!6}{\textbf{0.58}} & \cellcolor{gray!6}{0.237} & \cellcolor{gray!6}{0.561} & \cellcolor{gray!6}{\textbf{0.579}} & \cellcolor{gray!6}{0.276} & \cellcolor{gray!6}{0.571} & \cellcolor{gray!6}{\textbf{0.587}}\\
100 & 250 & 0.114 & \textbf{0.546} & 0.545 & 0.121 & 0.507 & \textbf{0.525} & 0.137 & 0.518 & \textbf{0.535} & 0.170 & 0.494 & \textbf{0.516}\\
\cellcolor{gray!6}{100} & \cellcolor{gray!6}{500} & \cellcolor{gray!6}{0.081} & \cellcolor{gray!6}{0.473} & \cellcolor{gray!6}{\textbf{0.487}} & \cellcolor{gray!6}{0.086} & \cellcolor{gray!6}{0.479} & \cellcolor{gray!6}{\textbf{0.489}} & \cellcolor{gray!6}{0.100} & \cellcolor{gray!6}{0.47} & \cellcolor{gray!6}{\textbf{0.487}} & \cellcolor{gray!6}{0.118} & \cellcolor{gray!6}{0.444} & \cellcolor{gray!6}{\textbf{0.463}}\\
\addlinespace
250 & 10 & 0.426 & 0.597 & \textbf{0.609} & 0.462 & 0.592 & \textbf{0.601} & 0.472 & 0.585 & \textbf{0.591} & 0.494 & 0.581 & \textbf{0.588}\\
\cellcolor{gray!6}{250} & \cellcolor{gray!6}{50} & \cellcolor{gray!6}{0.256} & \cellcolor{gray!6}{0.594} & \cellcolor{gray!6}{\textbf{0.607}} & \cellcolor{gray!6}{0.284} & \cellcolor{gray!6}{0.586} & \cellcolor{gray!6}{\textbf{0.607}} & \cellcolor{gray!6}{0.309} & \cellcolor{gray!6}{0.585} & \cellcolor{gray!6}{\textbf{0.609}} & \cellcolor{gray!6}{0.341} & \cellcolor{gray!6}{0.586} & \cellcolor{gray!6}{\textbf{0.609}}\\
250 & 100 & 0.173 & 0.574 & \textbf{0.583} & 0.201 & 0.578 & \textbf{0.589} & 0.246 & 0.569 & \textbf{0.59} & 0.271 & 0.569 & \textbf{0.598}\\
\cellcolor{gray!6}{250} & \cellcolor{gray!6}{250} & \cellcolor{gray!6}{0.106} & \cellcolor{gray!6}{\textbf{0.562}} & \cellcolor{gray!6}{0.558} & \cellcolor{gray!6}{0.130} & \cellcolor{gray!6}{0.544} & \cellcolor{gray!6}{\textbf{0.555}} & \cellcolor{gray!6}{0.140} & \cellcolor{gray!6}{0.536} & \cellcolor{gray!6}{\textbf{0.546}} & \cellcolor{gray!6}{0.161} & \cellcolor{gray!6}{0.529} & \cellcolor{gray!6}{\textbf{0.558}}\\
250 & 500 & 0.081 & 0.499 & \textbf{0.507} & 0.079 & 0.517 & \textbf{0.531} & 0.092 & 0.5 & \textbf{0.524} & 0.105 & 0.469 & \textbf{0.508}\\
\addlinespace
\cellcolor{gray!6}{500} & \cellcolor{gray!6}{10} & \cellcolor{gray!6}{0.395} & \cellcolor{gray!6}{0.597} & \cellcolor{gray!6}{\textbf{0.606}} & \cellcolor{gray!6}{0.436} & \cellcolor{gray!6}{0.593} & \cellcolor{gray!6}{\textbf{0.604}} & \cellcolor{gray!6}{0.461} & \cellcolor{gray!6}{0.574} & \cellcolor{gray!6}{\textbf{0.583}} & \cellcolor{gray!6}{0.482} & \cellcolor{gray!6}{0.57} & \cellcolor{gray!6}{\textbf{0.575}}\\
500 & 50 & 0.261 & 0.592 & \textbf{0.61} & 0.279 & 0.589 & \textbf{0.613} & 0.314 & 0.585 & \textbf{0.616} & 0.330 & 0.581 & \textbf{0.609}\\
\cellcolor{gray!6}{500} & \cellcolor{gray!6}{100} & \cellcolor{gray!6}{0.236} & \cellcolor{gray!6}{0.581} & \cellcolor{gray!6}{\textbf{0.582}} & \cellcolor{gray!6}{0.222} & \cellcolor{gray!6}{0.579} & \cellcolor{gray!6}{\textbf{0.589}} & \cellcolor{gray!6}{0.242} & \cellcolor{gray!6}{0.573} & \cellcolor{gray!6}{\textbf{0.593}} & \cellcolor{gray!6}{0.264} & \cellcolor{gray!6}{0.571} & \cellcolor{gray!6}{\textbf{0.598}}\\
500 & 250 & 0.133 & \textbf{0.544} & 0.522 & 0.137 & \textbf{0.542} & 0.532 & 0.157 & \textbf{0.537} & 0.533 & 0.183 & 0.537 & \textbf{0.543}\\
\cellcolor{gray!6}{500} & \cellcolor{gray!6}{500} & \cellcolor{gray!6}{0.078} & \cellcolor{gray!6}{\textbf{0.519}} & \cellcolor{gray!6}{0.499} & \cellcolor{gray!6}{0.087} & \cellcolor{gray!6}{\textbf{0.494}} & \cellcolor{gray!6}{0.481} & \cellcolor{gray!6}{0.100} & \cellcolor{gray!6}{\textbf{0.502}} & \cellcolor{gray!6}{0.495} & \cellcolor{gray!6}{0.108} & \cellcolor{gray!6}{0.506} & \cellcolor{gray!6}{\textbf{0.515}}\\
\addlinespace
1000 & 10 & 0.413 & 0.593 & \textbf{0.605} & 0.430 & 0.59 & \textbf{0.602} & 0.440 & 0.571 & \textbf{0.583} & 0.469 & 0.559 & \textbf{0.565}\\
\cellcolor{gray!6}{1000} & \cellcolor{gray!6}{50} & \cellcolor{gray!6}{0.234} & \cellcolor{gray!6}{0.589} & \cellcolor{gray!6}{\textbf{0.604}} & \cellcolor{gray!6}{0.297} & \cellcolor{gray!6}{0.594} & \cellcolor{gray!6}{\textbf{0.624}} & \cellcolor{gray!6}{0.309} & \cellcolor{gray!6}{0.586} & \cellcolor{gray!6}{\textbf{0.612}} & \cellcolor{gray!6}{0.326} & \cellcolor{gray!6}{0.584} & \cellcolor{gray!6}{\textbf{0.608}}\\
1000 & 100 & 0.204 & 0.576 & \textbf{0.587} & 0.229 & 0.576 & \textbf{0.581} & 0.243 & 0.578 & \textbf{0.596} & 0.269 & 0.575 & \textbf{0.598}\\
\cellcolor{gray!6}{1000} & \cellcolor{gray!6}{250} & \cellcolor{gray!6}{0.135} & \cellcolor{gray!6}{\textbf{0.523}} & \cellcolor{gray!6}{0.515} & \cellcolor{gray!6}{0.167} & \cellcolor{gray!6}{\textbf{0.505}} & \cellcolor{gray!6}{0.497} & \cellcolor{gray!6}{0.185} & \cellcolor{gray!6}{0.504} & \cellcolor{gray!6}{\textbf{0.509}} & \cellcolor{gray!6}{0.187} & \cellcolor{gray!6}{0.526} & \cellcolor{gray!6}{\textbf{0.536}}\\
1000 & 500 & 0.094 & \textbf{0.458} & 0.452 & 0.100 & \textbf{0.481} & 0.478 & 0.125 & \textbf{0.441} & 0.44 & 0.134 & 0.461 & \textbf{0.467}\\
\addlinespace
\cellcolor{gray!6}{2500} & \cellcolor{gray!6}{10} & \cellcolor{gray!6}{0.412} & \cellcolor{gray!6}{0.591} & \cellcolor{gray!6}{\textbf{0.606}} & \cellcolor{gray!6}{0.431} & \cellcolor{gray!6}{0.581} & \cellcolor{gray!6}{\textbf{0.595}} & \cellcolor{gray!6}{0.444} & \cellcolor{gray!6}{0.568} & \cellcolor{gray!6}{\textbf{0.575}} & \cellcolor{gray!6}{0.461} & \cellcolor{gray!6}{0.546} & \cellcolor{gray!6}{\textbf{0.557}}\\
2500 & 50 & 0.228 & 0.594 & \textbf{0.604} & 0.294 & 0.595 & \textbf{0.616} & 0.304 & 0.59 & \textbf{0.61} & 0.329 & 0.584 & \textbf{0.596}\\
\cellcolor{gray!6}{2500} & \cellcolor{gray!6}{100} & \cellcolor{gray!6}{0.228} & \cellcolor{gray!6}{\textbf{0.572}} & \cellcolor{gray!6}{0.566} & \cellcolor{gray!6}{0.229} & \cellcolor{gray!6}{0.574} & \cellcolor{gray!6}{\textbf{0.582}} & \cellcolor{gray!6}{0.252} & \cellcolor{gray!6}{0.576} & \cellcolor{gray!6}{\textbf{0.59}} & \cellcolor{gray!6}{0.266} & \cellcolor{gray!6}{0.574} & \cellcolor{gray!6}{\textbf{0.587}}\\
2500 & 250 & 0.134 & 0.506 & \textbf{0.516} & 0.188 & 0.468 & \textbf{0.478} & 0.191 & 0.475 & \textbf{0.503} & 0.206 & 0.486 & \textbf{0.518}\\
\cellcolor{gray!6}{2500} & \cellcolor{gray!6}{500} & \cellcolor{gray!6}{0.105} & \cellcolor{gray!6}{0.45} & \cellcolor{gray!6}{\textbf{0.459}} & \cellcolor{gray!6}{0.125} & \cellcolor{gray!6}{0.415} & \cellcolor{gray!6}{\textbf{0.45}} & \cellcolor{gray!6}{0.135} & \cellcolor{gray!6}{0.396} & \cellcolor{gray!6}{\textbf{0.437}} & \cellcolor{gray!6}{0.150} & \cellcolor{gray!6}{0.397} & \cellcolor{gray!6}{\textbf{0.444}}\\
\bottomrule
\textbf{\footnotesize Note:} & \multicolumn{13}{l}{\footnotesize Higher is better. Best method for each $N$,$K$,$d$, in bold. Values are proportion of times that the true value of $theta_{ij}$ falls within}\\
& \multicolumn{13}{l}{ \footnotesize the 95\% Credible Interval generated by the posterior draws of each estimated latent factor. Proportions are computed over $N$}\\
& \multicolumn{13}{l}{\footnotesize units separately for each dimension of $\btheta$, resulting values are averaged across dimensions and 50 simulations. For }\\
& \multicolumn{13}{l}{\footnotesize  All results from Bayesian models are computed from 4000 posterior samples obtained from 4 parallel MCMC chains}\\
& \multicolumn{13}{l}{\footnotesize after 2000 burn-in iterations.}
\end{tabular}}
\end{table}

Tables \ref{tab:thetagew} and \ref{tab:thetarhat} display convergence of the compared models as measured by the Geweke \citep{geweke1992} and adjusted $\hat{R}$ \citep{vehtari2021rank} statistics. Both tables show that IRT-M generally converges faster than classical IRT. Table \ref{tab:thetarhat} shows that all IRT-M models display $\hat{R}$ coefficients below 1.1, the recommended threshold of \cite{vehtari2021rank}, and in most cases these coefficients are very close to 1.0, indicating almost optimal convergence.
\begin{table}
\caption{Geweke convergence for $\btheta$.}
\label{tab:thetagew}
\centering
\resizebox{0.98\textwidth}{!}{%
\begin{tabular}{rr|lll|lll|lll|lll}
\toprule
N & K &  \multicolumn{3}{c|}{d=2} & \multicolumn{3}{c|}{d=3} & 
\multicolumn{3}{c|}{d=5} & \multicolumn{3}{c}{d=8} \\
\midrule
&  & IRT & IRT-M & IRT-M & IRT & IRT-M  & IRT-M & IRT & IRT-M  & IRT-M  & IRT  & IRT-M  & IRT-M \\
\multicolumn{2}{c|}{Correlated $\btheta$?} & No & No & Yes & No & No & Yes & No & No & Yes & No & No & Yes \\
\midrule
\cellcolor{gray!6}{10} & \cellcolor{gray!6}{10} & \cellcolor{gray!6}{0.413} & \cellcolor{gray!6}{\textbf{0.292}} & \cellcolor{gray!6}{0.3} & \cellcolor{gray!6}{0.355} & \cellcolor{gray!6}{0.307} & \cellcolor{gray!6}{\textbf{0.306}} & \cellcolor{gray!6}{\textbf{0.287}} & \cellcolor{gray!6}{0.355} & \cellcolor{gray!6}{0.342} & \cellcolor{gray!6}{--} & \cellcolor{gray!6}{--} & \cellcolor{gray!6}{--}\\
10 & 50 & 0.512 & 0.384 & \textbf{0.375} & 0.526 & \textbf{0.373} & 0.373 & 0.519 & \textbf{0.418} & 0.424 & -- & -- & --\\
\cellcolor{gray!6}{10} & \cellcolor{gray!6}{100} & \cellcolor{gray!6}{0.474} & \cellcolor{gray!6}{0.412} & \cellcolor{gray!6}{\textbf{0.39}} & \cellcolor{gray!6}{0.506} & \cellcolor{gray!6}{0.417} & \cellcolor{gray!6}{\textbf{0.41}} & \cellcolor{gray!6}{0.53} & \cellcolor{gray!6}{0.44} & \cellcolor{gray!6}{\textbf{0.438}} & \cellcolor{gray!6}{--} & \cellcolor{gray!6}{--} & \cellcolor{gray!6}{--}\\
10 & 250 & 0.474 & 0.471 & \textbf{0.427} & 0.527 & 0.475 & \textbf{0.444} & 0.549 & 0.468 & \textbf{0.461} & -- & -- & --\\
\cellcolor{gray!6}{10} & \cellcolor{gray!6}{500} & \cellcolor{gray!6}{0.508} & \cellcolor{gray!6}{\textbf{0.452}} & \cellcolor{gray!6}{0.469} & \cellcolor{gray!6}{0.528} & \cellcolor{gray!6}{0.484} & \cellcolor{gray!6}{\textbf{0.475}} & \cellcolor{gray!6}{0.558} & \cellcolor{gray!6}{0.476} & \cellcolor{gray!6}{\textbf{0.466}} & \cellcolor{gray!6}{--} & \cellcolor{gray!6}{--} & \cellcolor{gray!6}{--}\\
\addlinespace
50 & 10 & 0.525 & \textbf{0.249} & 0.252 & 0.546 & \textbf{0.261} & 0.263 & 0.525 & \textbf{0.298} & 0.306 & 0.468 & \textbf{0.327} & 0.331\\
\cellcolor{gray!6}{50} & \cellcolor{gray!6}{50} & \cellcolor{gray!6}{0.524} & \cellcolor{gray!6}{\textbf{0.289}} & \cellcolor{gray!6}{0.298} & \cellcolor{gray!6}{0.594} & \cellcolor{gray!6}{\textbf{0.293}} & \cellcolor{gray!6}{0.293} & \cellcolor{gray!6}{0.607} & \cellcolor{gray!6}{\textbf{0.302}} & \cellcolor{gray!6}{0.302} & \cellcolor{gray!6}{0.593} & \cellcolor{gray!6}{\textbf{0.33}} & \cellcolor{gray!6}{0.338}\\
50 & 100 & 0.486 & \textbf{0.353} & 0.377 & 0.566 & \textbf{0.341} & 0.342 & 0.603 & 0.353 & \textbf{0.347} & 0.602 & 0.368 & \textbf{0.361}\\
\cellcolor{gray!6}{50} & \cellcolor{gray!6}{250} & \cellcolor{gray!6}{0.584} & \cellcolor{gray!6}{\textbf{0.433}} & \cellcolor{gray!6}{0.436} & \cellcolor{gray!6}{0.586} & \cellcolor{gray!6}{0.429} & \cellcolor{gray!6}{\textbf{0.419}} & \cellcolor{gray!6}{0.593} & \cellcolor{gray!6}{\textbf{0.398}} & \cellcolor{gray!6}{0.403} & \cellcolor{gray!6}{0.598} & \cellcolor{gray!6}{\textbf{0.4}} & \cellcolor{gray!6}{0.405}\\
50 & 500 & 0.631 & 0.501 & \textbf{0.482} & 0.615 & \textbf{0.444} & 0.455 & 0.616 & 0.433 & \textbf{0.426} & 0.609 & \textbf{0.434} & 0.436\\
\addlinespace
\cellcolor{gray!6}{100} & \cellcolor{gray!6}{10} & \cellcolor{gray!6}{0.512} & \cellcolor{gray!6}{\textbf{0.205}} & \cellcolor{gray!6}{0.211} & \cellcolor{gray!6}{0.577} & \cellcolor{gray!6}{\textbf{0.228}} & \cellcolor{gray!6}{0.232} & \cellcolor{gray!6}{0.585} & \cellcolor{gray!6}{\textbf{0.259}} & \cellcolor{gray!6}{0.268} & \cellcolor{gray!6}{0.536} & \cellcolor{gray!6}{\textbf{0.3}} & \cellcolor{gray!6}{0.307}\\
100 & 50 & 0.495 & 0.236 & \textbf{0.235} & 0.561 & \textbf{0.241} & 0.243 & 0.616 & \textbf{0.247} & 0.25 & 0.619 & \textbf{0.273} & 0.279\\
\cellcolor{gray!6}{100} & \cellcolor{gray!6}{100} & \cellcolor{gray!6}{0.491} & \cellcolor{gray!6}{0.319} & \cellcolor{gray!6}{\textbf{0.315}} & \cellcolor{gray!6}{0.559} & \cellcolor{gray!6}{\textbf{0.286}} & \cellcolor{gray!6}{0.296} & \cellcolor{gray!6}{0.611} & \cellcolor{gray!6}{\textbf{0.291}} & \cellcolor{gray!6}{0.292} & \cellcolor{gray!6}{0.621} & \cellcolor{gray!6}{0.305} & \cellcolor{gray!6}{\textbf{0.302}}\\
100 & 250 & 0.558 & \textbf{0.434} & 0.443 & 0.587 & 0.384 & \textbf{0.375} & 0.596 & \textbf{0.362} & 0.367 & 0.603 & 0.356 & \textbf{0.353}\\
\cellcolor{gray!6}{100} & \cellcolor{gray!6}{500} & \cellcolor{gray!6}{0.611} & \cellcolor{gray!6}{\textbf{0.487}} & \cellcolor{gray!6}{0.509} & \cellcolor{gray!6}{0.615} & \cellcolor{gray!6}{0.45} & \cellcolor{gray!6}{\textbf{0.445}} & \cellcolor{gray!6}{0.641} & \cellcolor{gray!6}{0.401} & \cellcolor{gray!6}{\textbf{0.398}} & \cellcolor{gray!6}{0.625} & \cellcolor{gray!6}{0.386} & \cellcolor{gray!6}{\textbf{0.382}}\\
\addlinespace
250 & 10 & 0.446 & 0.171 & \textbf{0.16} & 0.531 & 0.192 & \textbf{0.191} & 0.582 & \textbf{0.233} & 0.235 & 0.563 & \textbf{0.272} & 0.29\\
\cellcolor{gray!6}{250} & \cellcolor{gray!6}{50} & \cellcolor{gray!6}{0.400} & \cellcolor{gray!6}{\textbf{0.17}} & \cellcolor{gray!6}{0.173} & \cellcolor{gray!6}{0.509} & \cellcolor{gray!6}{\textbf{0.17}} & \cellcolor{gray!6}{0.181} & \cellcolor{gray!6}{0.569} & \cellcolor{gray!6}{\textbf{0.183}} & \cellcolor{gray!6}{0.194} & \cellcolor{gray!6}{0.592} & \cellcolor{gray!6}{\textbf{0.207}} & \cellcolor{gray!6}{0.221}\\
250 & 100 & 0.433 & \textbf{0.228} & 0.235 & 0.500 & 0.224 & \textbf{0.222} & 0.562 & \textbf{0.212} & 0.221 & 0.596 & \textbf{0.222} & 0.228\\
\cellcolor{gray!6}{250} & \cellcolor{gray!6}{250} & \cellcolor{gray!6}{0.506} & \cellcolor{gray!6}{0.381} & \cellcolor{gray!6}{\textbf{0.359}} & \cellcolor{gray!6}{0.536} & \cellcolor{gray!6}{0.347} & \cellcolor{gray!6}{\textbf{0.341}} & \cellcolor{gray!6}{0.59} & \cellcolor{gray!6}{0.297} & \cellcolor{gray!6}{\textbf{0.28}} & \cellcolor{gray!6}{0.609} & \cellcolor{gray!6}{0.276} & \cellcolor{gray!6}{\textbf{0.273}}\\
250 & 500 & 0.559 & 0.473 & \textbf{0.441} & 0.613 & 0.421 & \textbf{0.402} & 0.621 & 0.359 & \textbf{0.336} & 0.622 & 0.318 & \textbf{0.314}\\
\addlinespace
\cellcolor{gray!6}{500} & \cellcolor{gray!6}{10} & \cellcolor{gray!6}{0.363} & \cellcolor{gray!6}{0.142} & \cellcolor{gray!6}{\textbf{0.14}} & \cellcolor{gray!6}{0.488} & \cellcolor{gray!6}{\textbf{0.16}} & \cellcolor{gray!6}{0.169} & \cellcolor{gray!6}{0.552} & \cellcolor{gray!6}{\textbf{0.203}} & \cellcolor{gray!6}{0.218} & \cellcolor{gray!6}{0.542} & \cellcolor{gray!6}{\textbf{0.25}} & \cellcolor{gray!6}{0.271}\\
500 & 50 & 0.343 & \textbf{0.134} & 0.14 & 0.434 & \textbf{0.139} & 0.148 & 0.505 & \textbf{0.148} & 0.164 & 0.556 & \textbf{0.166} & 0.19\\
\cellcolor{gray!6}{500} & \cellcolor{gray!6}{100} & \cellcolor{gray!6}{0.358} & \cellcolor{gray!6}{0.186} & \cellcolor{gray!6}{\textbf{0.18}} & \cellcolor{gray!6}{0.427} & \cellcolor{gray!6}{\textbf{0.172}} & \cellcolor{gray!6}{0.181} & \cellcolor{gray!6}{0.503} & \cellcolor{gray!6}{\textbf{0.167}} & \cellcolor{gray!6}{0.184} & \cellcolor{gray!6}{0.535} & \cellcolor{gray!6}{\textbf{0.177}} & \cellcolor{gray!6}{0.199}\\
500 & 250 & 0.411 & 0.292 & \textbf{0.272} & 0.463 & 0.271 & \textbf{0.244} & 0.509 & 0.235 & \textbf{0.233} & 0.536 & \textbf{0.226} & 0.227\\
\cellcolor{gray!6}{500} & \cellcolor{gray!6}{500} & \cellcolor{gray!6}{0.517} & \cellcolor{gray!6}{0.407} & \cellcolor{gray!6}{\textbf{0.395}} & \cellcolor{gray!6}{0.530} & \cellcolor{gray!6}{0.354} & \cellcolor{gray!6}{\textbf{0.331}} & \cellcolor{gray!6}{0.57} & \cellcolor{gray!6}{0.318} & \cellcolor{gray!6}{\textbf{0.285}} & \cellcolor{gray!6}{0.596} & \cellcolor{gray!6}{0.275} & \cellcolor{gray!6}{\textbf{0.259}}\\
\addlinespace
1000 & 10 & 0.316 & \textbf{0.125} & 0.125 & 0.419 & \textbf{0.14} & 0.147 & 0.479 & \textbf{0.175} & 0.198 & 0.502 & \textbf{0.222} & 0.254\\
\cellcolor{gray!6}{1000} & \cellcolor{gray!6}{50} & \cellcolor{gray!6}{0.249} & \cellcolor{gray!6}{\textbf{0.115}} & \cellcolor{gray!6}{0.124} & \cellcolor{gray!6}{0.363} & \cellcolor{gray!6}{\textbf{0.118}} & \cellcolor{gray!6}{0.129} & \cellcolor{gray!6}{0.431} & \cellcolor{gray!6}{\textbf{0.125}} & \cellcolor{gray!6}{0.145} & \cellcolor{gray!6}{0.492} & \cellcolor{gray!6}{\textbf{0.142}} & \cellcolor{gray!6}{0.174}\\
1000 & 100 & 0.282 & \textbf{0.136} & 0.149 & 0.345 & \textbf{0.135} & 0.157 & 0.425 & \textbf{0.138} & 0.159 & 0.467 & \textbf{0.147} & 0.174\\
\cellcolor{gray!6}{1000} & \cellcolor{gray!6}{250} & \cellcolor{gray!6}{0.316} & \cellcolor{gray!6}{0.23} & \cellcolor{gray!6}{\textbf{0.228}} & \cellcolor{gray!6}{0.345} & \cellcolor{gray!6}{\textbf{0.202}} & \cellcolor{gray!6}{0.208} & \cellcolor{gray!6}{0.417} & \cellcolor{gray!6}{\textbf{0.187}} & \cellcolor{gray!6}{0.192} & \cellcolor{gray!6}{0.460} & \cellcolor{gray!6}{\textbf{0.18}} & \cellcolor{gray!6}{0.195}\\
1000 & 500 & 0.368 & 0.334 & \textbf{0.304} & 0.421 & 0.283 & \textbf{0.275} & 0.443 & 0.255 & \textbf{0.248} & 0.471 & 0.229 & \textbf{0.224}\\
\addlinespace
\cellcolor{gray!6}{2500} & \cellcolor{gray!6}{10} & \cellcolor{gray!6}{0.240} & \cellcolor{gray!6}{\textbf{0.107}} & \cellcolor{gray!6}{0.115} & \cellcolor{gray!6}{0.332} & \cellcolor{gray!6}{\textbf{0.122}} & \cellcolor{gray!6}{0.135} & \cellcolor{gray!6}{0.393} & \cellcolor{gray!6}{\textbf{0.143}} & \cellcolor{gray!6}{0.18} & \cellcolor{gray!6}{0.417} & \cellcolor{gray!6}{\textbf{0.178}} & \cellcolor{gray!6}{0.231}\\
2500 & 50 & 0.193 & \textbf{0.1} & 0.108 & 0.257 & \textbf{0.104} & 0.121 & 0.345 & \textbf{0.112} & 0.136 & 0.395 & \textbf{0.125} & 0.163\\
\cellcolor{gray!6}{2500} & \cellcolor{gray!6}{100} & \cellcolor{gray!6}{0.210} & \cellcolor{gray!6}{\textbf{0.112}} & \cellcolor{gray!6}{0.128} & \cellcolor{gray!6}{0.252} & \cellcolor{gray!6}{\textbf{0.112}} & \cellcolor{gray!6}{0.13} & \cellcolor{gray!6}{0.322} & \cellcolor{gray!6}{\textbf{0.117}} & \cellcolor{gray!6}{0.142} & \cellcolor{gray!6}{0.375} & \cellcolor{gray!6}{\textbf{0.127}} & \cellcolor{gray!6}{0.161}\\
2500 & 250 & 0.218 & \textbf{0.153} & 0.169 & 0.270 & \textbf{0.14} & 0.172 & 0.312 & \textbf{0.138} & 0.175 & 0.354 & \textbf{0.141} & 0.182\\
\cellcolor{gray!6}{2500} & \cellcolor{gray!6}{500} & \cellcolor{gray!6}{0.247} & \cellcolor{gray!6}{\textbf{0.201}} & \cellcolor{gray!6}{0.215} & \cellcolor{gray!6}{0.278} & \cellcolor{gray!6}{\textbf{0.189}} & \cellcolor{gray!6}{0.22} & \cellcolor{gray!6}{0.323} & \cellcolor{gray!6}{\textbf{0.175}} & \cellcolor{gray!6}{0.207} & \cellcolor{gray!6}{0.352} & \cellcolor{gray!6}{\textbf{0.17}} & \cellcolor{gray!6}{0.201}\\
\bottomrule
\textbf{\footnotesize Note:} & \multicolumn{13}{l}{\footnotesize\textbf{Lower is better}. Values are proportion of times that a Geweke test of convergence resulted in a p-value less  than 0.5.}\\
& \multicolumn{13}{l}{\footnotesize   Proportions are taken over 100 simulations times $d$ parameters at each $N$ and $K$ value. }\\
& \multicolumn{13}{l}{\footnotesize  Convergence statistics are based on 1000 MCMC samples after 2000 burnin iterations with no thinning. }\\
& \multicolumn{13}{l}{\footnotesize  All results from Bayesian models are computed from 4000 posterior samples obtained from 4 parallel MCMC chains}\\
& \multicolumn{13}{l}{\footnotesize after 2000 burn-in iterations.}
\end{tabular}}
\end{table}

\begin{table}
\caption{Rhat convergence for $\btheta$.}
\label{tab:thetarhat}
\centering
\resizebox{0.98\textwidth}{!}{%
\begin{tabular}{rr|lll|lll|lll|lll}
\toprule
N & K &  \multicolumn{3}{c|}{d=2} & \multicolumn{3}{c|}{d=3} & 
\multicolumn{3}{c|}{d=5} & \multicolumn{3}{c}{d=8} \\
\midrule
&  & IRT & IRT-M & IRT-M & IRT & IRT-M  & IRT-M & IRT & IRT-M  & IRT-M  & IRT  & IRT-M  & IRT-M \\
\multicolumn{2}{c|}{Correlated $\btheta$?} & No & No & Yes & No & No & Yes & No & No & Yes & No & No & Yes \\
\midrule
\cellcolor{gray!6}{10} & \cellcolor{gray!6}{10} & \cellcolor{gray!6}{1.110} & \cellcolor{gray!6}{1.031} & \cellcolor{gray!6}{\textbf{1.03}} & \cellcolor{gray!6}{1.074} & \cellcolor{gray!6}{1.034} & \cellcolor{gray!6}{\textbf{1.032}} & \cellcolor{gray!6}{1.037} & \cellcolor{gray!6}{1.038} & \cellcolor{gray!6}{\textbf{1.034}} & \cellcolor{gray!6}{--} & \cellcolor{gray!6}{--} & \cellcolor{gray!6}{--}\\
10 & 50 & 1.521 & 1.069 & \textbf{1.066} & 1.513 & \textbf{1.068} & 1.07 & 1.431 & 1.09 & \textbf{1.088} & -- & -- & --\\
\cellcolor{gray!6}{10} & \cellcolor{gray!6}{100} & \cellcolor{gray!6}{1.612} & \cellcolor{gray!6}{1.096} & \cellcolor{gray!6}{\textbf{1.088}} & \cellcolor{gray!6}{1.657} & \cellcolor{gray!6}{1.1} & \cellcolor{gray!6}{\textbf{1.095}} & \cellcolor{gray!6}{1.742} & \cellcolor{gray!6}{1.111} & \cellcolor{gray!6}{\textbf{1.106}} & \cellcolor{gray!6}{--} & \cellcolor{gray!6}{--} & \cellcolor{gray!6}{--}\\
10 & 250 & 1.795 & 1.131 & \textbf{1.124} & 1.893 & \textbf{1.125} & 1.131 & 2.096 & 1.145 & \textbf{1.144} & -- & -- & --\\
\cellcolor{gray!6}{10} & \cellcolor{gray!6}{500} & \cellcolor{gray!6}{1.842} & \cellcolor{gray!6}{\textbf{1.155}} & \cellcolor{gray!6}{1.163} & \cellcolor{gray!6}{2.066} & \cellcolor{gray!6}{1.154} & \cellcolor{gray!6}{\textbf{1.151}} & \cellcolor{gray!6}{2.336} & \cellcolor{gray!6}{1.157} & \cellcolor{gray!6}{\textbf{1.156}} & \cellcolor{gray!6}{--} & \cellcolor{gray!6}{--} & \cellcolor{gray!6}{--}\\
\addlinespace
50 & 10 & 1.254 & 1.019 & \textbf{1.018} & 1.216 & 1.021 & \textbf{1.02} & 1.147 & 1.027 & \textbf{1.026} & 1.087 & 1.03 & \textbf{1.028}\\
\cellcolor{gray!6}{50} & \cellcolor{gray!6}{50} & \cellcolor{gray!6}{1.627} & \cellcolor{gray!6}{\textbf{1.028}} & \cellcolor{gray!6}{1.028} & \cellcolor{gray!6}{1.639} & \cellcolor{gray!6}{\textbf{1.029}} & \cellcolor{gray!6}{1.03} & \cellcolor{gray!6}{1.589} & \cellcolor{gray!6}{\textbf{1.036}} & \cellcolor{gray!6}{1.037} & \cellcolor{gray!6}{1.507} & \cellcolor{gray!6}{\textbf{1.051}} & \cellcolor{gray!6}{1.051}\\
50 & 100 & 1.671 & \textbf{1.045} & 1.046 & 1.760 & 1.043 & \textbf{1.041} & 1.800 & \textbf{1.047} & 1.047 & 1.750 & 1.064 & \textbf{1.062}\\
\cellcolor{gray!6}{50} & \cellcolor{gray!6}{250} & \cellcolor{gray!6}{1.799} & \cellcolor{gray!6}{1.089} & \cellcolor{gray!6}{\textbf{1.083}} & \cellcolor{gray!6}{1.920} & \cellcolor{gray!6}{\textbf{1.071}} & \cellcolor{gray!6}{1.072} & \cellcolor{gray!6}{2.014} & \cellcolor{gray!6}{1.074} & \cellcolor{gray!6}{\textbf{1.069}} & \cellcolor{gray!6}{2.021} & \cellcolor{gray!6}{\textbf{1.085}} & \cellcolor{gray!6}{1.092}\\
50 & 500 & 1.994 & \textbf{1.123} & 1.123 & 2.102 & 1.087 & \textbf{1.084} & 2.175 & 1.083 & \textbf{1.082} & 2.195 & \textbf{1.103} & 1.112\\
\addlinespace
\cellcolor{gray!6}{100} & \cellcolor{gray!6}{10} & \cellcolor{gray!6}{1.313} & \cellcolor{gray!6}{\textbf{1.013}} & \cellcolor{gray!6}{1.013} & \cellcolor{gray!6}{1.275} & \cellcolor{gray!6}{\textbf{1.015}} & \cellcolor{gray!6}{1.015} & \cellcolor{gray!6}{1.208} & \cellcolor{gray!6}{1.022} & \cellcolor{gray!6}{\textbf{1.021}} & \cellcolor{gray!6}{1.137} & \cellcolor{gray!6}{\textbf{1.024}} & \cellcolor{gray!6}{1.024}\\
100 & 50 & 1.635 & \textbf{1.018} & 1.019 & 1.672 & \textbf{1.019} & 1.019 & 1.618 & \textbf{1.022} & 1.023 & 1.567 & \textbf{1.031} & 1.033\\
\cellcolor{gray!6}{100} & \cellcolor{gray!6}{100} & \cellcolor{gray!6}{1.742} & \cellcolor{gray!6}{1.032} & \cellcolor{gray!6}{\textbf{1.03}} & \cellcolor{gray!6}{1.816} & \cellcolor{gray!6}{\textbf{1.028}} & \cellcolor{gray!6}{1.029} & \cellcolor{gray!6}{1.849} & \cellcolor{gray!6}{\textbf{1.031}} & \cellcolor{gray!6}{1.031} & \cellcolor{gray!6}{1.802} & \cellcolor{gray!6}{\textbf{1.039}} & \cellcolor{gray!6}{1.039}\\
100 & 250 & 2.031 & \textbf{1.064} & 1.064 & 2.000 & 1.053 & \textbf{1.052} & 2.057 & 1.049 & \textbf{1.048} & 2.079 & \textbf{1.054} & 1.054\\
\cellcolor{gray!6}{100} & \cellcolor{gray!6}{500} & \cellcolor{gray!6}{2.079} & \cellcolor{gray!6}{1.109} & \cellcolor{gray!6}{\textbf{1.103}} & \cellcolor{gray!6}{2.158} & \cellcolor{gray!6}{1.085} & \cellcolor{gray!6}{\textbf{1.079}} & \cellcolor{gray!6}{2.250} & \cellcolor{gray!6}{\textbf{1.062}} & \cellcolor{gray!6}{1.062} & \cellcolor{gray!6}{2.253} & \cellcolor{gray!6}{\textbf{1.066}} & \cellcolor{gray!6}{1.067}\\
\addlinespace
250 & 10 & 1.329 & \textbf{1.009} & 1.009 & 1.294 & \textbf{1.013} & 1.013 & 1.260 & \textbf{1.015} & 1.017 & 1.194 & \textbf{1.021} & 1.024\\
\cellcolor{gray!6}{250} & \cellcolor{gray!6}{50} & \cellcolor{gray!6}{1.649} & \cellcolor{gray!6}{\textbf{1.011}} & \cellcolor{gray!6}{1.011} & \cellcolor{gray!6}{1.660} & \cellcolor{gray!6}{\textbf{1.011}} & \cellcolor{gray!6}{1.012} & \cellcolor{gray!6}{1.643} & \cellcolor{gray!6}{\textbf{1.013}} & \cellcolor{gray!6}{1.015} & \cellcolor{gray!6}{1.580} & \cellcolor{gray!6}{\textbf{1.018}} & \cellcolor{gray!6}{1.02}\\
250 & 100 & 1.853 & \textbf{1.017} & 1.017 & 1.866 & \textbf{1.016} & 1.018 & 1.806 & \textbf{1.017} & 1.019 & 1.797 & \textbf{1.021} & 1.023\\
\cellcolor{gray!6}{250} & \cellcolor{gray!6}{250} & \cellcolor{gray!6}{2.065} & \cellcolor{gray!6}{1.044} & \cellcolor{gray!6}{\textbf{1.043}} & \cellcolor{gray!6}{2.074} & \cellcolor{gray!6}{1.037} & \cellcolor{gray!6}{\textbf{1.035}} & \cellcolor{gray!6}{2.147} & \cellcolor{gray!6}{1.031} & \cellcolor{gray!6}{\textbf{1.029}} & \cellcolor{gray!6}{2.130} & \cellcolor{gray!6}{1.032} & \cellcolor{gray!6}{\textbf{1.031}}\\
250 & 500 & 2.006 & 1.085 & \textbf{1.079} & 2.269 & 1.064 & \textbf{1.06} & 2.292 & 1.048 & \textbf{1.043} & 2.310 & 1.044 & \textbf{1.042}\\
\addlinespace
\cellcolor{gray!6}{500} & \cellcolor{gray!6}{10} & \cellcolor{gray!6}{1.375} & \cellcolor{gray!6}{\textbf{1.006}} & \cellcolor{gray!6}{1.007} & \cellcolor{gray!6}{1.318} & \cellcolor{gray!6}{\textbf{1.009}} & \cellcolor{gray!6}{1.009} & \cellcolor{gray!6}{1.283} & \cellcolor{gray!6}{\textbf{1.014}} & \cellcolor{gray!6}{1.016} & \cellcolor{gray!6}{1.225} & \cellcolor{gray!6}{\textbf{1.02}} & \cellcolor{gray!6}{1.023}\\
500 & 50 & 1.652 & \textbf{1.008} & 1.008 & 1.679 & \textbf{1.008} & 1.009 & 1.617 & \textbf{1.01} & 1.011 & 1.569 & \textbf{1.012} & 1.015\\
\cellcolor{gray!6}{500} & \cellcolor{gray!6}{100} & \cellcolor{gray!6}{1.772} & \cellcolor{gray!6}{\textbf{1.012}} & \cellcolor{gray!6}{1.012} & \cellcolor{gray!6}{1.844} & \cellcolor{gray!6}{\textbf{1.011}} & \cellcolor{gray!6}{1.013} & \cellcolor{gray!6}{1.806} & \cellcolor{gray!6}{\textbf{1.012}} & \cellcolor{gray!6}{1.013} & \cellcolor{gray!6}{1.765} & \cellcolor{gray!6}{\textbf{1.015}} & \cellcolor{gray!6}{1.018}\\
500 & 250 & 1.975 & 1.029 & \textbf{1.027} & 2.087 & 1.025 & \textbf{1.024} & 2.078 & 1.022 & \textbf{1.021} & 2.030 & \textbf{1.022} & 1.023\\
\cellcolor{gray!6}{500} & \cellcolor{gray!6}{500} & \cellcolor{gray!6}{2.181} & \cellcolor{gray!6}{1.056} & \cellcolor{gray!6}{\textbf{1.053}} & \cellcolor{gray!6}{2.241} & \cellcolor{gray!6}{1.047} & \cellcolor{gray!6}{\textbf{1.041}} & \cellcolor{gray!6}{2.319} & \cellcolor{gray!6}{1.038} & \cellcolor{gray!6}{\textbf{1.033}} & \cellcolor{gray!6}{2.326} & \cellcolor{gray!6}{1.033} & \cellcolor{gray!6}{\textbf{1.03}}\\
\addlinespace
1000 & 10 & 1.368 & \textbf{1.006} & 1.006 & 1.312 & \textbf{1.007} & 1.01 & 1.287 & \textbf{1.012} & 1.016 & 1.235 & \textbf{1.02} & 1.024\\
\cellcolor{gray!6}{1000} & \cellcolor{gray!6}{50} & \cellcolor{gray!6}{1.733} & \cellcolor{gray!6}{\textbf{1.007}} & \cellcolor{gray!6}{1.007} & \cellcolor{gray!6}{1.629} & \cellcolor{gray!6}{\textbf{1.007}} & \cellcolor{gray!6}{1.008} & \cellcolor{gray!6}{1.617} & \cellcolor{gray!6}{\textbf{1.008}} & \cellcolor{gray!6}{1.01} & \cellcolor{gray!6}{1.566} & \cellcolor{gray!6}{\textbf{1.01}} & \cellcolor{gray!6}{1.013}\\
1000 & 100 & 1.796 & \textbf{1.009} & 1.011 & 1.791 & \textbf{1.009} & 1.012 & 1.815 & \textbf{1.01} & 1.012 & 1.729 & \textbf{1.011} & 1.014\\
\cellcolor{gray!6}{1000} & \cellcolor{gray!6}{250} & \cellcolor{gray!6}{2.038} & \cellcolor{gray!6}{\textbf{1.019}} & \cellcolor{gray!6}{1.019} & \cellcolor{gray!6}{1.997} & \cellcolor{gray!6}{\textbf{1.016}} & \cellcolor{gray!6}{1.019} & \cellcolor{gray!6}{1.973} & \cellcolor{gray!6}{\textbf{1.015}} & \cellcolor{gray!6}{1.02} & \cellcolor{gray!6}{1.992} & \cellcolor{gray!6}{\textbf{1.016}} & \cellcolor{gray!6}{1.02}\\
1000 & 500 & 2.180 & \textbf{1.035} & 1.036 & 2.280 & \textbf{1.03} & 1.032 & 2.210 & \textbf{1.026} & 1.03 & 2.187 & \textbf{1.024} & 1.026\\
\addlinespace
\cellcolor{gray!6}{2500} & \cellcolor{gray!6}{10} & \cellcolor{gray!6}{1.330} & \cellcolor{gray!6}{\textbf{1.005}} & \cellcolor{gray!6}{1.006} & \cellcolor{gray!6}{1.299} & \cellcolor{gray!6}{\textbf{1.008}} & \cellcolor{gray!6}{1.008} & \cellcolor{gray!6}{1.289} & \cellcolor{gray!6}{\textbf{1.015}} & \cellcolor{gray!6}{1.022} & \cellcolor{gray!6}{1.245} & \cellcolor{gray!6}{\textbf{1.02}} & \cellcolor{gray!6}{1.03}\\
2500 & 50 & 1.676 & \textbf{1.006} & 1.008 & 1.619 & \textbf{1.007} & 1.011 & 1.564 & \textbf{1.008} & 1.009 & 1.541 & \textbf{1.009} & 1.012\\
\cellcolor{gray!6}{2500} & \cellcolor{gray!6}{100} & \cellcolor{gray!6}{1.770} & \cellcolor{gray!6}{\textbf{1.008}} & \cellcolor{gray!6}{1.009} & \cellcolor{gray!6}{1.784} & \cellcolor{gray!6}{\textbf{1.009}} & \cellcolor{gray!6}{1.012} & \cellcolor{gray!6}{1.740} & \cellcolor{gray!6}{\textbf{1.008}} & \cellcolor{gray!6}{1.011} & \cellcolor{gray!6}{1.714} & \cellcolor{gray!6}{\textbf{1.009}} & \cellcolor{gray!6}{1.015}\\
2500 & 250 & 1.977 & \textbf{1.012} & 1.018 & 1.944 & \textbf{1.011} & 1.021 & 1.947 & \textbf{1.011} & 1.023 & 1.901 & \textbf{1.011} & 1.02\\
\cellcolor{gray!6}{2500} & \cellcolor{gray!6}{500} & \cellcolor{gray!6}{2.090} & \cellcolor{gray!6}{\textbf{1.019}} & \cellcolor{gray!6}{1.023} & \cellcolor{gray!6}{2.169} & \cellcolor{gray!6}{\textbf{1.018}} & \cellcolor{gray!6}{1.053} & \cellcolor{gray!6}{2.180} & \cellcolor{gray!6}{\textbf{1.016}} & \cellcolor{gray!6}{1.032} & \cellcolor{gray!6}{2.110} & \cellcolor{gray!6}{\textbf{1.017}} & \cellcolor{gray!6}{1.026}\\

\bottomrule
\textbf{\footnotesize Note:} & \multicolumn{13}{l}{\footnotesize Lower is better. Best method for each $N$,$K$,$d$, in bold. Values are Rhat averaged over $N$ units, $d$ dimensions, }\\
&\multicolumn{13}{l}{\footnotesize and 50 simulations. Rhat is a statistic that outputs an adjusted autocorrelation between MCMC}\\
&\multicolumn{13}{l}{\footnotesize posterior samples. Here Rhat is computed over 10000 posterior samples. }\\
& \multicolumn{13}{l}{\footnotesize  All results from Bayesian models are computed from 4000 posterior samples obtained from 4 parallel MCMC chains}\\
& \multicolumn{13}{l}{\footnotesize after 2000 burn-in iterations.}
\end{tabular}}
\end{table}





\section{Appendix C: Roll Call Coding}

We specified our coding rules in stages. The first set of rules was written prior to examining bills. We specified five latent dimensions, chosen to be theoretically distinguishable and encompass many different policy areas. The dimensions are Defense/Security, Economic Development, Civil Rights/Social equality,  Entitlements/Redistribution/Welfare, and Socio-cultural. The second set of latent dimensions was created after having coded bills for the first set. It comprises six latent dimensions: Economic Policy, Foreign Policy, Public Distribution, Redistribution, Power, and Civil Rights. One guiding principle of the second set was to have Civil Rights as its own latent dimension, to enable clearer comparisons to DW-Nominate. For each set, in addition to analyzing the five- or six-dimensional latent space, we also combined theoretical concepts in different ways to create and analyze latent spaces with fewer dimensions. The three coding rules discussed in the paper's analysis were derived from three different such combinations. Coding rule A came from the first set of latent dimensions, while coding rules B and C came from the second set. Further discussion and justification of each set of latent dimensions, particularly the second set, can be found in the next two subsections.

Once we had specified latent dimensions, we then determined coding rules for how one would assign a value of $1,-1,$ or $0$ for each voting opportunity-latent dimension pair. Recall that a $1$ ($-1$) is assigned if larger (smaller) values of that latent dimension would theoretically predict a more likely yea (nay) vote, and a $0$ is assigned if that dimension's value would not theoretically help to predict a vote. The rules we used can be found in the next two subsections.

Finally, we coded all bills from both Congresses according to those coding rules. For the first set of dimensions we had two coders, who had both contributed to specifying the set of dimensions, independently code bills and then coordinate the final coding between them. For the second set, one coder specified the set of dimensions, while the other coded according to it. We note that our coding method leads to many bills not being used at all to compute latent dimensions. That occurs when none of our theoretically-derived dimensions are deemed to be relevant for predicting votes on that bill, implying that we effectively use fewer bills in determining latent positions than does a method such as DW-Nominate.

\subsection*{First Set}

-Defense/Security: votes that are intended to support initiatives that improve or increase the power of U.S. defense, further national security aims, improve counterterrorism capacity or improve the lives of veterans are coded as (+1). Actions that negatively affect any of these aims, weaken U.S. defense or national security, offer leniency to terrorism suspects or limit engagement for counterterrorism are coded with (-1). Votes on bills that commend, decry, or congratulate other nations’ actions (holding elections, celebrating a deceased leader, or expressing opposition to curbing of freedoms) are coded as (+1) as they presume positive engagement in the international community. 
Examples: - clerk session vote number 608 of the 109th House: HRES571 Express the Sense of the House of Representatives that the deployment of United States forces in Iraq be terminated immediately. Coded as (-1) because it negatively affects U.S. national security and defense.
-Clerk session vote number 611- HRES479:  Recognizing the 50th Anniversary of the Hungarian Revolution that began on October 23, 1956 and reaffirming the friendship between the people and governments of the United States and Hungary. Coded as (+1).

-Economic Development: votes on bills/questions that advance overall economic development by bringing in more activity, increased budget, support for small business initiatives or international trade initiatives/trade deals are coded as (+1).  Actions that decrease government spending and/or budget, decrease support for small businesses, or increase taxation on or eliminate subsidies for big business are coded as (-1).  
An example of this is clerk session vote number 232 of the 109th Senate: HR 2862 To prohibit weakening any law that provides safeguards from unfair foreign trade practices. Coded as (+1).

-Civil Rights/Social Equality: This category captures actions to protect historically oppressed or racial and ethnic minority groups, undocumented immigrants and other vulnerable groups. It also supports initiatives that improve access to basic needs. Actions that support their protections are coded as (+1) and actions that would strip them of rights or go against the protection of these groups or against the provision of basic needs are coded as (-1). 
Examples of this:
-clerk session vote number 270 in the 109th Senate: HR3010: To provide for appropriations for the Low-Income Home Energy Assistance Program. Coded as (+1)
-clerk session vote number 295 in the 109th Senate: S1932 To replace title VIII of the bill with an amendment to section 214(c) of the Immigration and Nationality Act to impose a fee on employers who hire certain non-immigrants. Coded as (-1)

-Entitlements/Redistribution/Welfare: Captures actions in which goods are distributed to the general public or a smaller vulnerable group. Actions that fund public-interest projects, that protect public health, food stamps/SNAP, support agriculture, and support to federal employees and veterans are coded as (+1). Actions that are against funding, supporting or regulating public goods and infrastructure are coded as (-1). An example is Vote 27 of the 85th Senate: HR7221, which stipulates that the appropriation for feed and seed in disaster areas can only be used by states that have matched it with a 25\% state appropriation is coded as (-1) because it denies funds to states that have not been able to match this amount.
Examples Include:
-Vote number 19 in the 85th House: HR6287 Cut unemployment compensation to federal employees. Coded as (-1).
-Vote number 130 in the 85th House: HR12065 Temporary Unemployment Compensation. Coded as (+1)

-Socio-Cultural: Captures improvements for education, research and development, public parks, community centers, coded as (+1). Actions that would reduce or eliminate funding for these projects, limit access to them, or eliminate existing programs are coded as (-1)
Examples include:
-Vote 39 in the 85th Senate: HR. 6500. FISCAL 1958 APPROPRIATIONS FOR D.C. AMENDMENT TO INCREASE FUNDS FOR TEACHING PERSONNEL IN D.C. PUBLIC SCHOOLS. Coded as (+1)
-Vote 40 in the 85th Senate: HR. 7441. FISCAL 1958 APPROPRIATIONS FOR AGRICULTURE DEPARTMENT. AMENDMENT TO ELIMINATE THE PROVISION LIMITING NATIONAL AVERAGE FOR CONSERVATION RESERVE PAYMENTS PER ACRE. Coded as (-1).

When the 5-factor collapses into the 4-factor, 3-factor and 2-factor arrangements, certain categories are grouped together (for example: Economic/Redistribution in the 4 factor or Social/Cultural/Civil Rights/Equality in the 2-factor). While there is certainly some overlap between certain categories (actions that promote redistribution of wealth to a historically oppressed group may also have some aspect of civil rights protection) these categories are coded based on the numeric values from the separate codes for each category. For example, if a bill is coded with a +1 for Economic policy but 0 for Distribution and Power, it will coded +1 based on the code for Economic Policy, not zero based on the codes for Distribution and Power. The non-zero value takes precedence in coding for all categories below the 5-factor categories. Similarly, where Civil Rights and Redistribution are both coded +1, then the combined Civil Rights/Redistribution will also be +1. If there are contradictory codes, and there is one predominant or primary category that can be assessed for the bill, the value of that category takes precedence.

\subsection*{Second Set}

6-factor
Economic Policy
    Distinguishes two general macroeconomic theories in U.S. politics -- hands on government intervention in the economy, or hands-off government in the economy. I assume that all members of Congress would be pro-economic development. The code captures ideological differences on perspectives for U.S. economic development. A (+1) indicates pro government spending, increasing the budget, increasing taxes for big businesses, in addition to pro- government intervention in businesses and the economy. A (-1) indicates actions that decrease government spending, decrease the budget, decrease taxes for big businesses, and against government intervention in business and the economy. 

Foreign Policy
    The code tries to generally capture two trends in foreign policy; however, it is not all encompassing. Perspectives on foreign policy change over time. The code attempts to capture “soft” and “hard” approaches to foreign policy. A (+1) indicates actions that are for soft U.S. intervention, diplomacy, and cooperation with other nations--e.g. Humanitarian assistance, trade regulation, and support for the United Nations. A (-1) indicates actions more representative of “hard” U.S. intervention--i.e. Against diplomatic measures, humanitarian assistance, but rather pro-military force. 
The code does not intend to capture for general actions taken to increase or decrease U.S. defense. Much like in economic policy--where I assume that members of Congress are for economic development and prosperity, and thus distinguish perspectives on economic policy--I assume that members of Congress support U.S. defense, especially at heightened periods of national security. In periods of national stability, however, there are different approaches to defense. Actions taken to decrease the defense budget, may be in effort to defund humanitarian assistance, more so than being against strong U.S. defense. General actions to increase or decrease the overall defense budget will be coded a 0, and rather captured in other categories such as Economic Policy. Actions pertaining to veterans will also be coded a 0, and rather captured in other codes such as Public Distribution. Veterans assistance does pertain to U.S. defense; however, it is not distinguished in different approaches to foreign policy. Actions that are unanimous and non-controversial in Foreign Policy--e.g. Senate votes to condemn terrorist attacks and sympathize with victims and their families--should be coded a 0, because they do not express ideological differences with respect to foreign policy. 

Public Distribution
    Captures actions in which goods, whose costs are collected from either a small group or the general public, are distributed to the general public. It does not include actions that distribute goods to a smaller group, such as the low income community. A (+1) includes actions that fund public-interest projects such as infrastructure, general public K-12 education--that is not specified as helping a smaller group such as low income, and furthering research and development. It also includes efforts to increase regulations that protect the environment and/or public health. Efforts to assist farmers and regulate agriculture--e.g. government funding to rotate crops-- is also included, because they protect long-term food supply for the general public. Actions taken to increase the wages, benefits, and work conditions for federal employees, including veterans, are also included, because they work for the public good. Disaster relief is also included, because although the disaster may be specific to one community at any given time, actions taken to increase disaster relief in one specific area prepares the nation for aid in any given part of the nation in the future, and can thus be thought of as a public good. A (-1) includes actions that are against funding, supporting, or increasing regulations for the public good. Actions related to active military personnel and military related research and technology should be coded a 0, because the distinction between different kinds of funding within defense is coded for in Foreign Policy. 

Redistribution
    Unlike Public Distribution, where the goods are distributed to the general public, Redistribution includes actions where costs are collected from a larger majority or a smaller group--e.g. the wealthy elite--and the benefits are distributed to a smaller group, especially low income or historically oppressed groups. A (+1) includes actions such as SNAP/food stamps, housing assistance for the poor, disability assistance. It also includes actions that try to increase the taxes on the small elite, and redistribute to low-income communities. A (-1) includes actions that are against redistributing goods to low income or other historically oppressed groups. It also includes actions taken to defend the wealth of the elite minority--e.g. tax breaks for the wealthy--against redistribution of wealth. 

Power
    Captures actions taken to increase or decrease federal government power. A (+1) includes actions that strengthen the power of the federal government. It includes actions that defend public interest over individual liberty, including business endeavors. A (-1) includes actions that decrease the power of the federal government. It includes actions that defend individual liberty over public interest. It does not, however, include defense of historically oppressed minorities, criminal or terrorist rights. Although these are protections of individual liberty, there are different conceptions of who is considered a “person” throughout U.S. history. The protection of historically oppressed groups is coded for in Civil Rights. Terrorist protections are accounted for in Foreign Policy and Civil Rights. Treatment of terrorists changes over time in relation to national security. 

Civil Rights
    Includes protection of historically oppressed groups. A (+1) includes actions that protect groups including but not limited to racial and ethnic minorities, women, sexual and gender minorities, indigenous communities, undocumented immigrants, criminals and terrorists. A (-1) includes actions that are against the protection of historically oppressed communities. 

4-factor
Economic Policy
    Same as in the 6-factor
Foreign Policy
    Same as in the 6-factor
Public Distribution/Power
    Encompasses all (+1) from Public Distribution and Power as a (+1), and both categories’ (-1) as (-1). Although there are differences between the two, there is reason to believe that there is overlap between both categories. If there are contradictory codes-e.g. an action that is a (-1) in Public Distribution and a (+1) in Power--the action should be coded 0. If there are combinations of 0s and values--e.g. (0) in Public Distribution and (+1) in Power-- the action should be coded with the non-zero value.

Civil Rights/Redistribution
    Encompasses all (+1) from Civil Rights and Redistribution as a (+1), and both categories’ (-1) as (-1). Although there are differences between the two, there is reason to believe that there is overlap between both categories. If there are contradictory codes-e.g. an action that is a (-1) in Civil Rights and a (+1) in Redistribution--the action should be coded 0. If there are combinations of 0s and values--e.g. (0) in Civil Rights and (+1) in Redistribution-- the action should be coded with the non-zero value.

4-factor
Economic Policy/ Power 
    Encompasses all (+1) from Economic Policy and Power as a (+1), and both categories’ (-1) as (-1). Although there are differences between the two, there is reason to believe that there is overlap between both categories. If there are contradictory codes-e.g. an action that is a (-1) in Economic Policy and a (+1) in Power--the action should be coded 0. If there are combinations of 0s and values--e.g. (0) in Economic Policy and (+1) in Power-- the action should be coded with the non-zero value.

Foreign Policy
    Same as in the 6-factor
Distribution
    Encompasses all (+1) from Public Distribution and Redistribution as a (+1), and both categories’ (-1) as (-1). Although there are differences between the two, there is reason to believe that there is overlap between both categories. If there are contradictory codes-e.g. an action that is a (-1) in Public Distribution and a (+1) in Redistribution--the action should be coded 0. If there are combinations of 0s and values--e.g. (0) in Public Distribution and (+1) in Redistribution-- the action should be coded with the non-zero value.

Civil Rights
    Same as in 6-factor

3-factor
Economic Policy 
    Same as in 6-factor

Distribution/Power/Civil Rights
    Encompasses all (+1) from Public Distribution, Redistribution, Power, and Civil Rights as a (+1), and all four categories’ (-1) as (-1). Although there are differences between the four, there is reason to believe that there is overlap between the four categories. If there are contradictory codes--e.g. an action that is a (-1) in Redistribution and a (+1) in Power--the action should be coded 0. If there are combinations of 0s and values--e.g. (0) in Redistribution and (+1) in Public Distribution and (+1) in Power-- the action should be coded with the non-zero value.

Foreign Policy
    Same as in 6-factor

3-factor
Economy/Distribution/Power
Encompasses all (+1) from Economic Policy, Public Distribution, Redistribution, and Power as a (+1), and all four categories’ (-1) as (-1). Although there are differences between the four, there is reason to believe that there is overlap between the four categories. If there are contradictory codes--e.g. an action that is a (-1) in Redistribution and a (+1) in Power--the action should be coded 0. If there are combinations of 0s and values--e.g. (0) in Redistribution and (+1) in Public Distribution and (+1) in Power-- the action should be coded with the non-zero value.

Civil Rights
    Same as in 6-factor

Foreign Policy
    Same as in 6-factor

2-factor
Economy/ Distribution/ Power
Encompasses all (+1) from Economic Policy, Public Distribution, Redistribution, and Power, as a (+1), and all four categories’ (-1) as (-1). Although there are differences between the four, there is reason to believe that there is overlap between the four categories. If there are contradictory codes--e.g. an action that is a (-1) in Redistribution and a (+1) in Power--the action should be coded 0. If there are combinations of 0s and values--e.g. (0) in Redistribution and (+1) in Public Distribution and (+1) in Power-- the action should be coded with the non-zero value.

Civil Rights
    Same as in 6-factor

2-factor 
Economy/Public Distribution/ Power
Encompasses all (+1) from Economic Policy, Public Distribution, and Power as a (+1), and all three categories’ (-1) as (-1). Although there are differences between the three, there is reason to believe that there is overlap between the three categories. If there are contradictory codes--e.g. an action that is a (-1) in Public Distribution  and a (+1) in Power--the action should be coded 0. If there are combinations of 0s and values--e.g. (0) in Economic Policy, (+1) in Public Distribution and (+1) in Power-- the action should be coded with the non-zero value.

Civil Rights/Redistribution
Encompasses all (+1) from Civil Rights and Redistribution as a (+1), and both categories’ (-1) as (-1). Although there are differences between the two, there is reason to believe that there is overlap between the two categories. If there are contradictory codes--e.g. an action that is a (-1) in Redistribution and a (+1) in Civil Rights--the action should be coded 0. If there are combinations of 0s and values--e.g. (0) in Civil Rights and (+1) in Redistribution - the action should be coded with the non-zero value.

\section{Appendix D: Results for Senate}

\begin{sidewaysfigure}[h!]
\centering

\begin{subfigure}[t]{.49\textwidth}
\centering
\includegraphics[width=\linewidth]{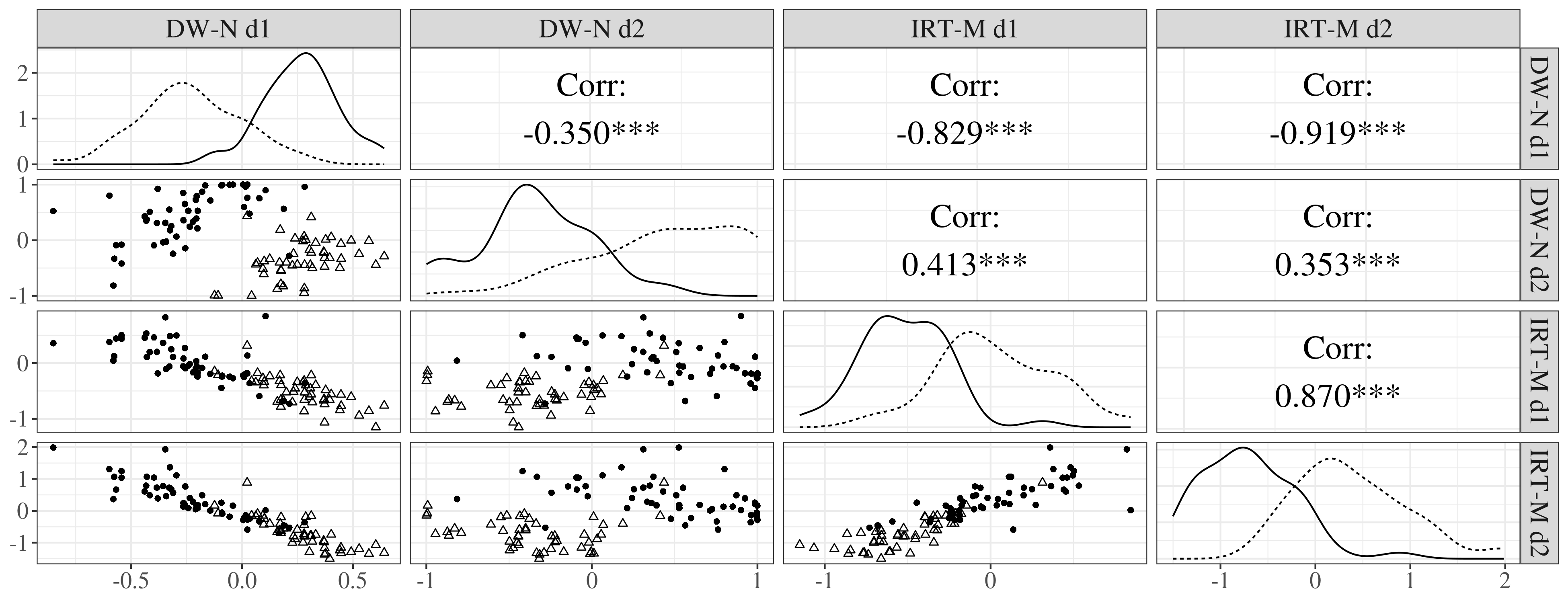}
        \caption{85$^{th}$ Senate, coding rule A}\label{figcod2:fig_a}
\end{subfigure}
\begin{subfigure}[t]{.49\textwidth}
\centering
\includegraphics[width=\linewidth]{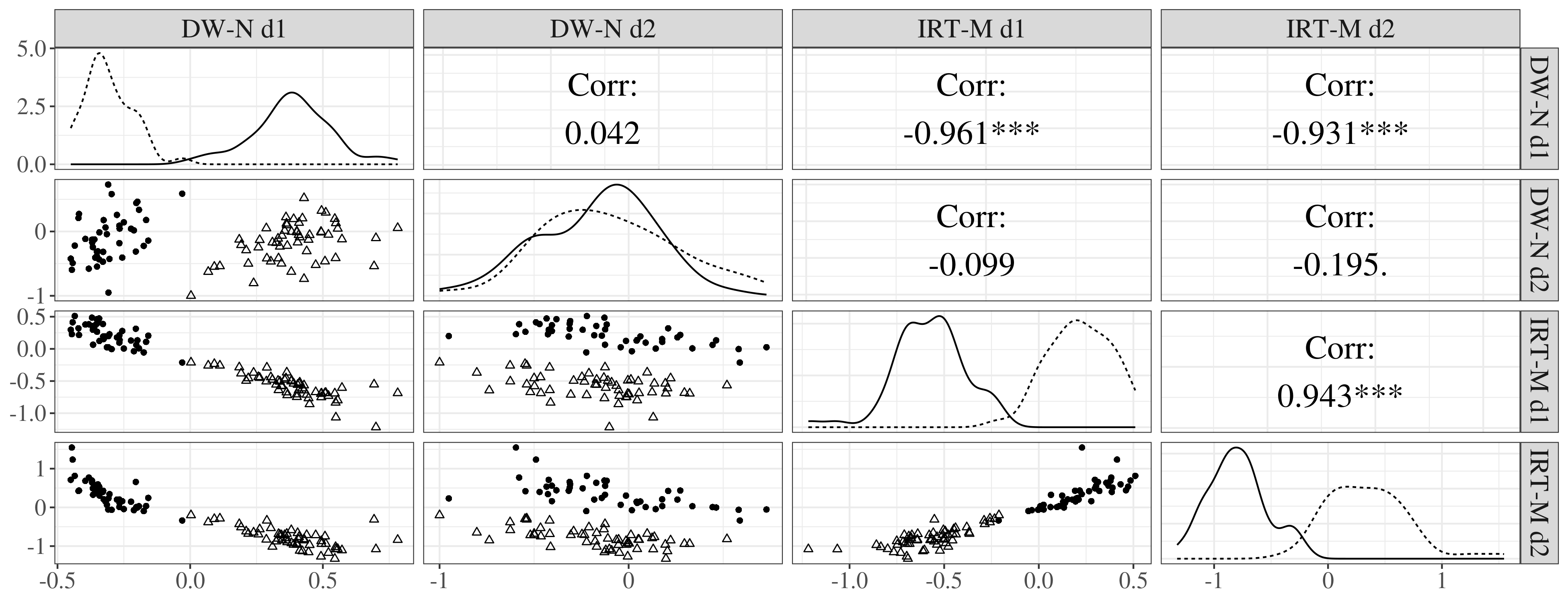}
\caption{109$^{th}$ Senate, coding rule A}\label{figcod2:fig_b}
\end{subfigure}

\medskip

\begin{subfigure}[t]{.49\textwidth}
\centering
\vspace{0pt}
\includegraphics[width=\linewidth]{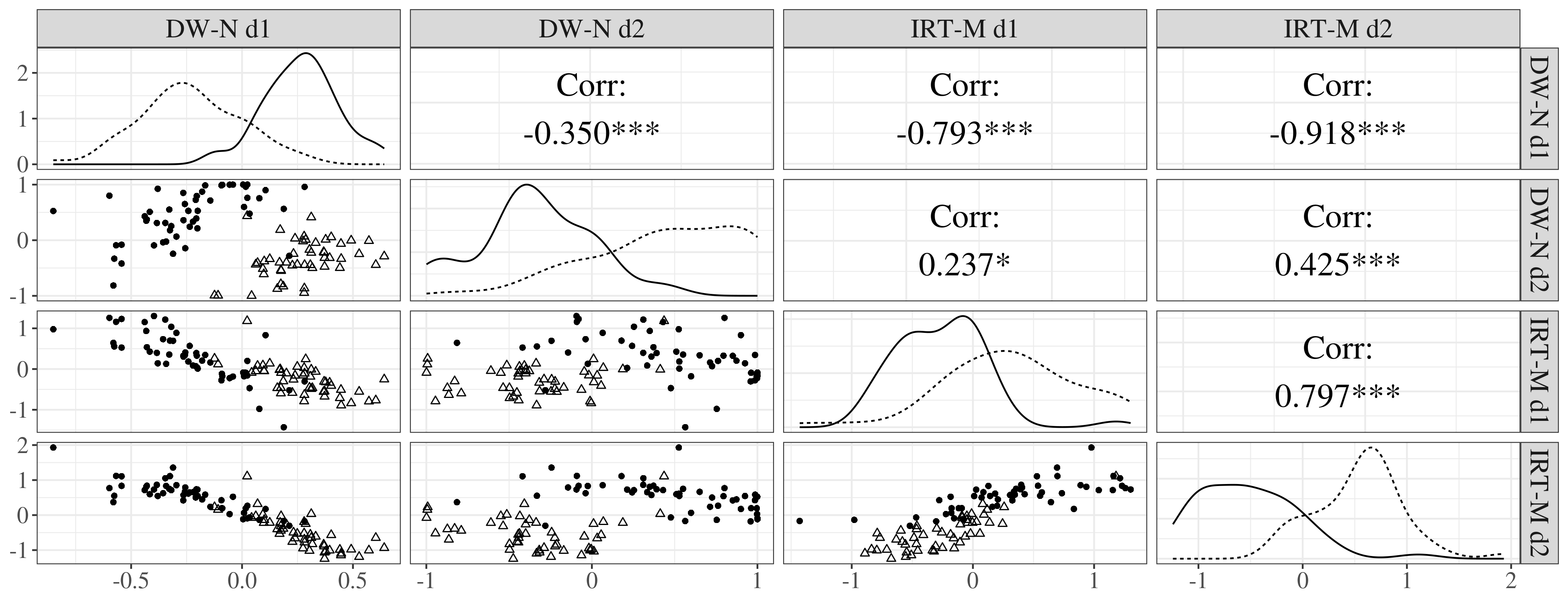}
\caption{85$^{th}$ Senate, coding  rule B}\label{figcod2:fig_c}
\end{subfigure}
\begin{subfigure}[t]{.49\textwidth}
\centering
\vspace{0pt}
\includegraphics[width=\linewidth]{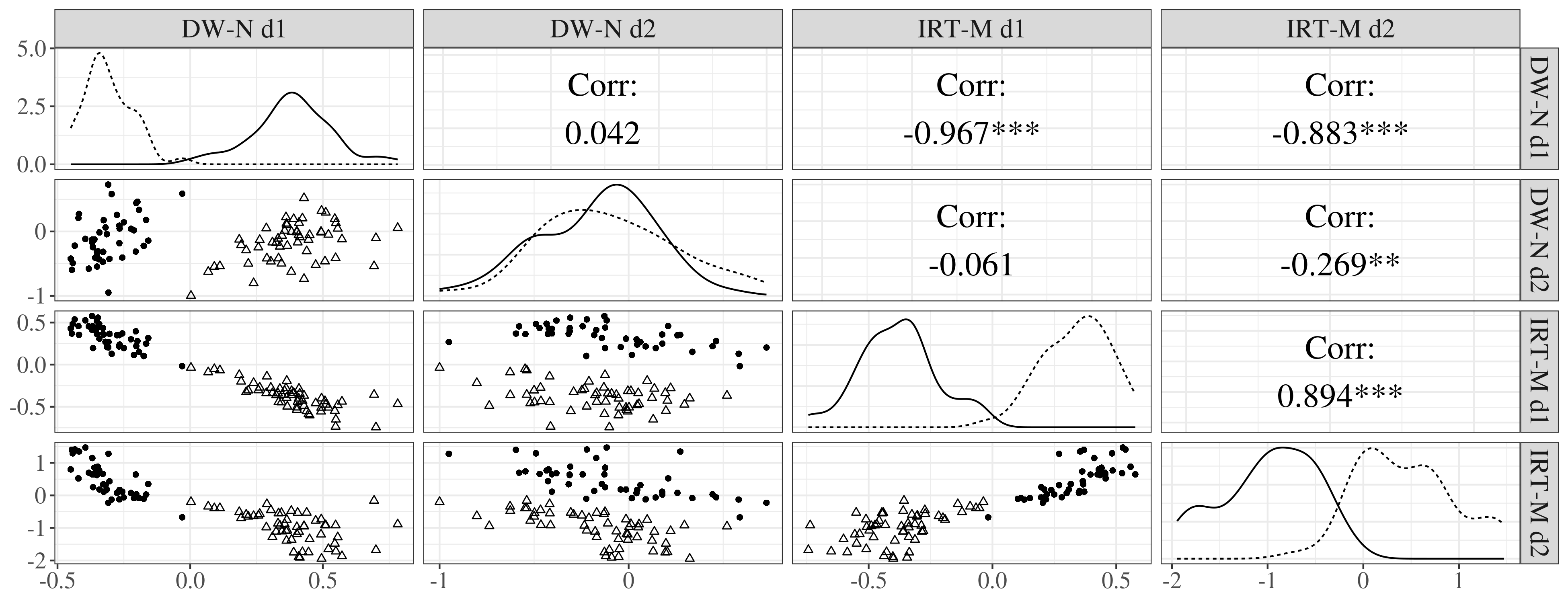}
\caption{109$^{th}$ Senate, coding rule B}\label{figcod2:fig_d}
\end{subfigure}

\medskip

\begin{subfigure}[t]{.49\textwidth}
\centering
\vspace{0pt}
\includegraphics[width=\linewidth]{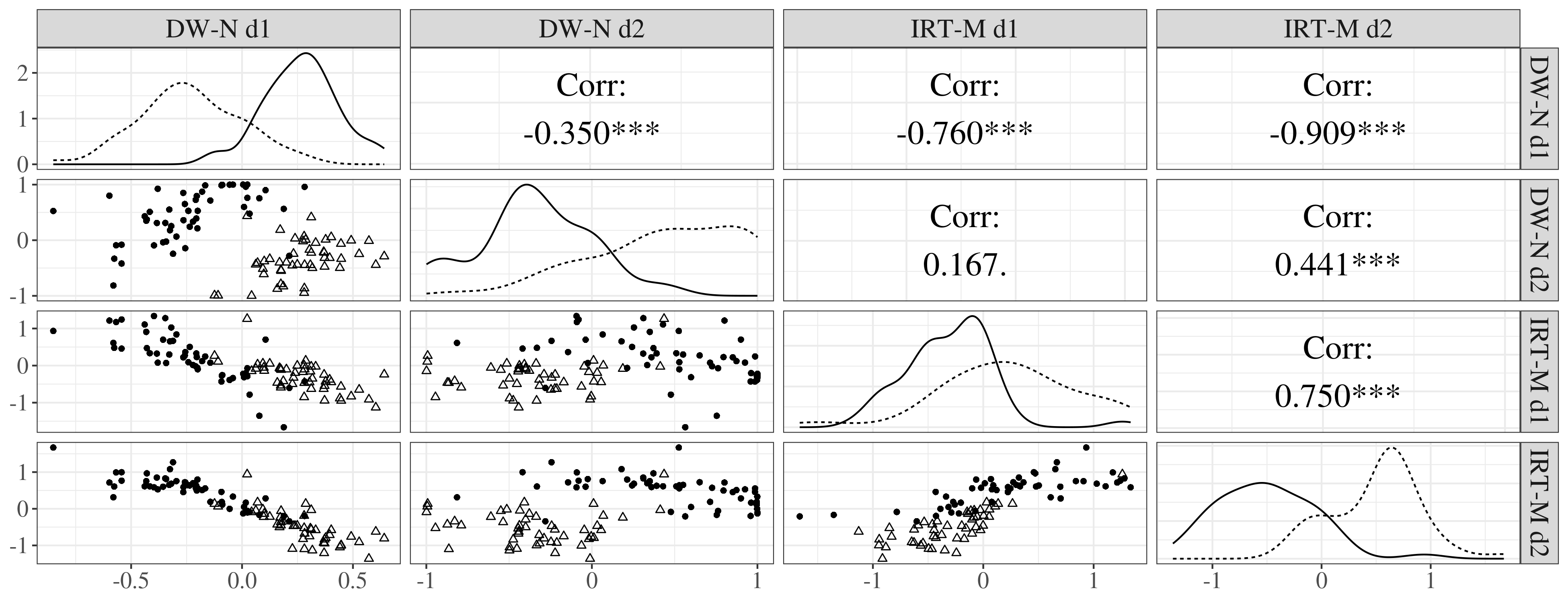}
\caption{85$^{th}$ Senate, coding rule C}\label{figcod2:fig_e}
\end{subfigure}
\begin{subfigure}[t]{.49\textwidth}
\centering
\vspace{0pt}
\includegraphics[width=\linewidth]{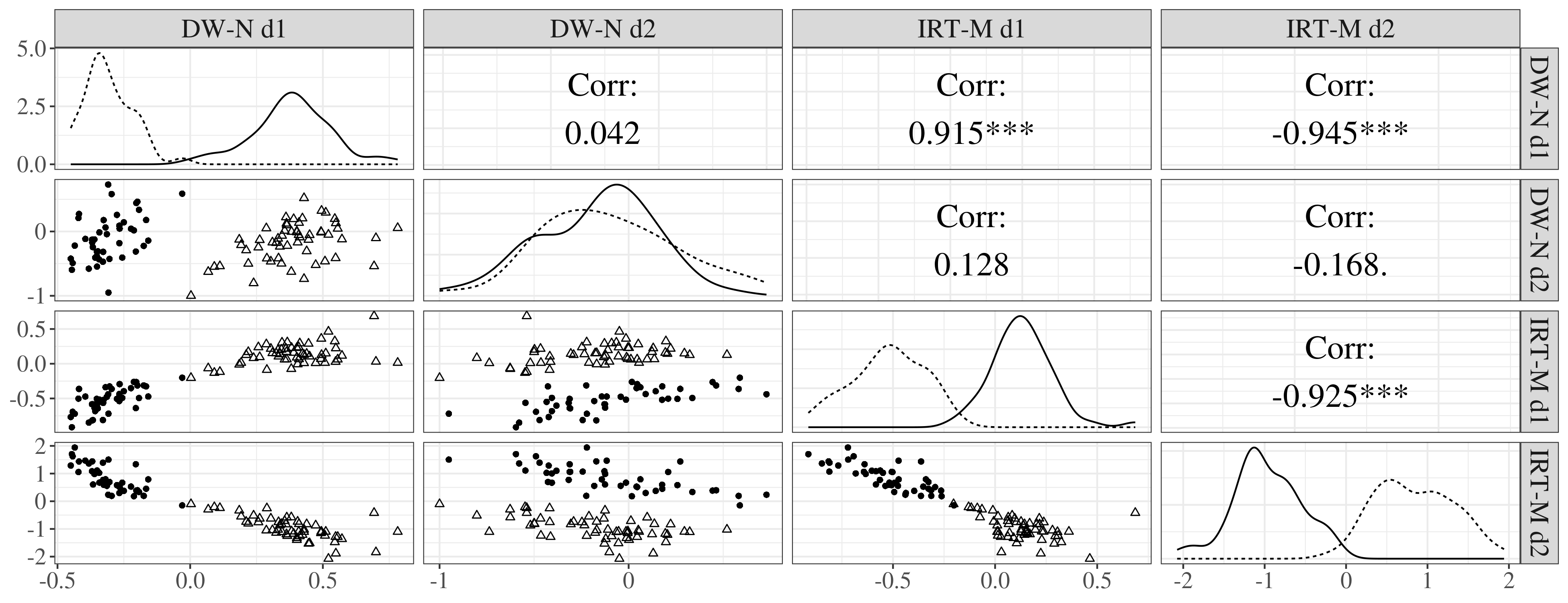}
\caption{109$^{th}$ Senate, coding rule C}\label{figcod2:fig_f}
\end{subfigure}

\footnotesize{\textbf{Note:} Each row/column within each subfigure is one of the latent dimensions estimated either by Nominate or IRT-M. The bottom triangle of each subfigure displays scatterplots with each pair of dimensions on each axis. The diagonal contains density plots for each pair of dimensions. The top triangle contains Spearman correlation coefficients for each pair of dimensions.}

\caption{Correlations between IRT-M and DW-NOMINATE ideal points in the Senate}
\label{fig:nomcorrS}
\end{sidewaysfigure}

\end{document}